\documentclass[12pt,aps,prd,plain,superscriptaddress,showpacs,preprint]{report}

\usepackage{graphicx}
\usepackage{epstopdf}
\usepackage{amsmath}
\usepackage{amsfonts} 
\usepackage{amssymb}
\usepackage{latexsym}
\usepackage{hyperref}
\usepackage{geometry}
\usepackage[T1]{fontenc}
\usepackage[english]{babel}
\usepackage[utf8]{inputenc}
\usepackage{xcolor}
\usepackage{bm}
\usepackage{float}

\begin{document}

\title{Investigations on Effective Electromagnetic and Gravitational Scenarios \\ \vspace{1.5cm} \Large Doctoral Thesis}

\author{Marcelo Granzotto Campos \and Supervisor: Prof. José Abdalla Helayël-Neto \and  Supervisor: Prof. Leonardo Ospedal Prestes Rosas \and Centro Brasileiro de Pesquisas Físicas - CBPF }
\date{June 2022}

\maketitle

\begin{figure}
    \vspace{-3.4cm}
    \hspace{-2.5cm}
    \includegraphics[width=200mm]{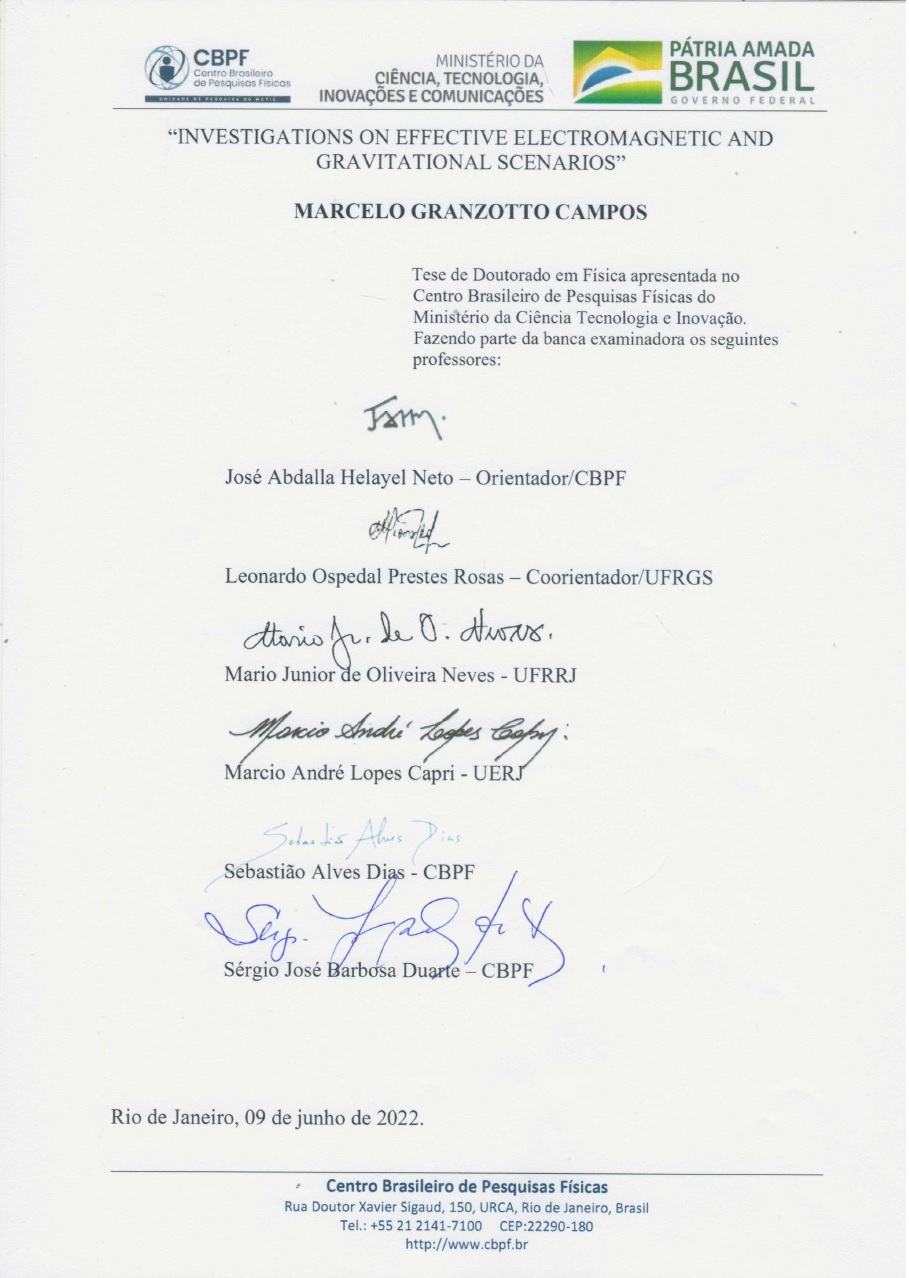}
\end{figure}


\newpage


\chapter*{\begin{center} Agradecimentos \end{center}}
\addcontentsline{toc}{chapter}{Agradecimentos}
\indent

À Amada/Ao Amado Ser, Consiência, Absoluto, que é amor, que é verdade, que tudo me deu e a todas, todos e tudo habita, agradeço pela experiência vivenciada, pelas pessoas que trouxeste até mim ao longo desse caminho, pelas situações proporcionadas, pelos aprendizados de toda ordem realizados, por tudo que ouvi, vi e saboreei, por cada momento, cada passo e cada respiração.  

Minha amada mãe Maristela, obrigado pelo amor e carinho presentes em cada instante. Suas palavras são emanações de Deus que se manifestam através do seu ser, pois provêm do fundo do seu coração. Suas palavras me fortaleceram durante esses oito anos de jornada, sempre me incentivaram e me reconfortaram.

Meu amado pai José Antônio, com o seu amor e esforço, o senhor se dedicou durante sua vida a assegurar um ensino libertador para mim. Veja-se como grande responsável pelo alcançado pelo seu filho, que se materializa através das linhas escritas nessa Tese.

Minha amada tia Dorita, que transcendeu desse plano recentemente, obrigado pelo carinho e amor com os quais, em incontáveis vezes, nos recebeu em sua casa e nos proporcionou momentos tão agradáveis, felizes e acolhedores.

Meu amado tio Orestes e minhas amadas primas-irmãs Andrea, Patrícia e Fernanda, juntos à tia Dorita, vocês estão em memórias repletas de felicidades e amor no meu coração. Tudo o que vivenciamos foi envolvido por esses sentimentos divinos, ainda que em circunstâncias difíceis, essas foram suavizadas pelo amor e pelo carinho. Obrigado a vocês. Amada Fernanda, obrigado pelo generoso auxílio com a tradução para a língua inglesa.

Minha amada Tatiana, você caminhou ao meu lado por toda essa estrada, compartilhou e conviveu cada curva, vale e cume. Por meio do amor e carinho você me apoiou e renovou, ajudando-me a seguir em frente. Agradeço a você por tanto e à/ao Ser por cruzar nossas existências nessa passagem terrena.

Amados Professores, Orientadores e amigos Helayël e Leo, obrigado por alicerça\-rem as bases dessa realização. Professor Helayël, o senhor foi a luz na minha trajetória do Doutorado. Com toda a sua suavidade, paciência, bondade e empatia abraçou a minha vida acadêmica desde o primeiro dia que fui ao CBPF conversar com o senhor para pleitear uma vaga sob sua preciosa Orientação, antes mesmo de ter ingressado na instituição. Ser seu orientando e aluno e receber seu valioso conhecimento científico e de vida é incalculável. Obrigado por tudo o que o senhor me proporcionou dentro e fora de sala de aula. Professor Leo, sua dedicação ao aluno e o seu tato para abordar as diversas situações que se interpõem são tesouros que você carrega em seu coração. Obrigado por cada diálogo, ensinamento, ajuda e paciência empreendidos nesses anos e pela amizade desenvolvida.

Aos amados Professores Accioly e Mário Junior, agradeço-lhes pelas contribuições construtivas e enriquecedoras agregadas ao projeto de Doutorado.

Ao amado Professor Sebastião, obrigado por nos enriquecer academicamente e enquanto pessoas através dos cursos ministrados e das conversas amigáveis sobre temas científicos e sociais.

Aos meus amados amigos de décadas (ou seriam séculos?) Bruno Homem, Daniel Prieto, Luiz Direito, Rafael Chapinoti, Raphael Cadete, Ricardo Brunet, Ruben Leon e Victor Fonseca, que são presentes ininterruptamente no meu coração e constituem o refúgio no qual atraco com segurança.

Às amadas servidoras e aos amados servidores do CBPF, através do Fabiano como o representante do serviço de guarda, da dona Maria como a representante pelo serviço de limpeza, do Francisco Leonardo pelo pessoal de apoio, da Bete e Cláudia Vinhas pelo administrativo e secretariado, a cada trabalhador não citado diretamente,  obrigado por se dedicarem ao pleno funcionamento da instituição. Obrigado pelas portas abertas, por nos receberem, por limparem os espaços, por assegurarem a iluminação, por suprirem os materiais utilizados, por realizarem as devidas manutenções nas instalações e equipamentos, pelos serviços administrativos, por cada hora de trabalho entregue, de modo a possibilitar o desenvolvimento da ciência em nosso país.

Obrigado às amadas brasileiras e aos amados brasileiros que me transportaram, pavimentaram as ruas e cimentaram as calçadas por onde caminhei, que limparam esses caminhos, que plantaram os alimentos que me nutriram, que cozinharam, que construíram a moradia e demais instalações que estive, que extraíram os materiais em sua forma bruta, processaram eles, armazenaram os mesmos, às e aos que costuraram as roupas vestidas, às e aos que me vacinaram, às e aos que me serviram nos postos de saúde e hospitais, às e aos porteira(o)s e demais funcionária(o)s dos prédios que habitei (Auguto, Pedro, seu Valdir, José, Valteír, David, seu Severino, Seu Gildo, Dona Maria), às e aos que garantem energia, gás e água e a toda sorte de serviço e trabalho que compuseram essa imensa interdependência social e proveram os meios para realização desse pequeno fruto científico brasileiro. Dedico esse às trabalhadoras, aos trabalhadores e aos marginalizados, pois, na verdade, delas e deles é esse trabalho. Eu apenas concluo algo que começou muito antes com elas e eles. Dessas trabalhadoras, desses trabalhadores e marginalizados, foi furtada a possibilidade de estarem em minha posição e por isso eu me comovo (e me envergonho), pois ocupo um lugar de privilégios e de exclusividades, sabotado e negado a elas e eles em grande maioria. Como posso dormir em paz se desfruto de condições que elas e eles não desfrutam e, no entanto, são igualmente merecedora(e)s e digna(o)s de tal. Saibam que fiz o melhor trabalho que pude. Espero que de algum meio ou de alguma forma isso possa se reverter em algum benefício sensível para vocês. Seguirei meu caminho servindo de coração a vocês, da melhor maneira possível.

\newpage

\chapter*{\begin{center} \Large{Abstract} \end{center}}
\addcontentsline{toc}{chapter}{Abstract}
\indent

The work aims effective and low-dimensional systems. Some different contexts involving gravitational and electromagnetic interactions are investigated. The electromagnetic one approaches bosonic and fermionic Effective Quantum Field Theories non-minimally coupled in three spacetime dimensions submitted to the expansion of Foldy-Wouthuysen Transformation, what generates (non-)relativistic corrections. A study of the effects of an external electromagnetic field derived from the Maxwell-Chern-Simons Electrodynamics on the obtained interactions are executed, as well as the impact produced by the dimensional reduction on expanded higher dimensional fermionic system in comparison to the low-dimensional one. In the scenario of gravitational effective model, scalar and fermionic particle scatterings reveal inter-particles interactions beyond monopole-monopole, leading to velocity and spin contributions, and the results are compared to a modified Electrodynamics effective model. A non-perturbative model resourcing to Casual Dynamics Triangulation data is adopted to serve as consistency check of the potentials resultants. Low-dimensional Maxwell-Higgs effective models with modified kinetic terms are studied, submitting them to a Bogomol'nyi prescription-type for calculation of inferior (non-trivial) bound energy and the self-dual equations. Vortex solutions for gauge field non-specified by an ansatz are achieved and their topological feature detailed.

\noindent
\textbf{Keywords:} low-dimensional phenomenology; non-minimum couplings; quantum gravity; vortices.

\newpage

\chapter*{\begin{center}\Large{Resumo}\end{center}}
\addcontentsline{toc}{chapter}{Resumo}
\indent

O trabalho é voltado para sistemas efetivos e de baixa dimensionalidade. Alguns diferentes contextos são investigados, incluindo Teorias de Campos Quânticos Efetivos bosônicos e fermiônicos não-minimamente acoplados em três dimensões espaco-temporais submetidos a expansão da Transformação de Foldy-Wouthuysen, o que gera correções (não-)relativísticas. Uma análise dos efeitos de um campo eletromagnético externo derivado da Eletrodinâmica de Maxwell-Chern-Simons sobre as interações obtidas é considerada, da mesma forma que os impactos produzidos por uma redução dimensional sobre o sistema fermiônico de dimensionalidade maior em comparação com o de baixa dimensionalidade. No cenário de modelo efetivo gravitacional, espalhamentos de partículas escalares e fermiônicas revelam interações entre partículas além de monopolo-monopolo, levando a contribuições de velocidade e spin, e os resultados são comparados aos de um modelo efetivo de Eletrodinâmica modificada. Um modelo não-perturbativo recorrendo a dados de Triangulação Dinâmica Causal é adotado para servir como cheque de consistência dos potenciais resultantes. Modelos efetivos de Maxwell-Higgs de baixa dimensão com termos cinéticos modificados são estudados, submetendo-os a uma prescrição tipo Bogomol`nyi para cálculo de limite mínimo (não-trivial) de energia e as equações auto-duais. Soluções de vórtices para campo de calibre não especificado através de palpite são alcançadas e as características topológicas deles detalhadas.

\noindent
\textbf{Palavras-chave:} fenômenos planares; acoplamentos não-mínimos; gravidade quântica; vórtices.

\newpage

\chapter*{\begin{center}Presentation and Contextualization\end{center}}
\addcontentsline{toc}{chapter}{Presentation and Contextualization}

\section*{\begin{center}A Tribute to Paulo Freire - One Brazilian Exponent: Educator, Intellectual and Militant\end{center}}
\addcontentsline{toc}{section}{A Tribute to Paulo Freire - One Brazilian Exponent: Educator, Intellectual and Militant}

\vspace{0.5cm}

\begin{flushright}
“Se não amo o mundo, \\
se não amo a vida, \\
se não amo os homens, \\
não me é possível o diálogo.” \\
(Paulo Freire)
\end{flushright}

\vspace{0.5cm}

The thesis starts with a brief and unassuming tribute to Paulo Reglus Neves Freire, the brilliant Brazilian Professor Paulo Freire, in his birth centenary. 
He was born in Recife, capital of Pernambuco state, in 1921, September $19^{th}$. He developed and elaborated a revolutionary methodology of education, centered in the idea of promoting the liberty of the students by the education, considering the education process as an exchanging of knowledge among teachers and students, developing a critical conscience in the students by self-inquiring and the capacity of reading the world that is around them, moving them away from the oppressed condition. In parallel, criticizing the conventional educational methodology (untitled by him as "bank account methodology")  which assumes the students as an empty box ("bank account"), inspired in the Aristotle and Locke`s philosophy of the "{\it tabula rasa}", where is deposited the information, so the students are only docile recipients of what is brought to them as ultimate truth. He used to say: "it is about to learn how to read the reality to then be able to rewrite this reality."
His endeavor to teach poor people how to read and write became a inspiration in Latin America and Africa. Paulo Freire and some mates, from 1960 to 1962, worked in the Brazilian North East region alphabetizing adults. In the famous case of Angicos, a small city in the Rio Grande do Norte state, where they alphabetized 300 persons in 40 hours, he registered:

"Three hundred persons were alphabetized in Angicos in less than 40 hours. Not only alphabetized. 300 persons were raising awareness in Angicos. Three hundred persons were learning how to read and write, and arguing Brazilian problems (...) Themes as regional and national development, base reformulation, which is among them the constitutional, nationalism, imperialism, profit remittance abroad, illiterate's vote, "colonelism", socialism (...) were debated with the participants of the Circles. We had the opportunity to watch some of those debates. Impressed me the attitude of decision they revealed (...) - "\textit{Ma'am do you know what is exploitation?} - asked certain visitor to one of the Circles participant (...) - \textit{Maybe you, whom is a rich young person} - said her - \textit {do not know. Me, who is a poor woman, know what exploitation is}." "\textit{I make shoes, and now I discover that I have the same value of the versed who makes books}." "\textit{The land only lives because the peasant works}." "\textit{The union results in force: if the designer draws the building, it is the workman who knows the brick that builds it, and both forces united that make the progress}." "  

In the begin of 1964, he was named the general coordinate of the National Plan of Alphabetization during João Goulart`s Presidency. But, few months later, with the military political coup, he was arrested for seventy days and, then, exiled. 
During the exiling in Chile he published his first book in Brazil ("Education: the Practice of the Freedom"), organized alphabetization planning for Africa countries, worked to the Movement of Agrarian Reform of the Christian Democracy and wrote in 1968 his most famous work - Pedagogy of Oppressed -, which was translated for over than 40 languages, third most cited reference in social science in the world. In 1979, he was amnestied, but returned to Brazil in 1980, affiliating to the Workers Party (Partido dos Trabalhadores) and assuming the Secretary of Education of São Paulo city from 1989 to 1991. Received 30 titles of Doctor of Honoris Causa and 6 {\it in memoriam}.    

We selected "The Concept of Technology", one excelsior Brazilian literature piece of his Master, Professor Álvaro Borges Vieira Pinto, which Paulo Freire used to refer to him as the "Brazilian Master", (in an independent translation version): 

"The role of the philosophers belonging to underdeveloped environment in the comprehension of their world, of the reasons of such conditions and in the propositions of guidelines and political and cultural actions capable of transforming the ambient reality is decisive. To do it so, however, is imperative, as initial point, to comprehend what signify being a philosopher in a poor and dependent country. The first requirement consists in to accept that it can not signify the same thing being a philosopher in a developed, imperialist and autonomous country and in one which vegetates in underdevelopment, in the ignorance of the literate knowledge and in the absence of sovereignty and capacity of determination and management of its own existence process as particular historical being. In the underdeveloped world and in the largest extension illiterate, the philosopher, to think authentically the reality, must be illiterate. It does not mean, evidently, to ignore the ability of reading and writing - nevertheless, we well know that it is not exclusively this lack what constitutes the illiteracy - but because firstly, in the attempting of conceiving and interpreting the world in its real conditions, among them is included being a world of illiterates. One will consider odd the culture accumulation and the multiple cogitations, past and present, known through the study of books, a subsidiary source, although indispensable, to the formation of the conscience of oneself. One will have to learn much more from what sees than from what reads. The philosophic conscience will only be legitimate if explains the state of its environment, not as an external passive reflection, even truthful, but by the apprehension of the social being essence in which the thinker is part of. The philosopher must identify himself with the illiterate masses, constitute the figure apparently paradoxical of the illiterate literate, to reach the foundations which support his thinking with maximized possibilities of legitimacy."          

After this masterful piece, which refreshes in our minds and hearts the role of the philosopher, as well as it extends to professors, scientists, teachers and a vast number of positions (possibly to all one) in the society, in an underdeveloped and dependent country, we initiate the contextualization and presentation of the thesis.

\section*{\begin{center}Back to Physics\end{center}}
\addcontentsline{toc}{section}{Back to Physics}
\indent

Maybe looking to mathematics as a tool conceded by God to the humanity to describe the world in which it is inserted, the ordinary physical world, one could accept that, as any regular tool, it accomplishes the finality what was designed for, but, in parallel, has its limitations, mainly performing a task different than the one it is project to. Thereby, for example, thinking the irrational feature of the pi number, one could interpret it as result of the tool (mathematical) limitation to describe the world of ideas, once the tool is made for the non-ideal world description. Additionally, maybe the infinity and continuum are essences of the ideas Universe (or mental world), and, if so, is comprehensive to expect some deviations produced by the tool, taking into account its foundations is in a discrete counting base, and artifices like the Dirac`s delta to comply with both Nature features. Therefore, imprecision occurs in attempt to cover a wide spectra of energy with a single theory. Effective theories restricts the covered energy range, what circumvents the obstacle of parametrization of the infinity and permits a more suitable application of the mathematics. Besides, the continuum is sufficiently reproduced, specially in theories with presence of higher spacetime derivatives, which delineates more accurately this feature, decreasing the intrinsic "lack" characteristic of the discrete system of counting (what is aggravated, for instance, by the description of the interaction system by point particles exchanging a mediated point particle). In a concurrent way, theories covering undefined spacetime dimensionality or carrying high dimension, when dimensional reduced, tends to diverge from one in low dimension due to the algebra inaccuracy to manipulate the continuum and infinity. Then, a tool, intrinsically based on discrete counting, describing continuity and infinity, two inseparable properties, is vulnerable to generate divergent and imprecise theories, mainly the ones that try to cover wide ranges, say, of energy, of space, of dimensions. Thus, theories restricted in its range action are an alternative to address both Natures imposition, in a way that the inaccuracies of the tool are not sensible, or at least, observable to the practiced measurements. 
After some reflection and to divagate about Nature aspects, the present thesis works on effective and low dimension theories analysing particles interactions in some scenarios, where is investigated gravitation in semi-classical arrangement, Maxwell electrodynamics and extensions. 

The chapter \ref{cap_FW} opens the debate about the gratifying work \cite{Ospedal_MPA} developed with Leonardo P.R. Ospedal, which is centred in effective low-dimensional systems in bosonic and fermionic context, non-minimally coupled. The expectation is, through a Foldy-Wouthuysen Transformation, to unfold these systems in Hamiltonian formalism and emerge (non)relativistic corrections. It is also questioned the influence of an external electromagnetic field on the interaction terms constituent of the Hamiltonian originated from the transformation, in this way, is proposed to alter the Maxwell Electrodynamics by the Maxwell-Chern-Simons one in the fermionic scenario. A second questioning is raised, in which is inquired the impacts on the interactions formulation when a Foldy-Wouthuysen transformed higher dimension system is dimensional reduced, and a comparative evaluation is carried out confronting it with the previous low-dimensional result. This chapter closes with some lines of conclusions and comments regarding to its content.

The chapters \ref{cap_grav_1} and \ref{cap_grav_2}, which are part of the work \cite{Brito_PRD} in collaboration with Gustavo P. de Brito, Judismar T. Guaitolini Jr., Leonardo P.R. Ospedal and Kim P.B. Veiga, move the effective theories inspection to the perspective of gravitational interaction. At this site, the physics is brought to the four spacetime dimensions and a gravitation model, with higher derivative terms containing form factors, is the starting point to calculation of inter-particle potentials beyond the monopole-monopole interaction, considering scalars and fermions. Then, the chapters are two parts of one body, where the chapter 2 is essentially theoretical-methodological approach presentation and the potentials resulted for scalar and fermionic scattering. In chapter 3 is discussed the latter chapter results and pushed them into some analysis. Firstly, comparing them to results of a modified effective Electrodynamics inter-particle potentials, published in the paper ref. \cite{Gustavo_PRD} of "Republic of Diracstan" collaborators and presented in the PhD Thesis \cite{Leonardo_tese}. Secondly is taking a non-perturbative model subsidized by Casual Dynamics Triangulation data \cite{Knorr_PRL} to verify the consistency of the potential results. This chapter is finalized with partial conclusions and perspectives.

The last chapter \ref{cap_vortice} is an investigation in progress, dedicated to low-dimensional Maxwell-Higgs models with modified kinetic term. The modification aggregates a strong electromagnetic background field to the systems. The Bogomol'nyi type prescription is adopted to express the inferior (non-trivial) bound energy and the self-dual equations. Hence, the results are obtained in vortex configuration, promoting an evaluation about the topological feature of them. Some effort to work with an unspecified gauge field is done, in a manner that, it is attempted to determine it naturally by the evolution of the calculations. It has been developed few variations of the models and progress have been made on it. At this stage is commented in the partial conclusions section the overall results so far and works on progress.

The general conclusions concerning the PhD program and incoming possibilities are synthesized in the Finals Conclusions topic \ref{cap_conclusao}. A Foldy-Wouthuysen Transformation review and Fourier transform integrals belonged to potential expressions in Chapters \ref{cap_grav_1} and \ref{cap_grav_2}  are in the appendixes \ref{apêndice_FWT} and \ref{apêndice_int}.


\tableofcontents

\chapter{Effective Theories and Non-minimal Couplings in Low-dimensional Systems}
\label{cap_FW}

\section{Introduction}

\subsection{A Brief Historical Review}
\indent

The low dimensional systems are active participants in the history of theoretical and experimental physics researches. Over the  years, physicists developed a brilliant group work in such arrangements, understanding its phenomenological particularities, grubbing new lands, formulating theories, discovering properties in materials and creating new ones.

It seems opportune to recapitulate some remarkable novels that compound the mentioned history. Here is recorded few moments of a period incredibly vast of fruitful research productions. There is no intention of covering minutely the theory and/or the history, even it was the case, the author has no knowledge enough to accomplish it. Sailing through the superconductors and superfluids sea, aiming to reach topological phases of matter, one returns to the second decade of twentieth century, precisely at 1911, when Heike Kamerlingh Onnes \cite{Onnes_Proc} studied the electric resistance of solid mercury at the cryogenic temperature of the liquid helium and observed that its electric property disappeared in such condition. Paul Dirac largely contributed to elucidated the conceit, interpretation and importance of symmetry to Quantum Mechanics. His perspective converges to the {\it principle of emergence} in condensed matter, which states that the properties of a material are mainly determined by how particles are organized in the material. One of his works in 1926 \cite{Dirac_PRSL1} and a later one in 1931 \cite{Dirac_PRSL2} guided the formulation of symmetry and topological perspectives, which probably influenced Landau later. In early of 1930, Lev Landau, a prominent theorist, for describing superconductivity, published his theory \cite{Landau_ZP} to describe quantizations of cyclotron orbits of charged particles, resulting in degenerated and discrete energy values, called Landau levels. Three years later, Meissner and Ochsenfeld \cite{Meiss_Die} found out that magnetic field could not deeply penetrate in lead cylinders cooled below a critical temperature, but only in an extreme thin region on surface (the London length \cite{London_PRS}), what was named the Meissner effect. London brothers in 1935 proposed a phenomenological Electrodynamics model \cite{London_PRS} to describe superconductivity based on Meissner findings and on two fluids theory of Hendrik Casimir e Cornelis Gorter \cite{Casimir_Phy}. Landau resurged in 1937 with the theory of second order phase transitions, introducing the concept of an order parameter that increases, starting from zero, for systems at the critical temperature and lower ones \cite{Landau_ZET}. He realized that the difference among the phases (or orders) is due to the fact they have different symmetries, thus a phase transition is a change of the symmetry. Before the closure of thirty´s, Pyotr Kapitza and, independently, J.F. Allen and A.D. Misener, with the papers published in sequence \cite{Kapitza_Nat,Allen_Nat}, discovered the superfluid $^4$He, in which is supposed that the transition to superfluid occurs via Bose-Einstein condensation \cite{Bose,Einstein}, and presents properties like zero viscosity.

The postulation of a phenomenological model for description of superconductivity, dividing the superconductors in two classes according to their behaviour in a magnetic field, was proposed by Vitaly Ginzburg and Lev Landau in 1950 \cite{Ginzburg_ZET}. The different phases of matter are classified by a pair of groups, which are the symmetry group of the system and the unbroken symmetry group of the equilibrium state. Through a parameter -- Ginzburg-Landau parameter --, which transforms non-trivially under symmetry transformation, the superconductors are distinguished in the ones which superconductivity and strong magnetic field coexist and the ones which not. Still this year, a German and a North American physicists, Hebert Fröhlich \cite{Frohlich_PR}  and John Bardeen \cite{Bardeen_PR}, independently elaborated the idea of the superconductor state as a result of the interaction among the electrons and the crystal atoms vibrations (phonons) -- the further interaction electron-phonon. After six years, Leon Cooper described the process of formation of electrons pairs strongly coupled to each other -- Cooper pairs -- due to their interaction with the crystals lattice of a Fermi gas \cite{Cooper_PR}. Finally, in 1957, the celebrated BCS theory \cite{BCS_PR} -- Bardeen, Cooper and Schrieffer --, a theory for microscopic superconductivity, was published and could completely explain the superconductivity resourcing to the Cooper pairs, which, with helium atoms, are spherically symmetric objects, forming isotropic superfluids on condensation. It is important to remark that, in the same period, Nikolay Bogoliubov also explained superconductivity through the Bogoliubov transformations \cite{Bogoliubov_1,Bogoliubov_2} and Philip Anderson presented his theory introducing the pseudospins \cite{Anderson_PR}. Motivated by the Ginzburg-Landau theory, still in 1957, the young soviet physicist A. A. Abrikosov predicted vortices forming a lattice in superconductors under strong magnetic field and their core overlapping, what suppresses the order parameter allover the superconductor material \cite{Abrikosov} and, hence, allows the coexistence of the field and superconductivity. Therefore, covering an non-previewed situation by the Meissner effect and confirming the new type of superconductor, the anisotropic one named type-II, already appeared in Ginzburg-Landau theory and non-contemplated by BCS theory. Later, it was showed by Gor'kov \cite{Gorkov_JETP} that the Ginzburg-Landau equations are passive of being deduced in an appropriated limit from BCS theory. The invention of MOSFET -- metal-oxide-semiconductor field-effect transistor -- in 1959 permitted closer analysis of the electrons behavior and quantum effects in a nearly ideal two-dimensional space gas, once conducted electrons travel in a thin surface layer.

In the begin of sixties, the graduating student Brian Josephson, inspired by Giaever tunnel experiments \cite{Giaver_PRL}, predicted new phenomena in superconductors -- the Josephson effects -- \cite{Josephson_PL}, for instance, a Cooper pair tunneling through a thin insulating barrier.

Ten years later (1972), the anisotropic superfluidity in $^3$He was discovered by David Lee, Douglas Osheroff and Robert Richardson \cite{David_PRL}. The condensation of Cooper pairs formed by $^3$He atoms are "quasiparticles" in a way that it is an emergent phenomenon involving many particles and oppositely to what occurs in $^4$He atoms, $^3$He generates fermions. A breakthrough in anisotropic superfluid theory was revealed by Anthony Leggett in that year \cite{Leggett_PRL} (see also \cite{Leggett_RMP} for reviewing), when he demonstrated that several breaks of symmetries can manifest simultaneously in condensed matter. In fact, the pairs in $^3$He are organized in spin-triplet states what incurs in broken of the rotational symmetry in spin space, equivalently, the spontaneous symmetry break of orbital rotation is provoked by the anisotropy of the Cooper pairs wave functions and the gauge symmetry is broken as well, then, three symmetries are broken in this superfluid. Three years passed to the Japanese researches Tsuneya Ando, Yukio Matsumoto and Yasutada Uemura predicted the quantization of the Hall conductance in planar system \cite{Ando_JPSJ}. Another Japanese team of researches, Jun-ichi Wakabayashi and Shinji Kawaji, observed the quantum Hall effect in MOSFETs \cite{Kawaji_JPSJ}. In three spacetime dimensions, theoretical particles may not be submitted to Fermi-Dirac or Bose-Einstein statistics, but to both at the same time, continuously intermingling between them as Jon M. Leinaas and Jan Myrheim detailed in \cite{Leinaas_NC}.

In the beginning of 1980, Klaus von Klitzing, working with samples of silicon-based MOSFET elaborated by Michael Pepper and Gerhard Dorda, verified the existence of exactly quantized Hall resistance, rising the current integer quantum Hall effect \cite{Klaus_PRL}. The Hall resistance manifested itself in plateaus as function of the electron and magnetic flux densities, what determine the filling factor. Once the factor assumes integer values, degenerated electron energy levels or Landau levels are formed in a two-dimensional space electron gas. One year after, Robert Laughlin proposed a mental experiment \cite{Laughlin_PRB} to explain the exactness of Hall conductance based on gauge invariance. Disposing of lower temperatures and magnetic fields of higher insensitivity, Störmer and Tsui, manipulating gallium arsenide heterostructures developed by Arthur Gossard, observed a Hall resistance plateau three times higher than the highest measured by Klitzing and they noted fractional values for the physical property. In this sense, in 1982 the fractional quantum Hall effect in a quantum fluid of electrons in three spacetime dimensions arrangement was observed \cite{Tsiu_PRL}. Frank Wilczek published two papers \cite{Wilczek_PRL1,Wilczek_PRL2}, being chronologically one before and the other after the discovery of fractional quantum Hall effect, where he expatiated the fractional statistics of quasiparticle in two space dimensions and coined the term "anyon". In the next year, Laughlin presented a phenomenological explanation to the fractional quantum Hall effect \cite{Laughlin_PRL}, where fractionally charged quasiparticles are created. Arovas, Schrieffer and Wilczek explicitly deduced the statistic of quasiparticles proving that it is governed by fractional statistics and, hence, that the particles presented in such planar systems are anyons \cite{Arovas_PRL}. In 1986, Bednorz and Müller  working with a new class of ceramics of copper oxides, encountered that its electric resistance was zero at the temperature around --238°C (35.1 K) \cite{Bednorz_ZPB}, the starting point in superconductivity research in high temperatures.

In a superficial manner, the core of the theoretical and experimental unfolding was illustrated. At this stage, it is convenient to branch the historical review in anyons, quantum spin liquid and topological phases of matter.  

The hunting for (Abelian and non-Abelian) anyons expended over forty years since Leinaas and Myrheim predicted theoretically the fractional statistics particles and Störmer and Tsui achieved the superconductivity phase in heterostructures and observed fractional quantum Hall effect. The construction of adequate instrumentation for measuring anyons has demanded great endeavour. Experimental attempts involving single-particles interferometers \cite{Chamon_PRB,Law_PRB,Camino_PRL,Ofek_PNA,McClure_PRL,Willett_PRL,Nakamura_Nat1,Willett_Arx} were made, but the results contained noises from Coulomb blockage and Aharanov-Bohm interference. Current noise from fractional quasiparticles pointed out that they carry fractional charge \cite{Picciotto_Nat,Saminadayar_PRL}. For non-Abelian anyons \cite{Moore_NPB} some indirect evidences of this state were measured \cite{Banerjee_Nat,Kasahara_Nat}. In 2020, two different teams of researches announced the detection of anyons by micro-interferometers \cite{Bartolomei_Sci,Nakamura_Nat2}.    

Philip Anderson \cite{Anderson_MRB} suggested the existence of a new phase of matter, the quantum liquid (the current quantum spin liquid), with a lattice formed by a frustrated geometry of the spins disposition, what would result in a superposition of singlet pairs of spins, named resonating valence bond. This model failed, however, in 1987, moved by the Bednorz and Müller discovery \cite{Bednorz_ZPB}, he connected the resonating valence bond with Cooper pairs \cite{Anderson_Sci} and originated the concept of spin-charge separation, in which the electron is transformed in two quasiparticles: a spinon with neutral charge and spin-1/2 and a holon electrically charged and spin-0. In the same year, topology was linked to quantum spin liquid by Kalmeyer and Laughlin \cite{Kalmeyer_PRL}, introducing the ‘chiral spin liquid’ state, and by Kivelson, Rokhsar and Sethna \cite{Kivelson_PRB}. Two years after, Wen proposed the modern notion of topological order, describing it by Chern-Simons theory \cite{Wen_PRB}. Around 2000, Senthil and Fisher pointed out that spinons could present fractional quantum number \cite{Senthil_PRB,Senthil_PRL}. In 2003, Kitaev published his toric code model \cite{Kitaev_AP1}, interrelating a two-dimensional anyonic quantum system with quantum computation. Only passed three years and he developed the honeycomb model disposing of quantum spin liquid ground states and a solvable Hamiltonian \cite{Kitaev_AP2}. Recently, in 2015, the first signatures of Majorana fermions in two-dimensional materials were detected, and the results matching with Kitaev honeycomb model for quantum spin liquids \cite{Banerjee_Nat2} (see also ref. \cite{Banerjee_Sci}). Quantum spin liquid is at the forefront current research, as discussed in refs. \cite{Laurell_Nat,Yamada_Nat,Liu_PRL_2020}.

In the third branch, Kosterlitz and Thouless, in 1972, came across with the determinant role of topological defects (vortices) in two dimensional solids phase transition \cite{Kosterlitz_1,Kosterlitz_2}. Ten years later, Thouless, Kohmoto, Nightingale and den Nijs explained the quantization of the Hall conductance of electron gases in two space dimensions in the absence of external magnetic field using topological concepts \cite{Thouless_PRL}, specified by an integer topological invariant -- TKNN invariant also called the first Chern number. Since that, passed six years, Haldane \cite{Haldane_PRL} suggested in a model of graphene \cite{Novoselov_PNAS} that a quantum Hall effect -- with energy bands not following Landau levels -- may also result from breaking of time-reversal symmetry in a three spacetime dimensional system with no net magnetic flux through it. The mentioned Chern topological index of Thouless {\it et al.}  was interpreted as a topological property of bands in two dimensional insulators without time reversal symmetry. This phase of matter was named Chern insulators and took twenty five years to be observed in particular thin films by Chang {\it et al.} \cite{Chang_Sci}. 

In 2004 Murakami, Nagaosa and Zhang theoretically predicted the spin Hall insulators \cite{Murakami_PRL}, describing some classes of band insulators presenting finite spin-Hall conductivity but no charged current. After one year, Kane and Mele \cite{Kane_1,Kane_2}, also working on a graphene model, but preserving the time-reversal symmetry, proposed a new insulating phase, the further called quantum spin Hall insulator. Their model contained a $\mathcal{Z}_2$ topological index which reflects the constraints imposed on the electronic states by the time-reversal symmetry, indicating that it is a symmetry-related topological property. Independently, few months later, the Bernevig and Zhang's two-dimensional semiconductors model \cite{Bernevig_PRL} also proposed the quantized spin Hall effect. The same former duo and Hughes elaborated a two space dimensional model to produce a $\mathcal{Z}_2$ topological phase, what resulted in a prediction of generation of quantum spin Hall effect in semiconductor quantum wells \cite{Bernevig_Sci}. In less than one year, electronic transport measurements executed by König {\it et al.} \cite{Konig_Sci} confirmed that a thin layer of mercury and tellurium compound is a topological insulator. In 2007, Novoselov {\it et al.} reported the detection of integer quantum Hall effect in graphene at room temperatures \cite{Novoselov_Sci}. Haldane and Raghu \cite{Raghu_PRL} described the unidirectional electromagnetic waves propagation effect in photonic crystals analogous to quantum Hall effect. Since this description, photonic topological phases \cite{Lu_Nat} has been aroused the attention of physicists, with the capacity to provide robust unidirectional channels for light propagation \cite{Ozawa_RMP}. Photonic topological insulators \cite{Khanikaev_Nat}, similar to quantum spin Hall states, were recently proposed in planar meta-surfaces \cite{Gorlach_Nat,Honari_Nan}. See also ref. \cite{Gupta_livro} for a more detailed material about the paragraph contents.

For last, a relatively recent predicted material by Wang {\it et al.} \cite{Wang_Nat,Wang_Nan} (see also \cite{Liu_PRL,Wang_PRL,Ni_Nan,Hernandez_Nan}) is the organic topological insulator, moving into a new research area. Organic materials are recognized by their abundant electronic and condensed structures, although, their relatively weak interactions are a challenge to produce highly ordered structures for identification of this type of insulator. 

This introduction tries to minimally elucidate how wide is the research in low-dimensional systems and how deeply it composes the forefront of physics researches. The number of ramified areas under investigation increases, so that, this simple introduction is far away from encompass all the branches of low dimensionality studying.


\subsection{Foldy-Wouthuysen Transformation and \\ Non-minimal Couplings}
\indent

The Foldy-Wouthuysen (FW) Transformation, which is also called Pryce-Tani-Foldy-Wouthuysen Transformation \cite{Pryce_PRSA,FW_PR,Tani_PTP,Foldy_PR}, is detailed in appendix \ref{apêndice_FWT}. This transformation enables to diagonalize the Hamiltonian in an approximate way or, in some cases, in an exact form (check some examples in \cite{Case_PR,Eriksen_PR,Nikitin_JPA,Silenko_JMP,Silenko_PRD_2014}). 
The Foldy and Wouthuysen methodology, initially developed to set the Dirac equation in a comparable form to the Pauli Hamiltonian \cite{FW_PR}, is applied to describe particles with different spin configurations \cite{Silenko_PRD_2014,Silenko_TMP,Silenko_PRD_2013,Silenko_PRD_2013_2,Silenko_PRD_2018,Kapusta_PRD}, as well as to analyse fundamental and excited states of atomic nuclei \cite{Guo_PRC}, some interactions with gravitational and torsion fields \cite{Ryder_Shapiro,Obukhov_PRL,Accioly,Silenko_Teryaev_grav,Obukhov_PRDs}. It was also considered in scenarios with Lorentz symmetry violation \cite{Belich_LV_2015,Goncalves_2009,Goncalves_2014_2019,Xiao_PRD_2018}, accelerated frames \cite{Hehl_Ni_1990,Chowdhury_Basu_2013_2014}, topological defects \cite{Wang_PRA_87,Chowdhury_Basu_2014} and  topological insulators \cite{Rothe_NJP_2010,Dayi_Ann_2012}. There are different approaches to the FW transformation, in a way that, in the present work is adopted the non-relativistic method, explained in the Appendix \ref{apêndice_FWT}. For details on relativistic methods and their validity, see \cite{Silenko_PRA_2015} and references therein.

The FW transformation is also a methodology suitable for systems in different spacetime dimensions rather than four. In a seminal paper by Binegar \cite{Binegar}, where he structured the irreducible representations of the Poincaré group in three-dimensions, the author also contemplated the correspondent FW transformation for a free particle. Here, it is considered the effects of non-minimal electromagnetic couplings in interactions of three-dimensional systems. The inclusion of these couplings contributes, among others, to effects such as the  Aharonov-Casher phase \cite{Aharonov_Casher}, quantum Hall effect \cite{Yoshioka,Ezawa} and high-temperature superconductivity \cite{Mavromatos_Momen}. Moreover, the Chern-Simons term and non-minimal couplings in three dimensions allow spinless particles to acquire  anomalous magnetic moments \cite{Kogan_PL,Stern_PL}. Therefore, investigations on FW transformation in electromagnetic systems non-minimally coupled have potential to reveal new interactions and relativistic corrections to the scientific community and contribute to the literature. 

This chapter starts with an introductory section about non-minimal coupling, contemplating definitions and conventions. The section \ref{Sistemas de Baixa Dimensionalidade} contains the resultant interactions of scalar and fermionic effective systems in low-dimensional scenario and submitted to the Foldy-Wouthuysen transformation. It is also analysed the impacts due to substitution of the Maxwell Electrodynamics by the Maxwell-Chern-Simons one, which are structuring the external electromagnetic field in the fermionic case, and some comments around this modification and their consequences are elaborated. In section \ref{redução_dimensional} is evaluated the influence of a dimensional reduction on a fermionic higher dimensional system FW transformed when compared to the low-dimensional result of the antecedent section. This chapter, in section \ref{Conclusão_FW}, finishes with the conclusions and perspectives.


\section{Non-minimal Electromagnetic Coupling} 
\label{acoplamento não-mínimo}

\indent

The local presence of an electromagnetic field at the vicinity of a charged field naturally provokes changes in the equation structures. These changes are manifested through several theories. One manifestation emerges from the local transformations of the charged field, demanding the insertion of the Abelian gauge field in order to assure the Lagrangian invariance due to the transformation and promoting the redefinition of the regular derivative to the covariant one due to the addition of the minimal coupling (see, for instance, \cite{Ryder_livro}). In the Hamiltonian formalism \cite{Cohen_livro,Landau_livro}, one observes the modification of the canonical moment with the presence of the gauge vector potential, as well as the gauge scalar potential, both yielding from Helmholtz Theorem \cite{Fitzpatrick_livro}, representing the consequences of the external electromagnetic field presence acting on a charged particle. A third manifestation comes from the Topological Theory \cite{Nakahara_livro}, where the electric and magnetic fields are agents responsible for varying the velocity of the particle in direction and module, implying a similar behaviour of gravitation geometry. Therefore, the appearance and a possible interpretation to the minimal coupling as an affine connection gains relevance and a common origin for electromagnetism and gravity. Then, the minimal coupling in four spacetime dimensions,

\begin{equation}
  D_{\mu} = \partial_\mu + iq A_\mu \, ,
\label{derivada_covariante} 
\end{equation}

\noindent
alters the conventional derivative $\partial_\mu$ to the covariant one $D_{\mu}$, which carries the gauge potential $ A^\mu \equiv A_{\,\bf a}^{\mu} X^{\bf a}$, where $A_{\bf a}^{\mu}$ are fields associated to each generator $ X^{\bf a}$ of the group symmetry. In the present case, the Abelian one, it works with the simplest scenario of $U(1)$ group, $X^{\bf a} = 1$ and one has $A^\mu = ( \phi, \vec{A} \, )$. The coupling constant $q$ assumes the electrical charge dimension in a system settled in natural unity $c = \hbar = 1$.

Whether the system under analyses is a planar one, the dual $\widetilde{F}_\mu$ of the field strength $F_{\mu \nu} = \partial_\mu A_\nu - \partial_\nu A_\mu$ is a rank-1 tensor, bringing it to a vector condition $ \widetilde{F}_\mu \equiv \frac{1}{2} \epsilon_{\mu \nu \kappa} F^{\nu \kappa}  $, with $\epsilon_{\mu \nu \kappa}$ representing the Levi-Civita symbol obeying the convention $\epsilon^{012} = \epsilon_{012} = +1$ and the Minkowski metric $\eta_{\mu \nu} = \textrm{diag}(+,-,-)$. Hence, this particular property of the planar configuration opens the possibility to write a generalized prescription to the covariant derivative

\begin{equation}
  \mathcal{\mathfrak{D}}_{\mu} = \partial_\mu + iq A_\mu + ig \widetilde{F}_\mu \, ,
\label{derivada_covariante_não_mínima} 
\end{equation}

\noindent
with the introduction of the non-minimal coupling represented by the dual tensor $\widetilde{F}_\mu$ and the new coupling constant $g$, which performs the anomalous magnetic dipole moment and points out the feature of effective field theory, once it acquires negative mass dimension, $[g]= -1/2$, implying in a non-renomalizable theory. It is worthy to observe that the non-minimal coupling conserves the gauge invariance of the system. Throughout the development of the calculations is defined the dual vector $\widetilde{\vec{f}}_i = \epsilon_{ij} \vec{f}_j$. In this context, it is implicit $\epsilon_{ij} \equiv \epsilon_{0ij}$, where Latin indexes denote  the purely spatial sector $i,j = (1,2)$, and, consequently, $\widetilde{\vec{f}}$ is perpendicular to $\vec{f}$.

The covariant derivative eq. \eqref{derivada_covariante_não_mínima} is rewritten in the electromagnetic prescription

\begin{subequations}
  \begin{align}
  i \partial_t \, \rightarrow \, \Pi^0 \equiv i \partial_t - q \phi + g B \, , 
\label{prescrição_não_mínima_temporal}
\end{align}

\begin{align}
  \vec{p}  \, \rightarrow \, \vec{\Pi} \equiv \vec{p} - q \vec{A} + g \widetilde{\vec{E}} \, ,
\label{prescrição_não_mínima_espacial} 
\end{align}
\end{subequations}

\noindent
it already reveals the anomalous magnetic dipole moment $gB$ and the generalized canonical moment $\vec{\Pi}$. The contributions from the non-minimal coupling, contrarily from the minimal one, are manifested in terms of the electric and magnetic fields themselves. In time, remembering that, in three spacetime dimensions, the electric and magnetic fields are given by $\vec{E} = - \vec{\nabla} \phi - \partial_t \vec{A}$ and $B = \vec{\nabla} \times \vec{A} \equiv \epsilon_{ij} \partial_i \vec{A}_j$.

The non-minimal coupling shows up in a wide variety of applications in the literature. One relevant appearance is in the fractional quantum Hall effect theme, precisely the magnetic field redefinition in Jain's model for composite fermionic objects \cite{Jain_PL} is effectively described by Helayël-Neto and Paschoal \cite{Helayel_Paschoal_PL} through means of a Maxwell–Chern–Simons (MCS) gauge field model non-minimally coupled to matter. It is also possible to achieve fractional spin resourcing to this kind of coupling term, which configures an alternative to the Chern-Simons one \cite{Carrington_PRD_95,Nobre_PL,Itzhaki_PRD}. Dalmazi (and Mendonça) analyzed the influence of the non-minimal coupling of Pauli-type in static potentials for planar scalar and fermionic Electrodynamics cases (see \cite{Dalmazi_PRD,Dalmazi_JP}, and references therein). There are also  investigations applying $\widetilde{F}_\mu$ in supersymmetry \cite{Helayel_IJMP,Paschoal_etal_PLA}. Moreover, in ref. \cite{Belich_etal_EPJ}, the authors demonstrated that neutral bosonic particles acquire magnetic properties when immersed in a scenario with Lorentz symmetry violation. Additionally, a comparative evaluation demonstrates a similar behaviour among inter-particle potentials in non-commutative Maxwell-Chern-Simons Electrodynamics minimally coupled to matter and in the MCS with non-minimal Pauli interaction \cite{Ghosh_2005}. 

The panorama about non-minimal Abelian gauge coupling and, derived from it, non-minimal electromagnetic prescription articulated above paves the road to explorer its application in some particular cases involving coupling to scalar and fermionic matter fields. These cases, since they become structured in Hamiltonians, are subjected to the FW transformation, what reveals (non)relativistic corrections to the systems. In the following two sections are developed these steps for scalar and fermionic fields, respectively.

\section{Low-dimensional Systems}
\label{Sistemas de Baixa Dimensionalidade}
\subsection{Scalar Field}

\indent
The scalar field is taken firstly. It reveals a distinct particularity in this arrangement of non-minimal coupling in planar system, that is the coupling of the scalar particle to the magnetic field, probing, through a Pauli-type term, magnetic dipole moment \cite{Kogan_PL,Stern_PL}. Moreover, there is a possibility of generating a pure Chern-Simons term by spontaneous symmetry breaking of a generalized Abelian Higgs non-minimally coupled model \cite{Paul_PL,Carrington_PRD_94}. The relevance of anyons in the phenomenology is notorious and its possibility of description resourcing to scalar fields and non-minimal coupling turned attentions to this kind of structure. 

A simpler formulation is chosen: a scalar field carrying charge and matter non-minimally coupled to the gauge field, {\it i.e.}, the Klein Gordon Lagrangian (density),  

\begin{eqnarray}
  \mathcal{L}_{KG} = \left( \mathcal{\mathfrak{D}}_{\mu} \varphi \right)^\ast \, \mathcal{\mathfrak{D}}^{\mu} \varphi - m^2 \varphi^\ast \varphi \, .
\label{Klein_Gordon} 
\end{eqnarray}

\noindent
Manipulating eqs. \eqref{derivada_covariante} and \eqref{derivada_covariante_não_mínima} permits one to write the Lagrangian in terms of the covariant derivative 

\begin{eqnarray}
  \mathcal{L}_{KG} &=& \left( D_{\mu} \varphi \right)^\ast  D^{\mu} \varphi - m^2 \varphi^\ast \varphi - g \, J^\mu \widetilde{F}_\mu + g^2 \varphi^\ast \varphi \widetilde{F}_\mu^2 \, ,
\label{Klein_Gordon_aberta} 
\end{eqnarray}

\noindent
and to define the three-current

\begin{equation}
  J^\mu = i \left[ \left( D^{\mu} \varphi \right) \varphi^\ast - \left( D^{\mu} \varphi \right)^\ast \varphi \right] \, .
\label{Klein_Gordon_corrente} 
\end{equation}

\noindent
As anticipated, the coupling constant assumes the physical property of the anomalous magnetic dipole moment and the term $g \, J^\mu \widetilde{F}_\mu$ evidences the occurrence of this effect coupled to the scalar field $\varphi$. The term $g^2 \varphi^\ast \varphi \widetilde{F}_\mu^2 $ is characteristic of the scalar system. Terms proportional to quadratic and cubic powers of $ \varphi^\ast \varphi $ are suitable to compose the Lagrangian \eqref{Klein_Gordon_aberta}, nonetheless, they are out of the research scope due to the interest in terms coupled to external electromagnetic field.

The FW procedure demands the Hamiltonian of the system under investigation. To determine it for the case expressed through eq. \eqref{Klein_Gordon_aberta}, one starts from the equation of motion 

\begin{equation} 
  ( \mathcal{\mathfrak{D}}_{\mu}^2 - m^2 ) \, \varphi = 0 \, , 
\label{Klein_Gordon_equação_de_campo} 
\end{equation}

\noindent
which is passive of being written in two first-order equations with the inclusion of the auxiliary field $\chi_\mu$

\begin{subequations}
\begin{align} 
  m \chi_\mu = \mathcal{\mathfrak{D}}_{\mu} \varphi \, , 
\label{Klein_Gordon_campo_auxiliar_eq1}
\end{align}
\begin{align}
  \mathcal{\mathfrak{D}}^{\mu} \chi_\mu = m \varphi \, .
\label{Klein_Gordon_campo_auxiliar_eq2} 
\end{align}
\label{Klein_Gordon_campo_auxiliar}
\end{subequations}

\noindent
However, taking the spatial sector of eq. \eqref{Klein_Gordon_campo_auxiliar_eq1} with $\mu =i$, the constraints $\chi_i = \frac{1}{m} \Pi_i \varphi$ is manifested, remaining only $\chi_0$ and $\varphi$ as dynamical fields. Working on the temporal sector of eqs. \eqref{Klein_Gordon_campo_auxiliar}, is possible to establish the relations 

\begin{subequations}
\begin{align} 
  i \partial_t \varphi = m \chi_0 + (q \phi - gB) \varphi  \, , 
\label{Klein_Gordon_equação_de_campo_temporal}
\end{align}
\begin{align}
  i \partial_t \chi_0 = m \varphi + (q \phi - gB) \chi_0 + \frac{\vec{\Pi}^2}{m}  \varphi  \,  ,
\label{Klein_Gordon_equação_de_campo_espacial} 
\end{align}
\end{subequations}

\noindent
and recast them in a Schrödinger-like equation, $ i \partial_t \rho = H_0 \rho $, with a two-component field

\begin{equation}  
  \rho = \left( \begin{array}{c} \rho_a \\ \rho_b \end{array} \right) = \frac{1}{2} \left( \begin{array}{c} \varphi + \chi_0 \\ \varphi - \chi_0 \end{array} \right) \, .
\end{equation}

\noindent
These algebraic manipulation described above allow to set the Hamiltonian

\begin{equation}
  H_0 = \frac{\vec{\Pi}^2}{2m} \left( \mathcal{R} + \mathcal{N} \right) + m \mathcal{N} + (q \phi - gB) \mathbb{I} \, , 
\label{Klein_Gordon_Hamiltoniana} 
\end{equation}

\noindent
where $\mathbb{I}$ denotes the identity $2 \times 2$, $\mathcal{R} = i \sigma_y$ and $\mathcal{N} = \sigma_z$, with $\sigma_y$ and $\sigma_z$ being the Pauli matrices. 

Recovering the formulation presented in Appendix \ref{apêndice_FWT}, more specifically in eq. \eqref{Hamiltoniana_operadores_par_ímpar}, $H_0 = \beta m + \mathcal{E} + \mathcal{O}$, the operators $\mathcal{E}$ and $\mathcal{O}$ are  

\begin{equation}
  \mathcal{O} = \frac{\vec{\Pi}^2}{2m}  \mathcal{R} \, \; , \, \; \mathcal{E} = \frac{\vec{\Pi}^2}{2m}  \mathcal{N}  + (q \phi - gB) \mathbb{I} \, ,
\label{Klein_Gordon_operadores_par_ímpar}
\end{equation}  

\noindent
noticing that, in three spacetime dimensions, $\beta \equiv \mathcal{N}$. Thus, the diagonalization process culminates in the expression  \eqref{Hamiltoniana_diagonalizada_final} which deliveries a diagonalized Hamiltonian $H_{FW,0}$ up to order $O(1/m^3)$, 

\begin{eqnarray}
  H_{FW,0} & \approx & m \mathcal{N}  +  \mathcal{E} + \mathcal{N} \frac{\mathcal{O}^2}{2m} - \frac{1}{8m^2} [ \mathcal{O} , \, [ \mathcal{O} , \mathcal{E}] + i \dot{\mathcal{O}} ] - \mathcal{N} \frac{\mathcal{O}^4}{8m^3}\, , 
\label{Klein_Gordon_hamiltoniana_diagonalizada} 
\end{eqnarray}

\noindent
where the leading terms of the positive energy solution are

\begin{eqnarray}
  H_{FW,0}  &\approx& m   + \frac{ ( \vec{p} - q \vec{A} + g \widetilde{\vec{E}} \, )^2}{2m} + q \phi -  gB - \frac{ ( \vec{p} - q \vec{A} + g \widetilde{\vec{E}} \, )^4}{8m^3} \, .
\label{Hamiltoniana_diagonalizada_spin0} 
\end{eqnarray}

This Hamiltonian contains the familiar non-relativistic kinetic term composed by the canonical momentum $ \vec{\Pi} = \vec{p} - q \vec{A}  + g \widetilde{\vec{E}} $. It is worthy to signalize this canonical momentum is capable of generating the Aharonov-Casher and Aharonov-Bohm phases \cite{Aharonov_Casher,Aharonov_Bohm,Carrington_PRD_95}. The electric potential term $q\phi$ is also present. In the sequence, the anomalous magnetic dipole interaction $g B$ and the last term contributes to the relativistic mass correction. In a heuristic way, this mass correction emerges from the Taylor expansion of the relativistic kinetic energy $\sqrt{\vec{\Pi}^2 + m^2}$. Finally, it is highlighted that there is no term proportional to $O(1/m^2)$, which is a particularity of the spin-0 system. 

The fermionic case is developed in the next subsection, following the same steps adopt for the spin-0 system, with the presentation of the Lagrangian, structuring of the Hamiltonian and the calculation of the Foldy-Wouthuysen transformation. Furthermore, the resulted Hamiltonian in eq. \eqref{Hamiltoniana_diagonalizada_spin0} is obtained again in the next subsection. It occurs, as will be demonstrated, due to the convergence of a fermionic system into a scalar one, when the former is exposed to a Maxwell-Chern-Simons external electromagnetic field in a particular settlement.

\subsection{Fermionic Field} 
\label{campo_fermiônico}
\indent

At this stage, the fermionic field system is investigated. The presence of spin, being a quantum number, widely enriches the physical content, aggregating the group representation manifested in the Pauli matrices, new degrees of freedom and the dynamic associated to it, also the spin-statistics theory and spin peculiar properties as its unpredictability presented in Stern-Gerlach experiment (see, for instance, \cite{Sakurai_livro}), or its states rising in discrete values and so on.

Taking the massive fermionic field non-minimally coupled to an external field, represented through the Dirac Lagrangian,

\begin{equation}
  \mathcal{L}_D = i \bar{\psi} \gamma^\mu \mathcal{\mathfrak{D}}_{\mu} \psi - m \bar{\psi} \psi \, ,
\label{Dirac} 
\end{equation}

\noindent
or in an unfolded expression resourcing to the minimal covariant derivative \eqref{derivada_covariante},

\begin{equation}
  \mathcal{L}_D = i \bar{\psi} \gamma^\mu D_{\mu} \psi - m \bar{\psi} \psi - g \, j^\mu \widetilde{F}_\mu \, , 
\label{Dirac_aberta} 
\end{equation}

\noindent
with $ j^\mu = \bar{\psi} \gamma^\mu \psi $ and $\bar{\psi} = \psi^\dagger \gamma^0$. It is convenient to introduce some definitions. The Dirac representation conducts the three dimensional spacetime matrix algebra, which is defined in terms of the Pauli matrices: $\gamma^0 = \sigma_z$, $\gamma^1 = i \sigma_x$ and $\gamma^2 = i \sigma_y$, respects the Clifford algebra $\left\{ \gamma^\mu , \gamma^\nu \right\} = 2 \eta^{\mu \nu} \mathbb{I} \,$ and satisfies the identity

\begin{equation} 
  \gamma^\mu \gamma^\nu = \eta^{\mu \nu} \mathbb{I} - i \epsilon^{\mu \nu \kappa} \gamma_\kappa \, . 
\label{matriz_gamma_identidade} 
\end{equation}

\noindent 
The traditional Dirac equation is written in function of $\mathcal{\mathfrak{D}}_\mu$, 

\begin{equation}
  \left( i \gamma^\mu \mathcal{\mathfrak{D}}_{\mu} - m \right) \psi = 0 \, ,
\label{Dirac_equação_campo} 
\end{equation}

\noindent
and defining $\beta \equiv \gamma^0$ and $\vec{\alpha} \equiv \beta \vec{\gamma} $, one is able to rearrange eq. \eqref{Dirac_equação_campo} into the form $ i \partial_t \psi = H_D \, \psi $ and to determine the Dirac Hamiltonian

\begin{equation}
  H_D = m \beta  + \vec{\alpha} \cdot \vec{\Pi} + \left( q \phi - g B \right) \mathbb{I} \, .
\label{Dirac_hamiltoniana}
\end{equation}

Since the Hamiltonian is found, one is in condition of beginning the FW procedure. Then, the terms of $H_D$ are grouped in the operators $\mathcal{E}$ and $\mathcal{O}$

\begin{equation}
  H_D = m \beta  + \mathcal{E} + \mathcal{O} \, ,
\label{Dirac_hamiltoniana_operadores_par_ímpar} 
\end{equation}

\begin{equation}
  \mathcal{O} = \vec{\alpha} \cdot \vec{\Pi} \, \; , \, \; \mathcal{E} = \left( q \phi - g B \right) \mathbb{I} \, ,
\label{Dirac_operadores_par_ímpar}
\end{equation} 

\noindent
remembering the commutation and anti-commutation relationships pointed out in Appendix \ref{apêndice_FWT}, $\mathcal{E} \beta =  \beta \mathcal{E} $ and $\mathcal{O} \beta = - \beta \mathcal{O}$, respectively. Proceeding to the last stage of the diagonalization of $H_D$ -- eq. \eqref{Dirac_hamiltoniana} --, the odd and even operators specified in \eqref{Dirac_operadores_par_ímpar} are substituted in the diagonalized Hamiltonian $H_{FW}$ -- eq. \eqref{Hamiltoniana_diagonalizada_final} --

\begin{eqnarray}
  H_{FW,D} & \approx &  m \beta  +  \mathcal{E} + \beta \frac{\mathcal{O}^2}{2m} - \frac{1}{8m^2} [ \mathcal{O} , \, [\mathcal{O} ,  \mathcal{E} ]+ i \dot{\mathcal{O}} ] - \beta \frac{\mathcal{O}^4}{8m^3} \, . 
\label{Dirac_hamiltoniana_operadores_ímpar_par_diagonalizada} 
\end{eqnarray}

\noindent
At this point is convenient to observe that the structures of the diagonalized Hamiltonians in three spacetime dimensions for the scalar eq. \eqref{Klein_Gordon_hamiltoniana_diagonalizada} and fermionic eq. \eqref{Dirac_hamiltoniana_operadores_ímpar_par_diagonalizada} cases are the same (recovering that $\beta = \mathcal{N} = \sigma_{z}$), where both derived from eq. \eqref{Hamiltoniana_diagonalizada_final}. The leading terms for positive energy solutions of the Hamiltonian resulted from the substitutions are

\begin{eqnarray}
  H_{FW,D} &\approx & m + \frac{ ( \vec{p} - q \vec{A} + g \widetilde{\vec{E}} \, )^2}{2m} + q \phi - g B  - \frac{q }{2m} B - \frac{g}{2m} \vec{\nabla} \cdot \vec{E} \nonumber \\ 
  &-& \frac{iq}{8m^2} \, \vec{\nabla} \times \vec{\mathbb{E}}  \, - \frac{q}{4m^2} \, \vec{\mathbb{E}} \times ( \, \vec{p} - q \vec{A} + g \widetilde{\vec{E}} \,) \,
  - \frac{q}{8m^2} \, \vec{\nabla} \cdot \vec{\mathbb{E}} \, ,
\label{Dirac_hamiltoniana_diagonalizada} 
\end{eqnarray}

\noindent
with

\begin{equation}
  \vec{\mathbb{E}} \equiv \vec{E} + \frac{g}{q} ( \, \vec{\nabla} B + \partial_t \widetilde{\vec{E}} \, ) \, .
\label{campo_elétrico_eficaz} 
\end{equation}

\noindent

Immediately at the first glance is verified the effects of the presence of the spin generating a Hamiltonian with a physical content more robust than the spin-0 scenario in $H_{FW,0}$ (eq. \eqref{Hamiltoniana_diagonalizada_spin0}). At the first line of $H_{FW,D}$ are the nonrelativistic contributions, then, comparing them with the ones in $H_{FW,0}$ is patent the existence of two new terms in eq. \eqref{Dirac_hamiltoniana_diagonalizada}. The first from left to right is the usual Pauli term $\, -\frac{q}{2m}B \,$ representing the dipole magnetic moment of spin, in which the spin projection $S_z = \sigma_z/2$ is not evident, and the next is a Darwin-type term $\, -\frac{g}{2m} \vec{\nabla} \cdot \vec{E} \,$ coming from the non-minimal coupling \cite{Schwabl_livro}. Moving to the relativistic contributions sector at the second line of $H_{FW,D}$ and starting from left to right, one comes across the spin-orbit term with the non-minimal corrections. At last, a Darwin term concerning to a correction on the electrostatic energy $q\phi$ due to the electron fluctuation position (for detailing the physical origin of the last two terms, see ref. \cite{Itzykson_livro}). Whether $g$ is taken trivial, $g=0$, hence $\, (\vec{\mathbb{E}} \rightarrow \vec{E}) \,$, the spin-orbit components and the Darwin term assume the familiar interaction form of the four dimensions system. The results of $H_{FW,D}$ are restricted to order $O(1/m^2)$, in a way that the relativistic mass correction to the kinetic term appeared in the spin-0 system, $- ( \vec{p} - q \vec{A} + g \widetilde{\vec{E}} \, )^4/8m^3 \,$, is unrevealed in spin-1/2 Hamiltonian eq. \eqref{Dirac_hamiltoniana_diagonalizada}, once it is a contribution proceeding from the $\mathcal{O}^4$ term (of order $O(1/m^3))$ in eq. \eqref{Dirac_hamiltoniana_operadores_ímpar_par_diagonalizada}. In the coming section \ref{redução_dimensional}, a dimensional reduction is performed on a FW transformed Hamiltonian for a massive fermionic system non-minimally coupled in four spacetime dimensions and the resulting interactions are compared to the ones of $H_{FW,D}$ in eq. \eqref{Dirac_hamiltoniana_diagonalizada}.  

So far, the whole discussion works with no specification of the Electrodynamics which dictates the external electromagnetic field behaviour. For a more refined and profound examination, probing subtle and particular information about the intrinsic physics interactions, is specified the external Electrodynamics, adopting Maxwell-Chern-Simons theory through the Lagrangian

\begin{eqnarray}
  \mathcal{L}_{MCS} = - \frac{1}{4} F_{\mu \nu}^2 + \frac{\lambda}{2} \epsilon^{\mu \alpha \beta} A_\mu \partial_\alpha A_\beta - J^\mu A_\mu \, ,
\label{MCS_lagrangiana} 
\end{eqnarray}

\noindent
where $J^\mu = (\rho, \vec{J})$  is an unspecified current and $\lambda$ denotes the Chern-Simons parameter. MCS theory, in a context of topologically massive gauge theory was extensively studied by Deser, Templeton and Jackiw in ref. \cite{Deser_AP}. The equations of motion from $\mathcal{L}_{MCS}$ are

\begin{subequations}
\begin{align} 
  \vec{\nabla} \times \vec{E} = - \partial_t B \, , 
\label{MCS_equações_de_movimento_1}
\end{align}
\begin{align}
  \vec{\nabla} \cdot \vec{E}  - \lambda B = \rho \, , 
\label{MCS_equações_de_movimento_2}
\end{align}
\begin{align}
  \widetilde{\vec{\nabla}} B - \lambda \widetilde{\vec{E}} = \vec{J} + \partial_t \vec{E} \, .
\label{MCS_equações_de_movimento_3}  
\end{align}
\label{MCS_equações_de_movimento} 
\end{subequations}

\noindent
An appropriated manipulation on eq.\eqref{MCS_equações_de_movimento_3} allows to  rewrite it as $\, \vec{\nabla} B  + \partial_t \vec{\widetilde{E}} = \lambda \vec {E} - \vec{\widetilde{J}} \,$, which is in a suitable form to be substituted in eq. \eqref{campo_elétrico_eficaz}, altering it to

\begin{equation}
  \vec{\mathbb{E}} = \left( 1 + \frac{\lambda g}{q} \right) \vec{E} - \frac{g}{q} \, \widetilde{\vec{J}}  \, .
\label{campo_elétrico_eficaz_2} 
\end{equation}

\noindent
This new configuration to $\vec{\mathbb{E}}$ suggests some inspections for particular values of the constants and current contented in the equation. Therefore, initially rescuing the Maxwell theory in vacuum, $\lambda = 0$ and $\widetilde{\vec{J}} = \vec{0}$, what leads to $\vec{\mathbb{E}} \rightarrow \vec{E}$ and, consequently, is verified that the only relativistic correction by non-minimal coupling remained in eq. \eqref{Dirac_hamiltoniana_diagonalizada} is an electric density energy $\frac{q}{4m^2} g \vec{E}^{\,\, 2}$. 
Within the approximations practiced in eq. \eqref{Dirac_hamiltoniana_diagonalizada}, the non-minimal relativistic interactions are relevant only in the presence of matter, where $\vec{J} \neq \vec{0}$. Zeldovich anapole moment in four spacetime dimensions also presents a close behaviour, vanishing in absence of matter \cite{Nowakowski_EJP}. A second inspection occurs setting $\widetilde{\vec{J}} = \vec{0}$ and $\rho = 0 $, giving the Maxwell-Chern-Simons theory in vacuum. An effective form for the electric field appears $\vec{\mathbb{E}} \rightarrow \vec{E}_{\textrm{eff}} = \left( 1 + \frac{\lambda g}{q} \right) \vec{E}$ and the same is applicable to the magnetic field $B_{\textrm{eff}} = \left( 1 + \frac{\lambda g}{q} \right) B$ just managing the Pauli and Darwin-type terms, situated at the non-relativistic contributions sector of $H_{FW,D}$, with the modified Gauss equation $\vec{\nabla} \cdot \vec{E} - \lambda B = 0$ (eq. \eqref{MCS_equações_de_movimento_2}). These effective forms for the electric and magnetic fields depend on the value of the coupling constant $g$, what permits a critical value to it, namely $g_c = -q/\lambda$. In the critical condition, the relativistic corrections for $H_{FW,D}$ are null and the Pauli and Darwin-type terms cancel each other, remaining   

\begin{equation}
  H_c  \approx  m + \frac{ ( \vec{p} - q \vec{A} + g_c \widetilde{\vec{E}} \, )^2}{2m} + q \phi - g_c B \, 
\label{Hamiltoniana_crítica} 
\end{equation}

\noindent
that is the convergence of the spin-1/2 Hamiltonian to the spin-0 one (eq. \eqref{Hamiltoniana_diagonalizada_spin0}) up to order $O(1/m^2)$. The critical situation appears in different physical contexts and is demanded to achieve specific effects, for example as a special condition to acquire first order vortex solutions \cite{Torres_PRD} or disappearance of one-loop quantum corrections to the photon mass \cite{Georgelin_MPL}.

\section{Dimensional Reduction Analysis} 
\label{redução_dimensional}
\indent

At this stage is brought up a philosophical reflection about the interpretation and connection among physical systems in different dimensionalities. Upon this point, by construction, systems in low dimensions were studied, thus, they carry implicitly the conception of a planar geometry as representative of the Nature. A second proposal is to start from higher dimensions, so as to describe a system as a whole and containing the aimed physical environment. Then, executing a dimensional reduction, it creates a specie of delimited view on the original system, establishes particular physical conditions to it, approximates the higher to the low dimensional environment and brings physical information from the original one. Therefore, it seems worthy to inspect what impact a physical system originally designed in higher dimensions suffer under a dimensional reduction and, also, what information is aggregated or, even, lost with the procedure. Witten, Qi and Zhang demonstrated theoretical advantages of structuring in higher dimensions topological superconductors \cite{Witten_PRB}, while in ref. \cite{Cocuroci_EPJ} the authors presented an emergence of a dark sector through a dimensional reduction from a five dimensional Electrodynamics coupled to 3-form gauge field to four dimensional. An investigation over the transformation of the spin properties under compactification of a spatial dimension is done in ref. \cite{Silenko_PRD} and Helayël-Neto and Ospedal calculated new effects for interparticle potentials initially in five dimensions and, then, reduced to four \cite{Leo_Helayel_PRD}. Analysis of effects of boundary conditions on fermionic and bosonic dimensionally reduced models are developed by Cavalcanti \textit{et al} \cite{erich}. The listed papers are some examples of how the dimension reduction, and its philosophy inherent, is a potential theme to reflect and debate. In the current research, is consider a fermionic field non-minimally coupled to an external electromagnetic field in four dimensions, which Hamiltonian is diagonalized by FW prescription and its result, (non)relativistic interactions, are submit to a dimensional reduction rendering a three spacetime dimensions physical system. The reduced Hamiltonian is analytically compared to the one originally construct in three dimensions \eqref{Dirac_hamiltoniana_diagonalizada}. Henceforth, the indexes nomenclature is $\hat{\mu} , \hat{\nu} = (0,1,2,3)$ and the Minkowski metric $ \eta_{\hat{\mu} \hat{\nu}} = \textrm{diag}(+,-,-,-) $ for four dimensional spacetime and the three dimensional space vectors are in bold, \textit{e.g.}, the gauge potential is denoted as $ A^{\hat{\mu}} = (\phi, \vec{\bm{A}}) $.

The described fermionic Lagrangian is the Dirac one with the minimal covariant derivative and the Pauli interaction as an extra term,

\begin{equation}
  \mathcal{L}_{DP} = i \bar{\psi} \gamma^{\hat{\mu}} D_{\hat{\mu}} \psi - m \bar{\psi} \psi + \frac{f}{2} \, \bar{\psi} \sigma^{\hat{\mu} \hat{\nu}} \psi \, F_{\hat{\mu} \hat{\nu}} \,  
\label{Lagrangian_Dirac_4D} 
\end{equation}

\noindent
with $ D_{\hat{\mu}} = \partial_{\hat{\mu}} + i q A_{\hat{\mu}} $ and $ \sigma^{\hat{\mu} \hat{\nu} } = \frac{i}{2} \left[ \gamma^{\hat{\mu}}, \gamma^{\hat{\nu}} \right] $. The non-minimal coupling constant $f$, as occurred in the low dimensional cases, has negative mass dimension $[f] =-1$, signalizing that the theory should be treated as effective. The Pauli interaction, recently investigated in the context of  spin Hall effect \cite{Turcati_et_al}, is passive of being connected to the non-minimal coupling in three dimensions $j^\mu \widetilde{F}_\mu$. The resulting manipulation of the previous coupling and of the eq. \eqref{matriz_gamma_identidade} is susceptible to be written as $\frac{1}{2} \, \bar{\psi} \sigma^{\mu \nu} \psi \, F_{\mu \nu} $, representing, in three spacetime dimension, the analogous to the Pauli term. In ref. \cite{Carrington_PRD_95}, the authors worked on some comparisons in the non-relativistic limit involving these couplings, and now, is intended to extended the analyses to the relativistic interactions, connecting them though dimensional reduction.

The equation of motion of the Lagrangian \eqref{Lagrangian_Dirac_4D} is

\begin{equation}
  \left( i \gamma^{\hat{\mu}} D_{\hat{\mu}} - m + \frac{f}{2} \,  \sigma^{\hat{\mu} \hat{\nu} } F_{\hat{\mu} \hat{\nu}}\right) \psi = 0  \, ,
\label{Dirac_equações_de_movimento_4D} 
\end{equation}

\noindent
where the gamma matrices are in the Dirac representation

\begin{equation} 
\gamma^0 = \left( \begin{array}{cc} \mathbb{I}  &  0 \\ 0  &  - \mathbb{I} \end{array} \right) \, \; , \, \;  \gamma^i = \left( \begin{array}{cc} 0  &  \sigma_i \\   - \sigma_i & 0 \end{array} \right) \, . 
\label{matriz_gamma_4D} 
\end{equation}

\noindent
Reorganizing the equation of motion \eqref{Dirac_equações_de_movimento_4D} and managing some algebra, is obtained the Dirac-Pauli Hamiltonian

\begin{eqnarray}
  H_{DP} &=& \beta m + \vec{\bm{\alpha}} \cdot \vec{\bm{\pi}} + q \phi \, \bm{I} - i f \, \beta \, \vec{\bm{\alpha}} \cdot \vec{\bm{E}} + f \, \beta \, \vec{\bm{\Sigma}} \cdot \vec{\bm{B}} \, ,
\label{Dirac_hamiltoniana_4D}
\end{eqnarray}

\noindent
in which $\bm{I}$ stands for the identity $4\times 4 \,$, $\vec{\bm{\pi}} \equiv \vec{\bm{p}} - q \vec{\bm{A}}$, $\beta \equiv \gamma^0$, $\vec{\alpha} \equiv \beta \vec{\gamma} $ and $\vec{\bm{\Sigma}}$ corresponds to the spin matrix 

\begin{equation} 
  \vec{\bm{\Sigma}} = \left( \begin{array}{cc} \vec{\sigma}  &  0 \\  0 & \vec{\sigma}   \end{array} \right) \, . 
\label{matriz_spin}
\end{equation}

\noindent
Exactly as proceeded in lower dimensional systems, the resulting Hamiltonian, in this case, the Dirac-Pauli one in eq. \eqref{Dirac_hamiltoniana_4D}, is set in terms of odd $(\mathcal{O})$ and even $(\mathcal{E})$ operators ($H = \beta m + \mathcal{E} + \mathcal{O}$), which are defined as 

\begin{eqnarray}
  \mathcal{O} &=& \vec{\bm{\alpha}} \cdot \vec{\bm{\pi}} - i f \, \beta \, \vec{\bm{\alpha}} \cdot \vec{\bm{E}} \, \; \,  , \, \; \, \mathcal{E} = q \phi \, \bm{I} + f \, \beta \, \vec{\bm{\Sigma}} \cdot \vec{\bm{B}} \, . 
\label{Dirac_4D_operadores_par_ímpar}
\end{eqnarray}

\noindent
Substituting $\mathcal{O}$ and $\mathcal{E}$ in eq. \eqref{Dirac_hamiltoniana_operadores_ímpar_par_diagonalizada} and considering the four dimensional gamma matrices eq. \eqref{matriz_gamma_4D}, one is prompted to the diagonalized Hamiltonian (for positive energy solutions) 

\begin{eqnarray}
  H_{FW,DP} &&\approx  m \, \mathbb{I} + \frac{1}{2m} \, \left( \vec{\bm{\pi}} \, \mathbb{I} - f \vec{\bm{E}} \times \vec{\sigma} \right)^2 + q \phi \, \mathbb{I} + \left( f - \frac{q}{2m} \right) \vec{\sigma} \cdot \vec{\bm{B}} - \frac{f^2}{2m} \vec{\bm{E}}^2 \, \mathbb{I} \nonumber \\
  &&+ \left( \frac{f}{2m} - \frac{q}{8 m^2} \right) \vec{\nabla} \cdot \vec{\bm{E}} \, \mathbb{I} - i \, \frac{q}{8 m^2} \, \vec{\sigma} \cdot \left( \vec{\nabla} \times \vec{\bm{E}} \right) -  \, \frac{q}{4 m^2} \, \vec{\sigma} \cdot \left( \vec{\bm{E}} \times \vec{\bm{\pi}} \right) \nonumber \\
  &&- \frac{q f}{4 m^2} \, \vec{\bm{E}}^2 \, \mathbb{I} - i \, \frac{f}{8 m^2} \, \left[ \vec{\nabla} \cdot \left( \partial_t \vec{\bm{E}} \right) \right] \, \mathbb{I} + \frac{f}{8 m^2} \, \vec{\sigma} \cdot \left( \vec{\nabla} \times \partial_t \vec{\bm{E}} \right) \nonumber \\
  && + i \, \frac{ f}{4 m^2} \, \left( \vec{\nabla} \cdot \vec{\bm{B}} \right) \vec{\sigma} \cdot \vec{\bm{\pi}} + \frac{ f}{4 m^2} \, \mathbb{I} \left[ \partial_t \vec{\bm{E}} - \vec{\nabla} \times  \vec{\bm{B}} \right] \cdot \vec{\bm{\pi}}   \nonumber \\
  && + \frac{ f}{8 m^2} \, \left[ \nabla^2 \left( \vec{\sigma} \cdot  \vec{\bm{B}} \right) \right] + i \, \frac{ f}{4 m^2} \, \left[ \left( \vec{\sigma} \cdot \vec{\nabla} \right) \vec{\bm{B}} \right] \cdot \vec{\bm{\pi}} - \frac{ f}{2 m^2} \, \vec{\bm{B}} \cdot \Bigl[  \Bigl( \vec{\sigma} \cdot \vec{\bm{\pi}}  \Bigr) \vec{\bm{\pi}}  \Bigr] \nonumber \\
  && + \frac{ f^2}{4 m^2} \, \vec{\sigma} \cdot \left( \vec{\bm{E}} \times \partial_t \vec{\bm{E}} \right) - \frac{ f^2}{4 m^2} \, \left[ \vec{\sigma} \cdot \vec{\nabla} \Bigl( \vec{\bm{E}} \cdot \vec{\bm{B}} \Bigr)  \right] +  \frac{ f^2}{4 m^2} \, \left[ \vec{\bm{E}} \cdot \vec{\nabla} \Bigl( \vec{\sigma} \cdot \vec{\bm{B}} \Bigr)  \right] \nonumber \\
  &&  - \frac{ f^2}{4 m^2} \, \vec{\bm{E}} \cdot \left[ \Bigl( \vec{\sigma} \cdot \vec{\nabla} \Bigr) \vec{\bm{B}} \right] - \frac{ f^2}{4 m^2} \, \left[ \vec{\bm{B}} \cdot \vec{\nabla} \Bigl( \vec{\sigma} \cdot \vec{\bm{E}} \Bigr)  \right] \nonumber \\
  && - \frac{ f^3}{2 m^2} \, \Bigl( \vec{\bm{E}} \cdot \vec{\bm{B}} \Bigr) \Bigl( \vec{\sigma} \cdot \vec{\bm{E}} \Bigr).
\label{DP_hamiltoniana_diagonalizada_4D}
\end{eqnarray}

This Hamiltonian with terms until order $O(1/m^2)$ contains new interactions and properties which demand some notes. Most of the (non)relativistic interactions are fruit of the non-minimal coupling, hence, a familiar expression is recovered for $f=0$ \cite{Schwabl_livro} with the habitual spin-dependent interactions  $\, \vec{\sigma} \cdot \vec{\bm{B}}$, $\, \vec{\sigma} \cdot (\vec{\nabla} \times \vec{\bm{E}}) \,$ and $\, \vec{\sigma} \cdot ( \vec{\bm{E}} \times \vec{\bm{\pi}}) \,$, noticing that $\vec{S} = \vec{\sigma}/2$. At the non-relativistic sector are the previewed anomalous magnetic moment $ \, f  \vec{\sigma} \cdot \vec{\bm{B}}$ and the correction to Darwin-type term $\, \frac{f}{2m} \vec{\nabla} \cdot \vec{\bm{E}}$. Along the Hamiltonian, at the fourth line, the expressions $ \, \vec{\nabla} \cdot \vec{\bm{B}} \,$ and $\, \partial_t \vec{\bm{E}} - \vec{\nabla} \times  \vec{\bm{B}} \,$ are maintained once the Electrodynamics is unspecified. In case of Maxwell theory in vacuum describing the external electromagnetic field, both contributions are identically null. In counterpart, $\, \vec{\nabla} \cdot \vec{\bm{B}} \,$ may perform contribution in the presence of magnetic monopole, for example, as suggested in ref. \cite{Castelnovo_Nat} for spin ice configuration. In parallel, extensions of Maxwell Electrodynamics theories such as non-linear ones \cite{EH,BI}, model contemplating Lorentz symmetry violation \cite{CFJ}, manifesting quantum gravity effects \cite{GP,Alfaro_PRD}, are subjected to manifest non-trivial contributions of Ampère-Maxwell equation, even in vacuum. The diagonalized Dirac-Pauli Hamiltonian is compound by new spin-dependent interactions of magnitude $f^2/m^2$ and $f^3/m^2$, involving magnetic and electric fields and preserving parity symmetry. A last pointing is the observation that Chen and Chiou \cite{Chen_PRA}, in regime of weak, homogeneous and static field, worked on the high-order contributions of the FW transformation for the Dirac-Pauli Hamiltonian and obtained some interactions similar to $H_{FW,DP}$. 

Following the script, the next step is to reduce the dimensionality of the system in eq. \eqref{DP_hamiltoniana_diagonalizada_4D}, generating a three spacetime dimension Hamiltonian (for a detailed historical review of reduction dimension approach, consult ref. \cite{Leonardo_tese}). The methodological approach adopted is similar to the Scherk-Schwarz reduction \cite{Scherk-Schwarz}, taken in a simple formulation. The core idea is to assume that the fields are independent of the reduced spatial coordinate -- "$z$" in the present text --, what leads to $ \partial_z ( \textrm{any field}) = 0 $. Facing this condition, the integration of the action over the reduced coordinate (normally taken compact) renders a length dimension factor. This factor is absorbed by the coupling constants and fields, what assures the suitable mass dimension in the three-dimensional spacetime. In this configuration, the gauge vector potential $\, \vec{\bm{A}} \,$ is decomposed in a planar vector $\, \vec{A} \,$ and a new scalar field $\, \varphi$, thus $\,  \vec{\bm{A}} = ( \vec{A}, \varphi)$. The calculus is conducted assuming $\, \varphi = 0$ (which is the simplest choice based on the purposes of this Thesis), producing $\, \vec{\bm{B}} \rightarrow (0,0,B_z) \,$, $\, \vec{\bm{E}} \rightarrow (\vec{E},0) \,$ and $\, \vec{\bm{\pi}}_z \rightarrow 0 \,$. Then, inserting these formulations for the electric and magnetic fields and for the canonical moment in Dirac-Pauli Hamiltonian eq. \eqref{DP_hamiltoniana_diagonalizada_4D} and manipulating it algebraically, one finds the reduced Hamiltonian

\begin{eqnarray}
  H_{\textrm{red}} && \approx  m \, \mathbb{I} + \frac{1}{2m} \, \left[ \left( \vec{p} - q \vec{A} \right) \, \mathbb{I} - f \, \widetilde{\vec{E}} \,\sigma_z  \right]^2 + q \phi \, \mathbb{I} \nonumber \\ 
  && + \left( f - \frac{q}{2m} \right) \, B_z \, \sigma_z  + \frac{f}{2m} \, \vec{\nabla} \cdot \vec{E} \, \mathbb{I} \nonumber \\
  &&  - i \, \frac{q}{8m^2} \, \sigma_z \, \vec{\nabla} \times \left[ \vec{E} \, \mathbb{I} - \frac{f}{q} \, \partial_t  \widetilde{\vec{E}}  \, \sigma_z \right] \nonumber \\
  && -  \frac{q}{8m^2} \, \vec{\nabla} \cdot \left[ \vec{E} \, \mathbb{I} - \frac{f}{q} \, \left( \vec{\nabla} B_z + \partial_t \widetilde{\vec{E}} \right) \, \sigma_z \right]  \nonumber \\
  &&- \frac{q}{4m^2} \, \sigma_z \left[ \vec{E} \, \mathbb{I} - \frac{f}{q} \, \left( \vec{\nabla} B_z + \partial_t \widetilde{\vec{E}} \right) \, \sigma_z \right] \times \left[ \left( \vec{p} - q \vec{A} \right) \, \mathbb{I} - f \, \widetilde{\vec{E}} \,\sigma_z \right] \, ,
\label{Hamiltoniana_reduzida}
\end{eqnarray}

\noindent
where is recognized $\epsilon_{3 \, ij} = \epsilon_{ij}$ to recover the dual electric field $\widetilde{\vec{E}}_i = \epsilon_{ij} \vec{E}_j$.

The eq. \eqref{Hamiltoniana_reduzida} precisely recovers the Hamiltonian set originally in three spacetime dimensions in eq. \eqref{Dirac_hamiltoniana_diagonalizada}. This verification is facilitated resourcing to the associations among the magnetic fields $\, B_z \rightarrow B \,$ and the non-minimal couplings $\, f \rightarrow - g\, $ and projecting the result on the upper state ($\psi_1$) of the two-components spinor $\, \psi = (\psi_1, \psi_2)^T \,$. In the end is demonstrated that the non-minimal interactions in eq. \eqref{Dirac_hamiltoniana_4D}, after a particular dimensional reduction, become the ones in \eqref{Dirac_hamiltoniana_diagonalizada}, suggesting an existence of connection through them.

Turn the attention to the matrix structure of both systems, one observes that the spin arrangement in $H_{red}$ (eq. \eqref{Hamiltoniana_reduzida}) is evident ($S_z = \sigma_z/2$) and hidden in $H_{FW,D}$ (eq. \eqref{Dirac_hamiltoniana_diagonalizada}). $H_{red}$ is generated from the Hamiltonian $H_{FW,DP}$ (eq. \eqref{Dirac_hamiltoniana_4D}) conceived in four spacetime dimensions, based on gamma matrices $4 \times 4$ and four-components spinors. Considering only the positive energy solutions of $H_{FW,DP}$, set this Hamiltonian in Pauli matrices and two-components spinors. On the other hand, the three-dimensional $H_{FW,D}$ is structured in Pauli matrices and two-components spinors, in a way that, projecting its positive energy solutions results in a subspace $1 \times 1$ and one-component spinor, hiding the spin representation.

It is important to emphasize that the external electromagnetic field is determinant to achieve the connection of the Hamiltonians through the dimensional reduction. For example, taking the Maxwell Electrodynamics, the Hamiltonians $H_{red}$ and $H_{FW,D}$ converge to the same expression. However, remembering the Maxwell-Chern-Simons Electrodynamics in eq. \eqref{MCS_lagrangiana}, the Levi-Civita symbol $\epsilon^{\mu \alpha \beta}$ is peculiar to three spacetime dimensions, whereas an extension of four-dimensional Maxwell Electrodynamics, when submitted to a dimensional reduction, results in a theory with non-trivial equations of motion.    


\section{Partial Conclusion}
\label{Conclusão_FW}
\indent

This work is dedicated to effective low-dimensional systems carrying non-minimal electromagnetic couplings expanded through the Foldy-Wouthuysen transformation, revealing (non)relativistic interactions in the Hamiltonian formalism. The procedure of expanding is applied directly on systems genuinely constructed in the low dimension scenario or in reduced ones originated in higher dimensions. 

In the Klein-Gordon Lagrangian with the non-minimal coupling in three spacetime dimensions, the FW Transformation until $\mathcal{O} (1/m^{3})$ presented the usual anomalous magnetic dipole interaction and the relativistic kinetic term contributing to corrections on mass.   

The three-dimensional fermionic system non-minimally coupled to electromagnetic field, after the transformation up to $\mathcal{O} (1/m^2)$, unveiled that relativistic interactions are relevant only in presence of matter $(\vec{J} \neq \vec{0})$, with the Maxwell Electrodynamics describing the external electromagnetic field. Exchanging the Maxwell Electrodynamics by the Maxwell-Chern-Simons one, results in non-trivial contributions even in vacuum. Besides, in vacuum configuration, the effective electric field (eq. \eqref{campo_elétrico_eficaz_2}) vanishes when pushed to a particular critical coupling constant ($g_c = -q/\lambda)$, converging Hamiltonians from bosonic and fermionic systems to the same one. Taking the Hamiltonian generated from four dimensional fermionic Lagrangian containing non-minimal Pauli coupling, and reducing it dimensionality through a method similar to Scherk-Schwarz procedure, one recovers the Hamiltonian obtained for the low-dimensional fermionic system. This result should be taken as a particular achievement and not be generalized, once this is not expected for systems with extension of Maxwell Electrodynamics. This opens possibilities of further investigations, in which are contemplated the influence of the different dimensional reduction methods in overall results. The fermionic higher dimension is also passive of being analysed in detail in the future due to its relativistic interactions manifested from the FW transformation (eq. \eqref{DP_hamiltoniana_diagonalizada_4D}).

One more study branch is the Landau-Lifshitz-Gilbert (LLG) equation, observing that the magnetic sector of eq. \eqref{DP_hamiltoniana_diagonalizada_4D} is capable of originating the corresponding extension to the LLG equation and its non-minimal relativistic interactions may result in torque or damping contributions to magnetization. Relativistic interactions in magnetization dynamics are matter of research (see refs. \cite{Mondal_PRB_2016_2018,Mondal_PRB_2017} and references therein) and may motivate further works.  


\chapter{Beyond Monopole-Monopole Gravitational Interactions}
\label{cap_grav_1}

\section{Introduction}
\indent

One of the front line researches is the attempting to conciliate the General Relativity and the Quantum Field Theory, fusing Riemannian geometry and particle interaction physics. To elaborate a quantum gravity theory conciliating unitarity and renormalizability (covering from infrared to ultraviolet regime) stands as incompatible, until nowadays. On the other hand, consistent unitary theories respecting a cutoff scale have been formulated as effective field theories \cite{Donoghue_EFT_QG} and gained relevance in the last forty years. 

The gravitation theories containing terms with higher order derivatives back to around the begin of twenties with Weyl \cite{Weyl} and Eddington \cite{Eddington}. Utiyama and De-Witt \cite{Utiyama} pointed the necessity of inclusion of high-order curvature terms in Einstein-Hilbert action to permit a renormalizable theory at one-loop. Such theories were promoted when Stelle \cite{Stelle_1977} (see also ref. \cite{Voronov}) demonstrated they are renormalizable in purely gravitational scenario or when coupled to a neutral massive scalar field, although their non-unitarity formulation, manifesting a complex pole in the modified graviton propagator \cite{Stelle_1978}. This aspect was reinforced by Johnston \cite{Johnston}, observing that the appearance of ghost pole is independent of the chosen gauge. In parallel, it was verified that, whether taken the General Relativity as a perturbative theory, this theory was non-renormalizable at one-loop condition coupled to neutral scalar particle \cite{tHooft}. However, the paper kept opened the possibility of the theory be renormalizable in a free particle scenario at two-loops. In the middle of eighties, it was refuted by Marc Goroff and Augusto Sagnotti \cite{Goroff}, proving the non-renormalizability of the Einstein gravity at two-loops. 

As the challenges around General Relativity renormalization and its harmonization with Quantum Field Theory have raised, modified gravitation theories containing high-order curvature terms have been unfolding in several models. Among them, one finds theories that switch the Ricci curvature scalar $R$ by a general function ${\it f(R)}$ in the Einstein-Hilbert action (see the review \cite{Soritiou_RMP}), massive three spacetime dimension and unitary at tree-level \cite{Bergshoeff_PRL} (for a review, see ref. \cite{Oda_JHEP}), non-perturbative methods considering $4-\epsilon$ dimensionality systems \cite{Peixoto_PRD}, approaches proposing the re-scaling of a mass parameter associated to the Einstein-Hilbert action, which leads the system to a strongly interacting regime type \cite{Holdom}, non-local $D$-dimensional gravity \cite{Modesto_NPB_2015} (see also ref. \cite{Tomboulis}), a Yang-Mills gauge theory assisting a weakly coupled quadratic gravity, in a way to provide a ultraviolet completion \cite{Donoghue_PRD}, super-renormalizable theory with complex poles \cite{Modesto_PLB} and propositions involving Lee-Wick prescription to construct an unitary S-matrix \cite{Modesto_NPB_2016}.

In effective field theories and some other methods, the calculation of gravitational quantum corrections for inter-particle potentials have been received attention ( {\it e.g.} refs. \cite{Muzinich_PRD,Hamber_PLB,Akhundov_PLB,Khriplovich_JETP_2002,Bjerrum_PRD_2002,Bjerrum_PRD_2003,Faller_PRD}) since pointed by Donoghue in ref. \cite{Donoghue_EFT_QG} that these corrections for Newtonian potential, derived from quantum gravity model based in Quantum Field Theory, compose an efficient evaluation of consistency of the model. The research on inter-particles potentials reaches beyond the traditional monopole-monopole interaction, delivering also velocity- and spin-dependent corrections. The authors in \cite{Gupta_PR} obtained spin contributions to a graviton exchanging between particles of varied spin representations, Khriplovich and Kirilin \cite{Khriplovich_JETP_2004} calculated "spin" (in the particular case, they defined as the internal angular momenta of rotating compound body) and velocity corrections to the Newtonian potential for scalar particles and, latter, Kirilin compared these results with spinorial system and calculated spin-orbit and spin-spin quantum corrections \cite{Kirilin_NPB}. A Dirac-Einstein system forming an effective field theory is studied in ref. \cite{Butt_PRD}, where the leading quantum corrections to gravitational coupling of a charged massive particle with spin-1/2 are achieved. Ross and Holstein \cite{Ross_JPA,Ross_Arx} analysed the non-relativistic gravitational scattering amplitudes for different spin combinations, obtaining classical and quantum corrections in spin-orbit and spin-spin sectors. They extended the similar investigation for amplitudes of electromagnetic-gravitational scattering systems \cite{Ross_Arx_QED}. For theoretical and experimental reviews about the spin role in gravity, see refs. \cite{Ni_RPP,Ni_IJMP}.    

The chapters \ref{cap_grav_1} and \ref{cap_grav_2} are part of the fruitful collaboration with Gustavo P. de Brito, Judismar T. Guaitolini Jr., Leonardo P.R. Ospedal and Kim P.B. Veiga, which resulted in the paper \cite{Brito_PRD}. They are organized in a total of seven sections, so that, four are in the chapter 2 and three in the third one. The former is divided in the section \ref{seção_QG}, containing some definitions and main structural tensors of the linearized gravity; followed by sec. \ref{seção_EA} with a very short oversight about effective action definition and formulation. The section \ref{seção_metodologia} describes the methodological approach and formalism that supported the calculation of the gravitational potentials and in the sec. \ref{seção_resultados} are obtained the inter-particles non-relativistic potentials for two scalars and two spinors exchanging one graviton. The next chapter is composed by the sec. \ref{seção_comparação_potenciais}, dedicated to detailing and comparison of the non-relativistic potentials sector by sector (velocity-velocity, orbit-spin and spin-spin) and in static limit; the subsequent section \ref{seção_comparação_EM} contains a comparative analysis among the obtained potentials and the results of ref. \cite{Gustavo_PRD}, where the authors, in a modified (effective) electrodynamics, calculated inter-particles potentials for spin-0, -1/2 and -1, achieving velocity and spin contributions. Then, the results are submitted to a praxis test in section \ref{seção_CDT_fatores_de_forma}, where is chosen a particular effective quantum gravity model for low-energy condition (based on Casual Dynamics Triangulation) \cite{Knorr_PRL} to support the definition of the form factors and, thereby, to calculate the particular inter-particle potentials. The last section (\ref{seção_conclusão_grav}) is dedicated for final comments and conclusions.


\section{Linearized Gravity}
\label{seção_QG}
\indent

The starting point is to set the building blocks of the linearized gravity. In this way, it is worked in a Riemannian four dimensional spacetime, where an approximation of a local flat spacetime is regarded, keeping the coordinate invariance. With this approximation is possible to work with the Riemannian metric ($g_{\mu\nu}$) equal to the Minkowski one ($\eta_{\mu\nu} = \text{diag} (1, -1, -1, -1)$ - convention adopted) summed to a smooth perturbation designated by the symmetric rank two tensor $h_{\mu\nu}$. Making use of the familiar choice 

\begin{equation}
  g_{\mu\nu} =  \eta_{\mu\nu} + \kappa h_{\mu\nu} ,
\label{métrica gravitacional covariante} 
\end{equation}

\noindent
with $\kappa$ being the Einstein constant ($\kappa =\sqrt{32 \pi \text{G}}$ , $\text{G}$ denotes the Newtonian constant) and also considering $|\kappa h_{\mu\nu}| \ll 1$. It is demanded $g_{\mu\alpha} g^{\alpha\nu} = \delta_{\mu}^{\,\,\, \nu}$, what results in

\begin{equation}
 g^{\mu\nu} = \eta^{\mu\nu} - \kappa h^{\mu\nu} + \kappa^2 h^{\mu\alpha} {h_{\alpha}}^{\nu} + ...  ,
\label{métrica gravitacional contravariante}
\end{equation}

\noindent
still in time, it is defined  

\begin{equation}
 \sqrt{-\text{det}(g_{\mu\nu})} \equiv \sqrt{-g} = 1+\frac{1}{2}\kappa h + O(h^2) ,
\label{determinante métrica gravitacional}
\end{equation}

\noindent
in which $h \equiv {h_{\mu}}^{\mu}$.

The next step is the linearization of the main geometric elements of the General Relativity. Thus, the Riemann-Christoffel connection $\Gamma$ ($\Gamma^{\alpha}_{\mu\nu} = \frac{1}{2} g^{\alpha\rho} (\partial_{\mu} g_{\nu\rho} + \partial_{\nu} g_{\mu\rho} - \partial_{\rho} g_{\mu\nu})$), after linearization, assumes the form

\begin{equation}
 \Gamma^{\alpha}_{\mu\nu} = \frac{1}{2}\kappa (\partial_{\mu}{h_{\nu}}^{\alpha} + \partial_{\nu}{h_{\mu}}^{\alpha} - \partial^{\alpha}h_{\mu\nu}) + O(\kappa^2) ,
\label{conexão afim}
\end{equation}

\noindent
in the same way, the Riemann-Christoffel tensor (${R^{\alpha}}_{\mu\nu\beta} = \partial_{\nu}\Gamma^{\alpha}_{\mu\beta} - \partial_{\beta}\Gamma^{\alpha}_{\mu\nu} + \Gamma^{\alpha}_{\rho\nu}\Gamma^{\rho}_{\mu\beta} - \Gamma^{\alpha}_{\rho\beta}\Gamma^{\rho}_{\mu\nu}$), the Ricci tensor ($R_{\mu\nu}={R^{\alpha}}_{\mu\nu\alpha}$) and the curvature scalar ($R=g^{\mu\nu}R_{\mu\nu}$) are linearized, respectively,

\begin{equation}
{R^{\alpha}}_{\mu\nu\beta} = \frac{1}{2}\kappa (\partial_{\mu}\partial_{\nu} {h_{\beta}}^{\alpha} + \partial^{\alpha}\partial_{\beta}h_{\mu\nu} - \partial_{\mu}\partial_{\beta}{h_{\nu}}^{\alpha}  - \partial_{\nu}\partial^{\alpha} {h_{\mu}}^{\beta} ) + O(\kappa^2),
\label{tensor Riemann}
\end{equation}

\begin{equation}
 R_{\mu\nu} = \frac{1}{2}\kappa (\Box h_{\mu\nu} + \partial_{\mu}\partial_{\nu}h - \partial_{\rho}\partial_{\mu}{h_{\nu}}^{\rho}  - \partial_{\rho}\partial_{\nu} {h_{\mu}}^{\rho} ) + O(\kappa^2) ,
\label{tensor Ricci}
\end{equation}

\begin{equation}
 R = \kappa (\Box h - \partial_{\mu}\partial_{\nu}h^{\mu\nu}) + O(\kappa^2)  ,
\label{escalar curvatura}
\end{equation}

\noindent
using the symbol $\Box = \partial_{\mu}\partial^{\mu} $ to represent the d`Alembertian differential operator.

Those are the main elements of the Riemannian geometry and their linearized form and the tensor $h_{\mu\nu}$, in this specific context, assumes the role of the graviton as a spin-2 field \cite{Gupta_PS1,Gupta_PS2,Gupta_RMP}.

\section{The Effective Action: A Brief Review}
\label{seção_EA}
\indent

The \textit{n-point} Green`s functions have a central role in the Quantum Field Theory, once they structure the S-Matrix elements. They are the expected vacuum value of the chronological ordered product of the field operators

\begin{equation}
  {G_n} = \langle 0|\textit{T} \Phi_1 \Phi_2 ... \Phi_n |0 \rangle .
\label{função de Green-matrizS}
\end{equation}

\noindent 
In the functional approach, one extracts the Green`s function $G_{n}(x)$ (where $x \equiv x_1, x_2, ..., x_n$) of a system from the differentiation of its generating functional $Z[J]$ by an external source $J(x)$

\begin{equation}
  i^{n} \, G(x_1, x_2,...,x_n) =\dfrac{\delta^{n} {Z[J]}}{\delta{J(x_{1})} \delta{J(x_{2})} ...\delta{J(x_{n})}} \, \bigg|_{J=0} .
\label{função de Green-fgerador}
\end{equation}

\noindent
In a similar way, the connected Green`s function ${G^c}(x)$ - "connected" refers to the Feynman diagrams that are impossible to separate the parts which are not directly joined by a line - is expressed by 

\begin{equation}
 i^n \, G^{c}_n (x_1, x_2,...,x_n) = {\dfrac{\delta^{n} {W[J]}}{\delta{J(x_{1})} \delta{J(x_{2})} ...\delta{J(x_{n})}}} \, \bigg|_{J=0} ,
\label{função de Green-conexa}
\end{equation}

\noindent
with $W[J]$ representing the generating functional of $G^{c}_{n}(x)$. The generating functional in eqs. \eqref{função de Green-fgerador} and \eqref{função de Green-conexa} are related by $Z[J] = \exp({iW[J]})$.

Now, defining the so called effective action as a functional Legendre transformation (for more details, see ref. \cite{livro_Shapiro})

\begin{equation}
  \Gamma[\Phi] = W[J] + \int d^{4}x \, \Phi(x) \, J(x),
\label{ação efetiva}
\end{equation}

\noindent
with $\Phi(x)$ representing a function of infinity class decreasing fast and $\Gamma[\Phi]$ being the generating functional of the \textit{n-point} vertex function $\Gamma_{n}(x)$, $n \geq 2$, 

\begin{equation}
  \Gamma_{n}(x) = \dfrac{\delta^{n} \Gamma[\Phi]}{\delta{\Phi(x_{1})} \delta{\Phi(x_{2})} ...\delta{\Phi(x_{n})}} \, \bigg|_{J=0} , 
\label{função de vértice}
\end{equation}

\noindent
noting that $\Gamma_{n}(x)$ involves strictly the \textit{n-point} connected Feynman diagrams that are one particle irreducible - means the diagram that is impossible, by breaking one of its line, to transform it in two - and have no external lines (named "amputated"). 

After this superficial and brief presentation of the functional effective action concept, which stands as the protagonist of this study, the focus is moved toward the description of the methodological approach adopted for calculation of the non-relativistic potentials.   

\section{Non-relativistic Potentials}
\subsection*{Methodological Approach and Mathematical Formalism}
\label{seção_metodologia}
\indent

It seems convenient firstly to introduce the fundamental equations, conventions and methodology chosen to calculate the potentials. They are achieved through the Fourier transform of the Born approximation to first order in $V(r)$ (see, for example, ref.\cite{livro_Maggiore})

\begin{equation}
  V(r)= -\int \frac{d^3 \vec{q}}{(2\pi)^3} \mathcal{M}_{_{\textmd{NR}}} (\vec{q}) \, e^{i \vec{q} \cdot \vec{r}} \, ,
\label{aproximação de Born} 
\end{equation}

\noindent
where $\mathcal{M}_{_{\textmd{NR}}} (\vec{q})$ denotes the scattering amplitude in the non-relativistic limit. The former amplitude can be obtained from the adequate normalized relativistic scattering amplitude ($\mathcal{M} (\vec{q})$)

\begin{equation}
  \mathcal{M}_{_{\textmd{NR}}} (\vec{q}) = {\lim}_{_{\textmd{NR}}} \prod_{i=1,2} (2E_i)^{-1/2} \prod_{j=1,2} (2E'_j)^{-1/2} \,  \mathcal{M} (\vec{q}) \, ,
\label{amplitude não relativística} 
\end{equation}

\noindent
noting that the normalization factor is composed by the product of the scattered particles energies $E_{i}$ and $E'_{j}$, which are resulted of the Taylor expansion around the rest mass limit of the in and out going particles, respectively.

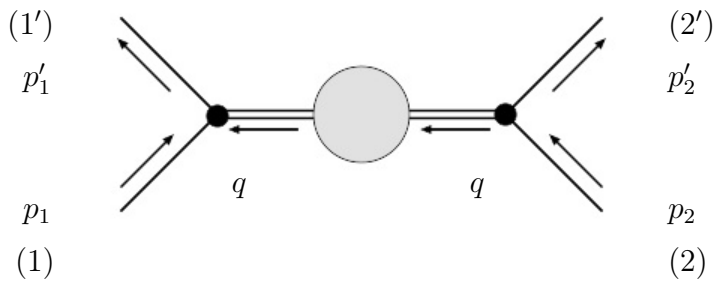
\begin{figure}[ht]
	\begin{center}
		\leavevmode
		\includegraphics[width=0.5\textwidth]{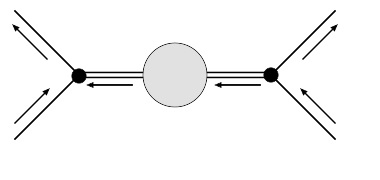}
		\put(0,0){(2)}
		\put(0,20){$p_2$}
		\put(-247,0){(1)}
		\put(-244,20){$p_1$}
		\put(0,90){($2^\prime$)}
		\put(0,70){$p_2^\prime$}
		\put(-250,90){($1^\prime$)}
		\put(-244,70){$p_1^\prime$}
		\put(-75,30){$q$}
		\put(-165,30){$q$}
	\end{center}
	\caption[Feynman diagram 3]{Diagrammatic representation of the 3-momentum convention and approximation adopted for the calculation of the scattering amplitude and potential.}
	\label{diagrama}
\end{figure}

An important remark on the physics approach methodology is the fact that the endeavour is destined exclusively to calculations of the graviton propagator quantum corrections. Therefore, the vertices are carried in their tree-level processes. This structure is pictorially expressed by the Feynman diagram in the fig. \ref{diagrama}, in which the centered circle represents the quantum corrections to the graviton propagator. Then, the relativistic scattering amplitude is expressed by

\begin{align}
  i\mathcal{M} = i\,T^{\mu\nu}(p_1,p_1^\prime) \, 
  \langle h_{\mu\nu}(-q) h_{\alpha\beta}(q)\rangle \, i\,T^{\alpha\beta}(p_2,p_2^\prime) \, ,
\label{amplitude relativística}
\end{align}

\noindent
being $T^{\mu\nu}$ (and $T^{\alpha\beta}$) the tree-level energy-momentum tensors of the scattered particles 1 and 2 (see fig. \ref{diagrama}) and $\langle h_{\mu\nu}(-q) h_{\alpha\beta}(q)\rangle$ the graviton propagator (accompanied by their quantum corrections).

The opted 3-momentum conventions to the scattering amplitudes and potentials calculations are represented in the fig. \ref{diagrama}. The algebra is developed in the center-of-mass (CM) reference frame, what allows to assume the following 3-momentum convention

\begin{equation}
  \vec{p}_1 = -\vec{p}_2 = \vec{p} - \frac{\vec{q}}{2} \, , \qquad   
  \vec{p'}_1 = -\vec{p'}_2 = \vec{p} + \frac{\vec{q}}{2} \,  ,
\label{momentum_CM} 
\end{equation}

\noindent
in which is made use of the 3-momentum arithmetic average $\vec{p}$, obtained from each pair $\vec{p}_{i}$ and $\vec{p'}_{i}$, and the momentum transfer $\vec{q}$. For the analyses of the system dynamics is considered an elastic scattering behaviour, what means the conservation of the system energy
$E_{final} = E_{initial}$. Expanding the energies around $\vec{p}_{i}=\vec{0}$ one has $E_{i} \approx m_{i} + \frac{{\vec{p}_{i}}^2}{2m_{i}}$, hence, applying the mentioned conservation of the energy system and resourcing to eqs. \eqref{momentum_CM}, the condition $\, \vec{q} \cdot \vec{p} = 0 \,$ is acquired. Due to the former condition, new relations among the energies come out $E_{1} = E'_{1}$ and $E_{2} = E'_{2}$. 

At this point, it is introduced the functional effective action formalism. The generating functional - conventionally named \textit{effective action} - has the general form $\Gamma = \bar{\Gamma} + \hat{\Gamma}$, in which, $\bar{\Gamma}$ expresses the sector containing diffeomorphism transformations invariance ($ \delta_{\text{diff}} \bar{\Gamma} = 0$) and $\hat{\Gamma}$ the sector that does not dispose of this property ($ \delta_{\text{diff}} \hat{\Gamma} \neq 0$). In terms of covariant gauge theory, the diffeomorphism transformations are interpreted as the gauge symmetry transformations. In this context of Quantum Gravity Field Theory scenario, with the field $h_{\mu\nu}$ being a small perturbation associated to the metric $g_{\mu\nu}$, the effective action assumes the form

\begin{align}
\Gamma[h;g] = \bar{\Gamma}[g] + \hat{\Gamma}[h] \,.
\label{ação efetiva-tqc}
\end{align}

\noindent
The eq. \eqref{ação efetiva-tqc} is composed by a gauge invariant term ($\bar{\Gamma}[g]$) for gauge transformations on $g_{\mu\nu}$ and by a non-invariant term ($\hat{\Gamma}[h]$), representing the gauge-fixing sector $\delta_{h_{\mu\nu}}\hat{\Gamma} \neq 0$. The portion $\bar{\Gamma}[g]$ represents

\begin{equation}
\bar{\Gamma}[g] = 
\frac{2}{\kappa^2} \int d^4 x \, \sqrt{-g} \left( -2 \Lambda -  R - \frac{1}{3} R F(\Box) R + C_{\mu \nu \alpha \beta} W(\Box) C^{\mu \nu \alpha \beta} \right) + \mathcal{O}(\mathcal{R}^3) \, ,
\label{ação efetiva-eq} 
\end{equation}

\noindent
with $\Lambda$ standing for the cosmological constant, $R$ the scalar of curvature, $C_{\mu\nu\alpha\beta}$ the Weyl tensor

\begin{eqnarray}
  C_{\mu \nu \alpha \beta} &=& R_{\mu\nu\alpha\beta} + \frac{1}{2} (g_{\mu\beta} R_{\alpha\nu} + g_{\nu\alpha} R_{\beta\mu} - g_{\mu\alpha} R_{\beta\nu} - g_{\nu\beta} R_{\alpha\mu}) \nonumber \\
  &+& \frac{1}{6} (g_{\mu\alpha} g_{\beta\nu} - g_{\mu\beta} g_{\alpha\nu})R ,
 \label{tensor_Weyl}
 \end{eqnarray}
 
 \noindent
 where $R_{\mu\nu\alpha\beta}$ and $R_{\mu\nu}$ are the Riemann and Ricci tensors. Each one of the mentioned tensors, including the constant $\sqrt{-g}$, will be linearized following the equations \eqref{escalar curvatura}, \eqref{tensor Riemann}, \eqref{tensor Ricci} and \eqref{determinante métrica gravitacional} respectively. $F(\Box)$ and $W(\Box)$ denote the form factors as functions of the covariant d`Alembertian operator and $\mathcal{O}(\mathcal{R}^3)$ includes terms contributions involving curvature invariant compositions of order higher than 2. The action \eqref{ação efetiva-eq} in a region around Minkowskian four dimensional background scenario, represents the more general torsion-free and parity invariant form that holds unitarity (ghost-free) \cite{Biswas_et_al-PRL, Biswas_et_al-CQG,Accioly_PRD}. $\bar \Gamma[g]$ contains explicitly the Einstein-Hilbert action term ($\Gamma_{\text{EH}} \sim \int d^4 x \, \sqrt{-g} R$) and two others geometrical ones quadratic in curvature. They are associated to form factors, which, in this approach, once expanded the Green`s functions, the quantum corrections of the interactions potentials are calculated in function of them. The effective action piece displayed above can be faced as a generalization of the van Nieuwenhuizen proposal \cite{Nieuwenhuizen_NP}, where $F(\Box)$ and $W(\Box)$ could assume the form of non-local arrangements \cite{Pais_PR}, as adopted, for instance, by Moffat, Barvinsky and Modesto \cite{Moffat_PRD, Barvinsky_PLB, Modesto_PRD_2012}, until the form of local polynomials functions (some references are available in Giacchini and Netto \cite{Giacchini_EPJ} and Accioly \textit{et al} \cite{Accioly_PRD}).
 
 Now, turning the attention to the expression $\hat{\Gamma}[h]$,
 
\begin{align}
  \hat{\Gamma}[h] = \frac{1}{2\alpha} \int d^4x 
  \sqrt{-{g}} \,{\eta}^{\mu\nu} F_\mu [h] F_\nu[h] \,,
\end{align}

\noindent 
where $\alpha$ is the gauge parameter and $F_\mu[h] = \bar{\nabla}^\nu h_{\mu\nu} - \frac{1}{2} \bar{\nabla}_\mu h $, with $\bar\nabla_{\mu}$ indicating the covariant derivative $\bar\nabla_{\mu} \equiv \partial_{\mu} + \bar\Gamma_{\mu} $, with $\bar\Gamma_{\mu}$ representing an appropriate connection. Basically, it exhibits the same appearance of the usual gauge-fixing term in the classical action. However, it is worked on the mass shell, what results in trivial Ward identities and makes the calculation gauge-independent.

From the eq. \eqref{ação efetiva-tqc} is calculated the Green`s functions for the graviton propagator, which are the ones described in Sec (\ref{seção_EA}) - diagrammatically represented by connected, amputated and one particle irreducible Feynman diagrams. This is accomplished taking the vertex function \eqref{função de vértice} and calculating the inverse of its two point function $ {\delta^2 \Gamma}/{{\delta h}^2} \, \big|_{h=0}$. In this way, the graviton propagator reads (in chapter 2 of ref. \cite{livro_Shapiro}, there is the detailing of this approaching)

\begin{eqnarray}
  \langle h_{\mu \nu} (-q) h_{\alpha \beta} (q) \rangle = 
  \frac{i}{q^2} \Bigg[ \frac{1}{Q_2 (q^2)} \mathcal{P}_{\mu\nu\alpha\beta}^{(2)}  - \frac{1}{2Q_0 (q^2)} \mathcal{P}_{\mu\nu\alpha\beta}^{(0)}   \Bigg] \, + \,
  i\Delta_{\mu \nu \alpha \beta}(q) \,,
\label{propagador_graviton} 
\end{eqnarray}

\noindent
where $Q_0$ and $Q_2$ are defined as

\begin{subequations}
\begin{align}
  Q_2 (q^2 ) = 1 +  \frac{2 \Lambda}{q^2} + 2 q^2 \, W(-q^2) \, ,
\label{definição_Q2}
\end{align}

\begin{align}
  Q_0 (q^2 ) = 1 +  \frac{2 \Lambda}{q^2} + 2 q^2 \, F(-q^2) \, ,
\label{definição_Q0}
\end{align}
\end{subequations}

\noindent
and the tensors $\mathcal{P}_{\mu\nu\alpha\beta}^{(2)} $ and $\mathcal{P}_{\mu\nu\alpha\beta}^{(0)} $ denote

\begin{subequations}
\begin{align}
\mathcal{P}_{\mu\nu\alpha\beta}^{(2)} =
\frac{1}{2} (\eta_{\mu \alpha} \eta_{\nu \beta}+ \eta_{\mu \beta} \eta_{\nu \alpha}) - \frac{1}{3}\eta_{\mu \nu}\eta_{\alpha \beta} \,,
\end{align}

\begin{align}
\mathcal{P}_{\mu\nu\alpha\beta}^{(0)} = \frac{1}{3}\eta_{\mu \nu}\eta_{\alpha \beta} \,.
\end{align}
\end{subequations}

\noindent
Additionally, the structure $\Delta_{\mu\nu \alpha \beta}$ encloses the terms that contracted to the energy-momentum tensors of the scattered particles and vanish. It seems relevant to point out the disconnection feature manifested by the propagator \eqref{propagador_graviton} through the "$Q$" factors (eqs. \eqref{definição_Q2} and \eqref{definição_Q0}). The form factor $W(-q^2)$ is restricted to $Q_2$ and, likely, $F(-q^2)$ is to the factor $Q_0$, what evidences their exclusive contribution by the tensor $\mathcal{P}_{\mu\nu\alpha\beta}^{(2)}$, in the case of $W$, and by $\mathcal{P}_{\mu\nu\alpha\beta}^{(0)}$ to $F$.

In the next subsection is evolved the calculations of the scattering amplitudes and potentials between two spin-0 scattered particles and two spin-1/2.

\subsection*{\Large{Results}}
\label{seção_resultados}

\subsection{Spin-0 External Particles}
\label{seção_potencial_spin0}
\indent

The start point is the calculation of the gravitational potential in an elastic scattering among massive spin-0 and spin-2 particles. The vertex is taken in the tree level. To suppress quantum corrections for the vertex and the external particles propagator is unfeasible when one tries to reach results experimentally reliable. Therefore, it is decided to work with this configuration as an approximation and a first stage for further calculations. In this way, the scalar particles dynamic assumes the form

\begin{eqnarray}
\Gamma_{\textmd{scalar}}[\phi,g] = \int{d^4x \, \sqrt{-g} \Bigg( \frac{1}{2} g^{\mu\nu} \partial_\mu \phi \partial_\nu \phi - \frac{1}{2}m^2 \phi^2} \Bigg) \, ,
\label{ação escalar}
\end{eqnarray} 

\noindent
observing that it derives from the classical Klein-Gordon Lagrangean density submitted to the principle of minimal gravitational coupling ($d^4x \rightarrow{d^{4}x} \, \sqrt{-g}$. In this case, the covariant derivative is identical to the regular derivative, $\nabla_{\mu} \equiv \partial_{\mu}$). Then, the energy-momentum tensor resulted from the interaction sector of eq. \eqref{ação escalar} is written in the momenta space, 

\begin{align}
  T_{\mu \nu}(p, p') = -\frac{\kappa}{2} \Big[\, p_\mu p'_\nu + p_\nu p'_\mu - \eta_{\mu \nu} \left(  p \cdot p' - m^2 \right) \Big] \, ,
\label{tensor_EM_escalar}
\end{align}

\noindent
noting that is considered terms up to first order in $\kappa$ and the "prime" symbol on momenta follows the definition expressed in fig. \ref{diagrama}, namely, the particle 4-momentum after the scattering. It is worthy to remember that $E_1 = E'_1$ and $E_2 = E'_2$ and the external particles in the vertex are on-shell, ${p_1}^2 = {p'_1}^2 = {m_1}^2$ and ${p_2}^2 = {p'_2}^2 = {m_2}^2$. The algebraic manipulation of eqs. \eqref{propagador_graviton} and \eqref{tensor_EM_escalar} based on eq. \eqref{amplitude relativística} provides the relativistic scattering amplitudes. Firstly, presenting the partial result   

\begin{eqnarray}
  i \mathcal{M}^{(s=0)} =  \frac{i}{q^2} \left[ \frac{1}{3 Q_2} \left( T_{1 \, \, \mu}^\mu T_{2 \, \, \beta}^\beta - \, 3 \, T_1^{\mu \nu} T_{2 \, \mu \nu} \right)  
  + \frac{1}{6 Q_0} T_{1 \, \, \mu}^\mu T_{2 \, \, \beta}^\beta\right] \, ,
\label{amplitude_R_escalar_compacta} 
\end{eqnarray}

\noindent
in which is adopted the notations $ T^{\mu \nu}_i \equiv T^{\mu \nu}(p_i, p'_i) $ and $Q_i \equiv Q_i (q^2)$. The expression for the amplitude in terms of the 4-momentum and masses is

\begin{eqnarray}
  \mathcal{M}^{(s=0)} &=& \frac{\kappa^2}{6 q^2 \, Q_2} \Big[ 2 m_1^2 m_2^2 - 3 (p_1 \cdot p_2) (p_1' \cdot p_2') - 3 (p_1 \cdot p_2') (p_1' \cdot p_2)  \nonumber \\
  &+&  2 (p_1 \cdot p_1') (p_2 \cdot p_2') - m_1^2 \, p_2 \cdot p_2' - m_2^2 \, p_1 \cdot p_1' \Big]  \nonumber \\
  &+& \frac{\kappa^2}{6 q^2 \, Q_0} \Big[ (p_1 \cdot p_1') (p_2 \cdot p_2') - 2 m_1^2 \, p_2 \cdot p_2' \nonumber \\ 
  &-& 2 m_2^2 \, p_1 \cdot p_1' + 4 m_1^2 m_2^2 \Big]  .
\end{eqnarray}

Following the methodology prescribed along the subsection \ref{seção_metodologia}, it is opted by the reference of center-of-mass and, hence, the 3-momentum transfer $\vec{q}$ and the average $\vec{p}$, both detailed in eq. \eqref{momentum_CM}. Thus, normalizing the relativistic scattering amplitude according to eq. \eqref{amplitude não relativística}, one finds the non-relativistic limit of the amplitude 

\begin{eqnarray}
  \mathcal{M}^{(s=0)}_{\textrm{NR}} &=& \frac{ \kappa^2 m_1 m_2 }{6 \, Q_2 \, \vec{q}^{\,2}} \, \left[ 1 +  \vec{p}^{\,2} \left( \frac{3}{m_1 m_2} + \frac{1}{m_1^2} + \frac{1}{m_2^2} \right)
  + \frac{ \vec{q}^{\,2} }{8} \,  \left( \frac{1}{m_1^2} + \frac{1}{m_2^2}  \right) + \mathcal{O}(3) \right]  \nonumber \\
  &-& \frac{ \kappa^2 m_1 m_2 }{24 \, Q_0 \, \vec{q}^{\, 2} } \, \left[ 1  - \frac{\vec{p}^{\, 2} }{2} \left( \frac{1}{m_1^2} + \frac{1}{m_2^2} \right) - \frac{5 \, \vec{q}^{\, 2}}{8}  \left( \frac{1}{m_1^2} + \frac{1}{m_2^2} \right) + \mathcal{O}(3)\right]\, ,
\label{amplitude_NR_escalar} 
\end{eqnarray}

\noindent
where $\mathcal{O}(3)$ indicates terms higher than second order in $|\vec{p}|/m_{1,2}$ and/or $|\vec{q}|/m_{1,2}$, which are neglected.

Then, once substituted the expression \eqref{amplitude_NR_escalar} in \eqref{aproximação de Born}, the Fourier transform of the non-relativistic scattering amplitude determines the non-relativistic inter-particle (gravitational) potential  

\begin{eqnarray}
  V^{(s=0)}(r) &=& - \frac{\kappa^2 m_1 m_2}{6}  \Bigg[ I_1^{(2)}(r) + \vec{p}^{\,2} \left( \frac{3}{m_1 m_2} + \frac{1}{m_1^2} + \frac{1}{m_2^2} \right) I_1^{(2)}(r) \nonumber \\
  &+&  \frac{1}{8} \left( \frac{1}{m_1^2} + \frac{1}{m_2^2} \right) I_0^{(2)}(r) \Bigg] 
  +  \frac{\kappa^2 m_1 m_2}{24}  \Bigg[  I_1^{(0)}(r) \nonumber \\
  &-& \frac{ \vec{p}^{\,2} }{2} \left(  \frac{1}{m_1^2} + \frac{1}{m_2^2} \right)  I_1^{(0)}(r) - \frac{5}{8} \left( \frac{1}{m_1^2} + \frac{1}{m_2^2} \right) I_0^{(0)}(r) \Bigg]   , 
\label{potencial_escalar}
\end{eqnarray}

\noindent
being the integrals $I_n^{(a)}(r)$ defined in Appendix \ref{apêndice_int}, eq. \eqref{apêndice_I_a_n}, with $n=0,1$ and $a=0,2$. The potential $V^{(s=0)}(r)$ reveals a structure compounded by terms beyond the monopole-monopole sector. This theme will be debated in the section \ref{seção_comparação_potenciais}, in which both potentials results (scalar and fermionic) are compared. In a section apart (sec. \ref{seção_comparação_EM}), the analyses are extended to a comparison with the electromagnetic potential cases. The next subsection is dedicated to the calculations of the fermionic non-relativistic gravitational potential.

\subsection{Spin-1/2 External Particles}
\label{seção_potencial_spin1/2}
\indent

In this subsection is established the non-relativistic gravitational potential involving two fermionic particles. As practiced in the last subsection \ref{seção_potencial_spin0} for the scalar case, the fermionic particles are taken on-shell, calculated quantum corrections for the graviton propagator and treated the fermion-graviton vertices at the classical level, suppressing their quantum corrections. 

The start point is the Dirac Lagrangean density, described by the functional action $\Gamma_{\textmd{ferm}}[\bar{\psi},\psi,g]$

\begin{eqnarray}
  \Gamma_{\textmd{ferm}}[\bar{\psi},\psi,g] = \int{d^4x \, \sqrt{-g} \left[ \frac{i}{2} (\bar{\psi} \, \gamma^\mu_g \, \nabla_\mu \psi - \nabla_\mu \bar{\psi} \, \gamma^\mu_g  \, \psi ) - m\bar{\psi} \psi \right]} \, ,
\label{ação_fermiônica}
\end{eqnarray}

\noindent
which is rewritten in the context of gravitational geometry scenario. It means to submit it to the prescription of gravitational minimal coupling, although, in this case, differently from what occurs in the scalar one \eqref{ação escalar}, the covariant derivative $\nabla_{\mu}$ carries the Lorentz connection $\Gamma_{\mu}$ responsible for restoring the local Lorentz symmetry \cite{livro_Gasperini}, thus $\nabla_{\mu} \psi = \partial_{\mu} \psi + \Gamma_{\mu} \psi$ or $\nabla_{\mu} \bar{\psi} = \partial_{\mu} \bar{\psi} - \bar{\psi} \Gamma_{\mu}$. One comes across with the gamma matrices $\gamma^{\mu}_{g}$ that obey to the Clifford`s algebra $\left\{\gamma^{\mu}_{g} , \gamma^{\nu}_{g}\right\} = 2 \, g_{\mu\nu}$, the symbols $\psi$, meaning the spin-1/2 field, and $\bar{\psi} = \psi^{\dagger} \gamma^0$, where $\gamma^0$ belongs to the usual gamma matrices in a flat background.

Once substituted eq. \eqref{determinante métrica gravitacional} in the action \eqref{ação_fermiônica}, one expands $\Gamma_{\textmd{ferm}}[\bar{\psi},\psi,g]$ in terms of the fluctuation field $h_{\mu\nu}$, up to its first order, and extracts the energy-momentum tensor relative to the fermion-graviton tree-level vertex   

\begin{align}
  T_{\mu \nu}(p, p') =& \,
  \frac{\kappa}{8} \Big\{ 2 \eta_{\mu \nu} \big[ (p +p')_\alpha \,\mathcal{J}^\alpha(p, p') - 2m \, \rho(p, p')\big]  \nonumber \\
  &\,\,-  (p + p')_\mu \mathcal{J}_\nu(p, p') - (p + p')_\nu \mathcal{J}_\mu(p, p') \Big\} \,.
\label{tensor_EM_fermion}
\end{align}

\noindent
The tensorial expression in eq. \eqref{tensor_EM_fermion} has the bi-linear $\rho(p, p') = \bar{u}(p') u(p)$ and $\mathcal{J}^{\mu}(p, p') = \bar{u}(p') \gamma^{\mu} u(p)$, where $u(p)$ denotes the free positive energy solution for the four-component spinor and $\bar{u}(p)={u}^\dagger(p) \gamma^0$. In detail

\begin{align}
  u(p) = \sqrt{E + m} \left( \begin{array}{c}\xi \\ \frac{\vec{\sigma} \cdot \vec{p}}{E + m} \, \xi\end{array}  \right) ,
\label{spinor}
\end{align}

\noindent
with $\xi$ representing the two component spinor eigenstates and $\vec{\sigma}$ the Pauli matrices. In on-shell condition, the Dirac equation reads $(\gamma^{\mu} \, p_{\mu} - m) \, u(p) = 0$.

Repeating the same procedure as done for the scalar case and described in section \ref{seção_metodologia}, specifically working on the eq. \eqref{amplitude relativística}, one calculates the relativistic scattering amplitude  

\begin{eqnarray}
  i \mathcal{M}^{(s=1/2)} = \frac{i}{q^2} \left[ \frac{1}{3 Q_2} \left( T_{1 \, \mu}^{\,\, \mu} T_{2 \, \beta}^{\,\, \beta} - 3 \, T_{1}^{\, \mu \nu} T_{2 \, \mu \nu}  \right) + \frac{1}{6 Q_0} T_{1 \, \mu}^{\,\, \mu} T_{2 \, \beta}^{\,\, \beta} \right] \, 
\label{amplitude_R_fermion_compacta}
\end{eqnarray}

\noindent
This equation has the same form of the scalar one \eqref{amplitude_R_escalar_compacta} with its appropriated energy-momentum tensors. After some algebraic manipulations and using a compact notation $\rho_j \equiv \rho (p_j, p_j')$ and $\mathcal{J}_j^{\mu} \equiv \mathcal{J}^{\mu} (p_j, p_j')$, noting that the particles are recognized by index $j=1,2$, $\mathcal{M}^{(s=1/2)}$ is obtained in the following result

\begin{eqnarray}
  \mathcal{M}^{(s=1/2)} &=& \frac{ \kappa^2}{q^2 Q_2} \Bigg\{ \frac{1}{16} (p_1 + p_1')_\mu (p_2 + p_2')_\nu \mathcal{J}_1^{\mu}  \mathcal{J}_2^{\nu}  - \frac{m_1}{8} \rho_1  (p_2 + p_2')_\mu \mathcal{J}_2^{\mu} \nonumber \\
  &-& \frac{m_2}{8} \rho_2  (p_1 + p_1')_\mu \mathcal{J}_1^{\mu} - \frac{1}{32} (p_1 + p'_1)^\nu (p_2 + p'_2)_\nu \mathcal{J}_1^{\mu}  
  \mathcal{J}_{2\mu}    \nonumber \\
  &-& \frac{1}{32}   (p_1 + p_1')_\mu (p_2 + p_2')_\nu \mathcal{J}_2^{\mu}  \mathcal{J}_1^{\nu}  + \frac{m_1 m_2}{3}  \rho_1 \rho_2 \Bigg\} \nonumber \\
  &+& \frac{ \kappa^2}{q^2 Q_0} \, \Bigg\{ \frac{3}{32}  (p_1 + p_1')_\mu (p_2 + p_2')_\nu \mathcal{J}_1^{\mu}  \mathcal{J}_2^{\nu}  + \frac{2 m_1 m_2}{3}  \rho_1 \rho_2  \nonumber \\
  &-&  \frac{m_1}{4} \rho_1  (p_2 + p_2')_\mu \mathcal{J}_2^{\mu}  - \frac{m_2}{4}  \rho_2(p_1 + p_1')_\mu \mathcal{J}_1^{\mu} \Bigg\} \, ,
\label{amplitude_R_fermion} 
\end{eqnarray}

\noindent
with $m_1$ and $m_2$ standing for the fermion masses. Then, moving toward the non-relativistic amplitude, one transfers the physical system to the center-of-mass referential writing the eq. \eqref{amplitude_R_fermion} in function of the momentum transfer $\vec{q}$ and the average $\vec{p}$ (see eq. \eqref{momentum_CM}). The bi-linear $\mathcal{J}_{j}^{\mu}$ and $\rho_j$ in non-relativistic approach are

\begin{subequations}
    \begin{align}
    \rho_{j}|_{\textmd{NR}}  = 2\,m_j \bigg[ 1 + \frac{1}{8m_j^2} \bigg( \vec{q}^{\,2} - 4i(\vec{q} \times \vec{p}\,) \cdot \vec{S}_j \bigg) + \mathcal{O}(3) \bigg] \,,
    \end{align}
    \begin{align}
    \mathcal{J}^0_{j}|_{\textmd{NR}}  = 2\,m_j \bigg[ 1 + \frac{1}{2m_j^2} \bigg( \vec{p}^{\,2} + i(\vec{q} \times \vec{p}\,) \cdot \vec{S}_j \bigg) + \mathcal{O}(3) \bigg]  \,,
    \end{align}
    \begin{align}
    \vec{\mathcal{J}}_{j}|_{\textmd{NR}} = 2\, \chi_j \bigg[ \vec{p} - i (\vec{q} \times  \vec{S}_j) \bigg]  \,.
    \label{bilinear_NR_sem_O3}
    \end{align} 
\label{bilinear_NR}
\end{subequations}

\noindent
The eqs. \eqref{bilinear_NR} are managed in order to present them in a compact form, hence, $\vec{S}_j = \xi'^{\dagger}_j \vec{\sigma} \xi_j$, the factors $\xi'^{\dagger}_j \xi_j$ are omitted, $\chi_1 = 1$, $\chi_2 = -1$ and terms in powers higher than quadratic order for $|\vec{p}|/m_{1,2}$ and/or $|\vec{q}|/m_{1,2}$ are neglected (referenced as $\mathcal{O}(3)$), highlighting that eq. (\ref{bilinear_NR_sem_O3}) is exact.

The non-relativistic elastic scattering amplitude is achieved after the substitution of the eqs. \eqref{bilinear_NR} in eq. \eqref{amplitude_R_fermion} and following the same procedure commented in the scalar case and described in sec. \ref{seção_metodologia}, precisely by eq. \eqref{amplitude não relativística}, resulting in

\begin{eqnarray}
  \mathcal{M}^{(s=1/2)}_{\textrm{NR}} &=& \frac{ \kappa^2 m_1 m_2 }{6 Q_2 \, \vec{q}^{\,2}} \, \Bigg\{ 1  +  \vec{p}^{\,2} \left( \frac{3}{m_1 m_2} + \frac{1}{m_1^2} + \frac{1}{m_2^2} \right)  \nonumber \\
  &+& i \left[ \left( \frac{1}{m_1^2} + \frac{3}{2} \frac{1}{m_1 m_2} \right) \vec{S}_1  
  + \left( \frac{1}{m_2^2} + \frac{3}{2} \frac{1}{m_1 m_2} \right) \vec{S}_2
  \right] \cdot \left( \vec{q} \times \vec{p} \, \right) \nonumber \\
  &-&  \frac{3}{4} \frac{\vec{q}^{\, 2}}{m_1 m_2}  \vec{S}_1 \cdot \vec{S}_2
  + \frac{3}{4} \frac{1}{m_1 m_2} \left( \vec{q} \cdot \vec{S_1} \right) \left( \vec{q} \cdot \vec{S_2} \right) + \mathcal{O}(3) \Bigg\} \nonumber \\
  &-& \frac{ \kappa^2 m_1 m_2 }{24 Q_0 \, \vec{q}^{\,2} } \, \Bigg\{ 1 - \frac{\vec{p}^{\,2}}{2} \left( \frac{1}{m_1^2} + \frac{1}{m_2^2} \right) \nonumber\\
  &-&  \frac{i}{2} \left[ \frac{1}{m_1^2} \vec{S}_1 + \frac{1}{m_2^2} \vec{S}_2
  \right] \cdot \left( \vec{q} \times \vec{p} \, \right) + \mathcal{O}(3) \Bigg\} \, ,
\label{amplitude_NR_fermion} 
\end{eqnarray}

\noindent
and, therefore, the gravitational potential for two spin-1/2 particles interacting via a graviton, resulted from the application of the first Born approximation \eqref{aproximação de Born} on the amplitude \eqref{amplitude_NR_fermion}, is 

\begin{align}
  &V^{(s=1/2)}(r) =  -\frac{\kappa^2 m_1 m_2}{6} \, \Bigg\{ I^{(2)}_1(r) + \vec{p}^{\,2} \left( \frac{3}{m_1 m_2} + \frac{1}{m_1^2} + \frac{1}{m_2^2} \right) I^{(2)}_1(r)    \nonumber\\
  &\quad+ \left[  \left( \frac{1}{m_1^2} +\frac{3}{2} \frac{1}{m_1 m_2} \right) \vec{S}_1 +
  \left( \frac{1}{m_2^2} +\frac{3}{2} \frac{1}{m_1 m_2} \right) \vec{S}_2
  \right] \cdot \frac{\vec{L}}{r} \frac{d}{dr} I^{(2)}_1(r) \nonumber \\
  &\quad- \frac{3}{4} \frac{\vec{S}_1 \cdot \vec{S}_2}{m_1 m_2}  \, I^{(2)}_0(r) +
  \frac{3}{4} \sum_{i,j=1}^{3} \frac{(\vec{S}_1)_i \, (\vec{S}_2)_j}{m_1 m_2} \, I_{ij}^{(2)}(r) \Bigg\} \nonumber \\
  &\quad+  \frac{\kappa^2 m_1 m_2}{24} \, \Bigg\{ I^{(0)}_1(r) - \frac{\vec{p}^{\,2}}{2} \left(  \frac{1}{m_1^2} + \frac{1}{m_2^2} \right) I_1^{(0)}(r)  
  -  \frac{1}{2} \left[ \frac{\vec{S}_1}{m_1^2} + \frac{\vec{S}_2}{m_2^2} \right] \cdot \frac{\vec{L}}{r} \frac{d}{dr} I^{(0)}_1(r) \Bigg\} \, ,
\label{potencial_fermion}
\end{align}

\noindent
observing that $\vec{L}= \vec{r} \times \vec{p} \,\,$ denotes the orbital angular momentum and the appearance of a derivative in the spin-orbit sector is due to the manipulation of particulars Fourier transform and the adoption of spherical coordinate to solve their angular components (for more details, see eq. \eqref{apêndice_int_A}). A last comment is about the anisotropic integral $I_{ij}^{(2)}(r) \,\,$ which is defined in eq. \eqref{apêndice_I_ij}. 

The analyses finalizes with the richness of the gravitational potentials results. The next chapter is dedicated to an exploration of their particularities and a comparison among both expressions eqs. \eqref{potencial_escalar} and \eqref{potencial_fermion}.


\chapter{Spin and Velocity Corrections in Gravitational Potentials}
\label{cap_grav_2}

\section{Scalar and Fermionic Gravitational Potentials: Aspects and Comparisons}
\label{seção_comparação_potenciais}
\indent

Continuing the analysis of the gravitational potentials for external spin-0 and spin-1/2 particles calculated in the last two subsections \ref{seção_potencial_spin0} and \ref{seção_potencial_spin1/2}, each one of the potentials is segregated according to the sectors that compound them. Beginning by the spin-0 scenario, one evidences its two sectors:

\vspace{0.5 cm}
$\bullet$ Monopole-monopole sector 
    \begin{subequations}
	\begin{eqnarray}
	  V_{\textmd{mon-mon}}^{(s=0)}(r) =  
	  &-&\frac{\kappa^2 m_1 m_2}{6} \Bigg[\bigg( I^{(2)}_1(r) - \frac{1}{4}  I^{(0)}_1(r) \bigg)  \nonumber \\
	  &+&\left( \frac{1}{m_1^2} + \frac{1}{m_2^2} \right) \bigg(\frac{1}{8}I^{(2)}_0(r) 
	  +\frac{5}{32}  I^{(0)}_0(r)\bigg)\Bigg] \, .
	\label{potencial_spin0_monopolo}
	\end{eqnarray}
	
\vspace{0.5 cm}
$\bullet$ Velocity-velocity sector
	\begin{eqnarray}
	V_{\textrm{vel-vel}}^{(s=0)}(r) =  &-&\frac{\kappa^2 m_1 m_2}{6}  \, \vec{p}^{\, 2} \Bigg[ \frac{3}{m_1 m_2} I^{(2)}_1(r)  \nonumber \\ 
	&+&\left( \frac{1}{m_1^2} + \frac{1}{m_2^2} \right) \left( I^{(2)}_1(r) + \frac{1}{8} I^{(0)}_1(r) \right) \Bigg] \, . 
	\label{potencial_spin0_velocidade} 
	\end{eqnarray}
\end{subequations}

\vspace{0.5 cm}
\noindent
In the same way, is presented the spin-1/2 sectors:

\vspace{0.5 cm}
$\bullet$ Monopole-monopole sector
\begin{subequations}
	\begin{eqnarray}
	V_{\textmd{mon-mon}}^{(s=1/2)}(r) =  
	-\frac{\kappa^2 m_1 m_2}{6} \bigg( I^{(2)}_1(r) - \frac{1}{4}  I^{(0)}_1(r) \bigg)  \, .
	\label{potencial_spin1/2_monopolo} 
	\end{eqnarray}
	
\vspace{0.5 cm}
$\bullet$ Velocity-velocity sector
	\begin{eqnarray}
	V_{\textrm{vel-vel}}^{(s=1/2)}(r) =  &-&\frac{\kappa^2 m_1 m_2}{6}  \, \vec{p}^{\, 2} \Bigg[ \frac{3}{m_1 m_2} I^{(2)}_1(r)  \nonumber \\
	&+& \left( \frac{1}{m_1^2} + \frac{1}{m_2^2} \right) \left( I^{(2)}_1(r) + \frac{1}{8} I^{(0)}_1(r) \right)
	 \Bigg] \, .
	\label{potencial_spin1/2_velocidade} 
	\end{eqnarray}
	
\vspace{0.5 cm}
$\bullet$ Spin-orbit sector
	\begin{eqnarray}
	V_{\textrm{spin-orbit}}^{(s=1/2)}(r)  = && -\frac{\kappa^2 m_1 m_2}{6}  \, 
	\Bigg\{ \Bigg[  \Bigg( \frac{1}{m_1^2} +\frac{3}{2} \frac{1}{m_1 m_2} \Bigg) \vec{S}_1 \nonumber \\
	&& + \left( \frac{1}{m_2^2} +\frac{3}{2} \frac{1}{m_1 m_2} \right) \vec{S}_2
	\Bigg] \cdot \frac{\vec{L}}{r} \frac{d}{dr} I^{(2)}_1(r) \nonumber \\
	&& + \frac{1}{8} \bigg(\frac{1}{m_1^2} \vec{S}_1+\frac{1}{m_2^2} \vec{S}_2\bigg)
	\cdot \frac{\vec{L}}{r} \frac{d}{dr} I^{(0)}_1(r) \Bigg\} \, .
	\label{potencial_spin1/2_spin-orbita}
	\end{eqnarray}

\vspace{0.5 cm}
$\bullet$ Spin-spin sector
	\begin{align}
	V_{\textrm{spin-spin}}^{(s=1/2)}(r) = -\frac{\kappa^2 m_1 m_2}{6}
	\bigg[ \!-\! \frac{3}{4} \frac{\vec{S}_1 \cdot \vec{S}_2}{m_1 m_2}  \, I^{(2)}_0(r) +
	\frac{3}{4} \sum_{i,j=1}^{3} \frac{(\vec{S}_1)_i \, (\vec{S}_2)_j}{m_1 m_2} \, I_{ij}^{(2)}(r) \bigg] \, .
	\label{potencial_spin1/2_spin}
	\end{align}
\end{subequations}

\noindent
The difference of the potential expressions related to both spin cases are evident and expected, since the spin-1/2 field is richer in terms of physical content. Observing the monopole-monopole sector for both scenarios, one verifies the universal potential term present in eqs. \eqref{potencial_spin0_monopolo} and \eqref{potencial_spin1/2_monopolo}. At the same time, it points out the extra term proportional to the Fourier transform eq. \eqref{apêndice_I_a_n} containing the form factors $W$ and $F$ in spin-0 potential (eqs. \eqref{definição_Q2} and \eqref{definição_Q0}). Once chosen a particular action which defines $W$ and $F$, this term may be managed to recover the universal potential due to its suppressed behaviour. This is exemplified for a specific choice of form factors in the subsection \ref{seção_CDT_fatores_de_forma}. On the other hand, the velocity dependence exists in the two cases under analysis and, comparing eqs. \eqref{potencial_spin0_velocidade} and \eqref{potencial_spin1/2_velocidade}, they have identical structure. The spin-1/2 particles potential reveals two exclusive sectors: the spin-orbit and spin-spin manifestations (eqs. \eqref{potencial_spin1/2_spin-orbita} and \eqref{potencial_spin1/2_spin}, respectively). Taking a closer look at $V_{\text{spin-orbit}}^{(s=1/2)}$, one realizes that both form factors are present in the mentioned potential, while in the sector $V_{\textrm{spin-spin}}^{(s=1/2)}$ there is participation only of $W$.
An important note is the fact that, even though the particularities of the scalar and fermionic fields as the existence of an extra term in monopole-monopole sector for spin-0 particle and the exclusive contributions of spin sectors in spin-1/2 case, the gravitational potential resulted from the imposition of the static limit restrictions ($\vec{p}\rightarrow{\vec{0}} \,$ and $\, m_{i}\rightarrow{\infty}$),

\begin{align}
\frac{1}{m_1 m_2} V_{\textmd{stat.}}^{(s)}(r) =  
\lim_{ \substack{\vec{p}\to \vec{0}\\m_i\to \infty} } \,\frac{1}{m_1 m_2} V^{(s)}(r) ,
\label{potencial_estático_limites}
\end{align}

\noindent
is equal for both type of particles,

\begin{align}
V_{\textmd{stat.}}^{(s)}(r) =  
-\frac{\kappa^2 m_1 m_2}{6} \bigg( I^{(2)}_1(r) - \frac{1}{4}  I^{(0)}_1(r) \bigg)  \, ,
\label{potencial_estático} 
\end{align}

\noindent
and recovers $V_{\textrm{mon-mon}}^{(s=1/2)}(r)$ \eqref{potencial_spin1/2_monopolo}. Furthermore, the classical Newtonian potential $V_{\textmd{stat.}}^{(s)}(r) = -\frac{\kappa^2 m_1 m_2}{32 \pi r} = -\frac{G m_1 m_2}{r}$ is recovered when the form factors and the cosmological constant tend to zero. 

One last observation should be done, the obtained results for spin-0 are passive of being extended to gravitational interaction of spin-0 particles in arbitrary dimensions as worked in refs.\cite{Accioly_CQG_2015} and \cite{Accioly_PRD_2018} for modified theories of gravity in monopole-monopole sector. Things are not so trivial for spin-1/2 particles. The spin arrangement assumes varied structure as the spacetime dimension vary. For instance, see refs. \cite{Dorey_NPB} and \cite{Leo_PRD} for electromagnetic cases, where spacetime with odd dimension are considered and the rising of new spin-dependent effects are discussed.

\section{Comparison with Electromagnetic Potentials}
\label{seção_comparação_EM}
\indent

At this stage, there is the opportunity of comparing the gravitational potentials results achieved in the last section \ref{seção_comparação_potenciais} with the electromagnetic potentials calculated in ref. \cite{Gustavo_PRD} for modified electrodynamics. In the mentioned paper, the authors adopted a very similar physical system configuration and approaching of the problem. The target was the electromagnetic potentials in the non-relativistic limit derived from spin-0, -1/2 and -1 particles interactions with the exchanging of one bosonic vector, what is summarized by the Feynman diagram fig. \ref{diagrama}. Turning the attention to spin-0 and -1/2 particles interactions, the same path chosen in sec. \ref{seção_metodologia} was gone through by the authors in the last mentioned reference, in which the first Born approximation \eqref{aproximação de Born}, amplitude expressions \eqref{amplitude relativística} and \eqref{amplitude não relativística}, center-of-mass reference frame, 3-momentum convention \eqref{momentum_CM} and elastic scattering behaviour were took in account identically. In addition, they worked until the second order in 3-momentum $\mathcal{O}\big(|\vec{p} \,|^{\,2} / {m^2} \big)$, so the spin and orbit couplings became notables. The electrodynamics effective action worked on was 

\begin{align}
\Gamma_{\textmd{EM}}[A] = -\frac{1}{4}\int d^4x \, F_{\mu\nu}(1+H(\Box))F^{\mu\nu} -
\frac{1}{2\alpha}\int d^4x \,(\partial_\mu A^\mu)^2 + \mathcal{O}(F^3) ,
\end{align}

\noindent
the structure $H(\Box)$ equally stood for a form factor as function of d'Alembertian operator. The additional integral was the gauge-fixing term. Vertices and the modified photon propagator were calculated at the tree level. The found resultant photon propagator was

\begin{align}
\langle A_\mu(-q) A_\nu(q) \rangle = -\frac{i}{q^2} \frac{1}{(1+H(-q^2))} \eta_{\mu\nu} + i\Delta_{\mu\nu}(q)\,,
\end{align}

\noindent
which presented similar configuration to eq. \eqref{propagador_graviton}, with $H(\Box)$ factoring the longitudinal and transversal projections operators contented in the Minkowskian metric $\eta_{\mu\nu}$ and $\Delta_{\mu \nu}(q)$ concentrating the terms that contracted with the vector current and vanish. For spin-0 case the 3-vertex originated from an Abelian vector and scalar particles, once they worked on-shell, was $V^{\mu}(p,p')=-ie(p'^{\mu}+p^{\mu})$. In parallel, for spin-1/2 situation the conserved vector current taking place of the vertex was $J^{\mu}(p, p')= e \bar{u}(p') \gamma^{\mu} u(p)$, being $u(p)$ the positive solutions for Dirac equation and $\bar{u}(p)$ its conjugated (the same terminology and convention described in subsection \ref{seção_potencial_spin1/2}) and "$e$" the electric charge.

This brief description of the elegant and rich content paper points out the proximity of physics scenarios and methodology applied among it and the present work, constituting a powerful voluntary to offer a valuable comparison between gravitation and electromagnetic modified theories settled in very close physics arrangements. Segregating the electromagnetic potentials, one has the monopole-monopole and velocity-velocity sectors for spin-0 particles interaction:

\vspace{0.5 cm}
$\bullet$ Monopole-monopole sector
\begin{subequations}
	\begin{align}
	V_{\textmd{EM, mon-mon}}^{(s=0)}(r) = e_1 e_2  I^{\textmd{EM}}_1(r) \, .
	\label{potencial_EM_spin0_monopolo}
	\end{align}

\vspace{0.5 cm}
$\bullet$ Velocity-velocity sector
	\begin{align}
	V_{\textmd{EM, vel-vel}}^{(s=0)}(r) = 
	\frac{e_1 e_2 }{m_1 m_2} \, \vec{p}^{\,2} I^{\textmd{EM}}_1(r) \, .
	\label{potencial_EM_spin0_velocidade}
	\end{align}
\vspace{0.5 cm}
\end{subequations}

\noindent
And monopole-monopole, velocity-velocity, spin-orbit and spin-spin for spin-1/2 particles:

\vspace{0.5 cm}
$\bullet$ Monopole-monopole sector
\begin{subequations}
    \begin{align}
	V_{\textmd{EM, mon-mon}}^{(s=1/2)}(r) =
	e_1 e_2  \left[ I^{\textmd{EM}}_1(r) - 
	\frac{1}{8} \left( \frac{1}{m_1^2} + \frac{1}{m_2^2} \right) I^{\textmd{EM}}_0(r) \right]\, .
    \label{potencial_EM_spin1/2_monopolo} 
    \end{align}

\vspace{0.5 cm}
$\bullet$ Velocity-velocity sector
    \begin{align}
	V_{\textmd{EM, vel-vel}}^{(s=1/2)}(r) =
	\frac{e_1 e_2 }{m_1 m_2} \, \vec{p}^{\,2} I^{\textmd{EM}}_1(r)\, .
    \label{potencial_EM_spin1/2_velocidade}
    \end{align}

\vspace{0.5 cm}
$\bullet$ Spin-orbit sector
    \begin{align}
	V_{\textmd{EM, spin-orbit}}^{(s=1/2)}(r) = e_1 e_2  \,
	\left[  \left( \frac{1}{2m_1^2}+\frac{1}{m_1 m_2} \right) \vec{S}_1 +
	\left( \frac{1}{2m_2^2} +\frac{1}{m_1 m_2} \right) \vec{S}_2
	\right] \cdot \frac{\vec{L}}{r} \frac{d}{dr} I_1^{\textmd{EM}}(r) \, .
    \label{potencial_EM_spin1/2_spin-orbita}
    \end{align}

\vspace{0.5 cm}
$\bullet$ Spin-spin sector
    \begin{align}
	V_{\textrm{EM, spin-spin}}^{(s=1/2)}(r) = e_1 e_2
	\bigg[ \!-\! \frac{\vec{S}_1 \cdot \vec{S}_2}{m_1 m_2}  \, I_0^{\textmd{EM}}(r) +
	\sum_{i,j=1}^{3} \frac{(\vec{S}_1)_i \, (\vec{S}_2)_j}{m_1 m_2} \, I_{ij}^{\textmd{EM}}(r) \bigg] \, .
    \label{potencial_EM_spin1/2_spin}
    \end{align}
\vspace{0.5 cm}
\end{subequations}

\noindent
The integrals $I_n^{\textmd{EM}}(r)$ and $I_{ij}^{\textmd{EM}}(r)$ assume the forms described in eqs. \eqref{apêndice_I_a_n} and \eqref{apêndice_I_ij} respectively, considering the substitution of $Q_a(q^2)$ by $1+H(-q^2)$. Comparing the results for gravitational and electromagnetic potentials, one realizes some correspondences of the sectors acquired in both spin-0 and spin-1/2. In eqs. \eqref{potencial_spin0_monopolo} and \eqref{potencial_EM_spin0_monopolo} are observed the universal terms (Newtonian and Coulombian), adequate to each interaction. For the electromagnetic situation, the integral $I_1^{\textmd{EM}}(r)$ "agglutinates" the contributions from the longitudinal (trivial) and transversal projections once is considered a massless photon and no anisotropy, while, for gravitational case, their projections appear separately through the integrals $I_1^{(2)}(r)$ and $I_1^{(0)}(r)$. The velocity sector manifests the same behaviour as the monopole-monopole one does, that is, contributions of $I_1(r)$. There is no quantum corrections attributed to $I_0^{\textmd{EM}}(r)$ inside the electromagnetic potential for spin-0 external particles. Dislocating towards spin-1/2 particles, the electromagnetic and gravitational potentials in monopole and velocity sectors perform a similar relation as obtained for spin-0. The monopole-monopole sector (eq. \eqref{potencial_EM_spin1/2_monopolo}) contains the $I_1^{\textmd{EM}}(r)$ term and, differently of occurred for spin-0, manifests a quantum correction with $I_0^{\textmd{EM}}(r)$, in contrast to gravitational potential monopole sector that has its result (eq. \eqref{potencial_spin1/2_monopolo}) concentrated in the integrals $I_1$. Moving to the velocity-dependent part, the expressions for both potentials are preserved in relation to spin-0 case. At last, the potentials dependencies on the spin-orbit and spin-spin contributions conserve close form. When $V^{(s=1/2)}_{\textrm{spin-orbit}}(r)$ \eqref{potencial_spin1/2_spin-orbita} is compared to $V^{(s=1/2)}_{\textrm{EM,spin-orbit}}(r)$ \eqref{potencial_EM_spin1/2_spin-orbita}, the rising of the equivalent terms are observed, which differ in factors of proportionality but maintain the major arrangement based on the spins $\vec{S}_1$ and $\vec{S}_2$ participation. One may extend the later analysis to the spin-spin sector, confirming the similarity of the eqs. \eqref{potencial_spin1/2_spin} and \eqref{potencial_EM_spin1/2_spin}. Noting that, for the gravitational scenario, only the term proportional to $\mathcal{P}_{\mu\nu\alpha\beta}^{(2)}$ contributes to this sector, remembering it is associated to $Q^{(2)}(q^2)$ in the propagator \eqref{propagador_graviton} and the same $Q^{(2)}$ is part of the integral $I_0^{(2)}(r)$.

\section{Form Factors Motivated by Quantum Gravity Model} 
\label{seção_CDT_fatores_de_forma} 
\indent

At this section the developed theory is submitted to analyses when applied to a model. The central reference that supports this model is an article by Knorr and Saueressig \cite{Knorr_PRL}, where the authors, through a non-pertubative method, established a reverse engineering procedure to determine a quantum effective action for gravity under low-energy conditions. 

The $\textit{matching-template formalism}$, as named by the authors, begun with an effective action in terms of parameters, that, in their case, is composed by one local and one non-local parts, respectively,

\begin{subequations}
    \begin{align}
	\Gamma^{\textmd{local}} = \frac{1}{16 \pi G_N} \int d^4x\, \sqrt{-g} \, \big( 2\Lambda - R \big) ,
	\label{ação_efetiva_CDT_local}
	\end{align}
	   
	\begin{align}
    \Gamma^{\textmd{NL}} = \frac{-1}{96 \pi G_N} \int d^4x\, \sqrt{-g} \, \big( b^{2} \, R \,\Box^{-2} R + \tilde{b}^2 \, C_{\mu \nu \alpha \beta} \,\Box^{-2} C^{\mu \nu \alpha \beta} \big) ,
    \label{ação_efetiva_CDT_não_local}
    \end{align}
\label{ação_efetiva_CDT}
\end{subequations}

\noindent
with $R$ being the regular Riemannian curvature scalar (see, for instance, \cite{livro_Gasperini}) and $C_{\mu \nu \alpha \beta}$ the Weyl tensor, eq. \eqref{tensor_Weyl}. The criteria to definition of the (non)local parts were the construction of an effective quantum action for gravity and make a connection with cosmological scales, in a way that the authors required the Einstein-Hilbert action and diffeomorphism-invariant contributions quadratic in the curvature tensors and presence of two inverse powers of the Laplacian $\Box^{-2}$ (keeping in mind that the spacetime coordinate system is the Euclidean one). The Ricci tensor $R_{\mu \nu}$ was absent, once is possible to write it in terms of the curvature scalar and the Weyl tensor. For the determination of the parameters $b$, $\tilde{b}$, $G_N$ and $\Lambda$, they worked with a methodology based on foliation structure \cite{Ambjorn_NP_98,Ambjorn_PRL}. This demanded the calculation of two-point autocorrelation functions derived from fluctuations of 3-volume around a toroidal geometry background \cite{Ambjorn_NP_17,Ambjorn_JHEP}. The algebra indicated no contribution of the mass-type parameter $\tilde{b}$, leaving only $b$. By matching the analytical approach with the lattice data from Monte Carlo simulation within Causal Dynamics Triangulation (CDT) (see the ref. \cite{Ambjorn_PR} for a wide review), the authors could reverse engineering the couplings through an establishment of a bind between continuum and lattice. The final non-local gravitational model

\begin{align}
  \Gamma = \frac{2}{\kappa^2} \int d^4x\, \sqrt{-g} 
  \Big( 2\Lambda - R - \frac{b^2}{6} R \,\Box^{-2} R \Big) \,
\label{ação_efetiva_CDT_resultante}
\end{align}

\noindent
is the central part of the Maggiore-Mancarella cosmological model \cite{Maggiore_Mancarella} (for a review of extended gravity models for cosmological scales involving non-local terms, see refs. \cite{Maggiore_JCAP_18,Maggiore_JCAP_20}), in which the authors studied a case with undefined number of spatial dimensions, hence, achieved an automatically dark energy manifestation and self-accelerating background evolution. In ref. \cite{Maggiore_JMIP} the authors studied a very similar non-local gravity model focusing in low energy cosmological dynamics, but altering the Einstein equation  $G_{\mu\nu} - m^2 \tilde{g_{\mu\nu}} = 8 \pi G T_{\mu\nu}$, where $\tilde{g_{\mu\nu}} \approx g_{\mu\nu} \Box^{-1} R$. The inclusion of the non-local term with a mass parameter $m$ generates a dynamical dark energy component, which reproduces the observed dark energy density with no introduction of the cosmological constant. According to the ref. \cite{Shapiro_JHEP}, the terms $R \, \Box^{-2}R$ and $C_{\mu \nu \alpha \beta} \Box^{-2} C^{\mu \nu \alpha \beta}$  appeared due to the decoupling phenomena in a renormalization group analysis. The Weyl tensor was also considered for cosmological perturbation and background evolution levels in non-local gravity in the ref. \cite{Maggiore_PRD}, indeed, performing an action in the mold of eq. \eqref{ação efetiva-eq}.

For completeness of the present study, the Weyl tensor term is also considered. Putting in practice the form factors methodology, one takes the sum of the effective actions $\Gamma^{\textrm{local}}$ and $\Gamma^{\textrm{NL}}$, eqs. \eqref{ação_efetiva_CDT}, as the guide to the analyses and defines

\begin{eqnarray}
    F(\Box) = - \frac{\rho_0}{\Box^2} \,  
    \qquad \textmd{and} \qquad
    W(\Box) = - \frac{\rho_2}{\Box^2} \, , 
\label{fator_de_forma_CDT}
\end{eqnarray}

\noindent
adopting $\rho_0$ and $\rho_2$ as positive parameters motivated by $b^2$ and $\tilde{b}^2$ from eq. \eqref{ação_efetiva_CDT_não_local} and imposing any restriction on them. It seems worthy to clarify that the reconstruction program content in ref. \cite{Knorr_PRL} is apart from the current example and the main contribution to it, besides the precious research delivered to the community and enlargement of its knowledge, is to supply with an effective action applicable (and not a toy model) to reproduce observed data. One more note should be done, the effective actions eqs. \eqref{ação_efetiva_CDT} and \eqref{ação_efetiva_CDT_resultante} are structured on Euclidean signature, and a transformation to Lorentzian one demands a Wick rotation, $\textit{e.g.}$ ref. \cite{Visser_Arxiv}, what is not addressed inside this work. From the form factors (\ref{fator_de_forma_CDT}) are determined the integrals $I_{0}^{(s)}(r)$, $I_{1}^{(s)}(r)$ and $I_{ij}^{(s)}(r)$ participants of the potentials eqs. \eqref{potencial_escalar} and \eqref{potencial_fermion}

\begin{subequations}
	\begin{align}
	I^{(s)}_1(r) = \int \frac{d^3\vec{q}}{(2 \pi)^3} \, \frac{1}{\vec{q}^{\,2} + \mu_s^2}\, e^{i \vec{q} \cdot \vec{r}} \,
	= \frac{e^{-\mu_s r}}{4 \pi r} ,
    \label{integral_CDT_1}
	\end{align}
	
	\begin{align}
	I^{(s)}_0(r) = \int \frac{d^3\vec{q}}{(2 \pi)^3} \, \frac{ \vec{q}^{\,2}  }{\vec{q}^{\,2} + \mu_s^2}\, e^{i \vec{q} \cdot \vec{r}} 
	= \delta^3 (\vec{r})- \mu_s^2 \, \frac{e^{-\mu_s r}}{4 \pi r}  ,
	\label{integral_CDT_0}
	\end{align}
	
	\begin{align}
	I_{ij}^{(s)}(r) &= \int \frac{d^3\vec{q}}{(2 \pi)^3} \, \frac{ \vec{q}_i \vec{q}_j}{  \vec{q}^{\,2} + \mu_s^2 } \,e^{i \vec{q} \cdot \vec{r}} \nonumber \\
	&= \frac{1}{3} \delta_{ij} \delta^3(\vec{r}) 
	+ \bigg[ (1 + \mu_s r) \delta_{ij}  
	-  (3 + 3 \mu_s r + \mu_s^2 r^2) \frac{x_i x_j}{r^2} \bigg] \frac{e^{-\mu_s r}}{4 \pi r^3} \, ,
	\label{integral_CDT_ij}
	\end{align}
\label{integral_CDT}
\end{subequations}

\noindent
where is defined $\mu_s^2 = 2(\rho_s - \Lambda)$ and $(\rho_s - \Lambda) > 0$, what assures that the non-local form factors introduce mass-type terms in the graviton propagator. At last, the particular gravitational potentials, derived from the integrals above (eqs. \eqref{integral_CDT}), are absent of Dirac deltas. Firstly, it is presented the case for spin-0 external particles:

\vspace{0.5 cm}
$\bullet$ Monopole-monopole sector

\begin{subequations}
	\begin{align}
	V_{\textmd{mon-mon}}^{(s=0)}(r) =  
	&-\frac{\kappa^2 m_1 m_2}{24 \pi \,r} \Bigg[ \left( e^{-\mu_{2} r} - \frac{1}{4}e^{-\mu_{0} r}\right) \nonumber \\
	&-\frac{1}{8} \bigg(\frac{1}{m_1^2} + \frac{1}{m_2^2}\bigg) \left( \mu_2^2 \, e^{-\mu_2 r} + \frac{5}{4} \mu_0^2 \,e^{-\mu_0 r} \right) \Bigg] \, .
	\label{potencial_CDT_spin0_monopolo}
	\end{align}
	
\vspace{0.5 cm}
$\bullet$ Velocity-velocity sector	
	\begin{align}
	V_{\textrm{vel-vel}}^{(s=0)}(r) =  -\frac{\kappa^2 m_1 m_2}{24\pi r} \, \vec{p}\,^{2} 
	\Bigg[ \bigg(\frac{1}{m_1^2} + \frac{1}{m_2^2}\bigg)  \left( e^{-\mu_2 r} + \frac{1}{8} e^{-\mu_0 r}\right)
	+ \frac{3}{m_1 m_2} e^{-\mu_2 r}  \Bigg] \, .
	\end{align}
	\label{potencial_CDT_spin0_velocidade}
\end{subequations}
\vspace{0.5 cm}

\noindent
In the sequence, for spin-1/2 external particles:

\vspace{0.5 cm}
$\bullet$ Monopole-monopole sector

\begin{subequations}
	\begin{align}
	V_{\textmd{mon-mon}}^{(s=1/2)}(r) =  
	-\frac{\kappa^2 m_1 m_2}{24 \pi \,r} \left( e^{-\mu_2 r} - \frac{1}{4}e^{-\mu_0 r}\right) \, .
	\label{potencial_CDT_spin1/2_monopolo}
	\end{align}
	
\vspace{0.5 cm}
$\bullet$ Velocity-velocity sector	

	\begin{align}
	V_{\textrm{vel-vel}}^{(s=1/2)}(r) =  -\frac{\kappa^2 m_1 m_2}{24\pi r} \, \vec{p}\,^{2}
	\Bigg[ \bigg(\frac{1}{m_1^2} + \frac{1}{m_2^2}\bigg)  \left( e^{-\mu_2 r} + \frac{1}{8} e^{-\mu_0 r}\right) + \frac{3}{m_1 m_2} e^{-\mu_2 r}  \Bigg] \, .
	\label{potencial_CDT_spin1/2_velocidade}
	\end{align}
	
\vspace{0.5 cm}
$\bullet$ Spin-orbit sector	

	\begin{align}
	V_{\textrm{spin-orbit}}^{(s=1/2)}(r) &= \frac{\kappa^2 m_1 m_2}{24\pi r^3} 
	\Bigg[ \bigg(\frac{1}{m_1^2} \vec{S}_1 \cdot\vec{L} + \frac{1}{m_2^2} \vec{S}_2\cdot\vec{L} 
	+ \frac{3(\vec{S}_1+\vec{S}_2)\cdot\vec{L}}{2\,m_1 m_2} \bigg) 
	 \,  (1+r\mu_2)e^{-\mu_2 r}    \nonumber\\
	&\,+\frac{1}{8} 
	\bigg(\frac{1}{m_1^2} \vec{S}_1 \cdot\vec{L} + \frac{1}{m_2^2} \vec{S}_2\cdot\vec{L}\bigg) 
	\, (1+r\mu_0)e^{-\mu_0 r} \Bigg] \, .
	\label{potencial_CDT_spin1/2_spin_orbita}
	\end{align}
	
\vspace{0.5 cm}
$\bullet$ Spin-spin sector	

	\begin{align}
	V_{\textrm{spin-spin}}^{(s=1/2)}(r) = &-\frac{\kappa^2 m_1 m_2}{32\pi \,r^3} \Bigg[ \frac{\vec{S}_1 \cdot \vec{S}_2}{m_1 m_2}\, (1+r\mu_2+r^2 \mu_2^2) e^{-\mu_2 r} \nonumber \\
	&-3\frac{(\hat{r}\cdot\vec{S}_1 )\, (\hat{r}\cdot \vec{S}_2)}{m_1 m_2}\, (1+r\mu_2+\frac{1}{3}r^2 \mu_2^2) e^{-\mu_2 r} \Bigg] \, .
	\label{potencial_CDT_spin1/2_spin}
	\end{align}
\label{potencial_CDT_spin1/2}
\end{subequations}
\vspace{0.5 cm}

\noindent
The equations are accompanied by exponential factors that generate a damping effect on the potentials sectors in general. This is consequence of the mass-type term presents in the graviton propagator. Concentrating on the respective equations \eqref{potencial_CDT_spin0_monopolo} and \eqref{potencial_CDT_spin1/2_monopolo} for $V_{\textmd{mon-mon}}^{(s=0)}(r)$ and $V_{\textmd{mon-mon}}^{(s=1/2)}(r)$, one observes that they are proportional to $r^{-1}$, what reaffirms the presence of the universal term. Furthermore, both monopole sectors -- for spin-0 and -1/2 -- carry the static limit potential (see eqs. \eqref{potencial_estático_limites} and \eqref{potencial_estático} to review the definitions and result), nonetheless, the spin-0 expression still contains an extra component, as already commented, in a manner that the latter referred potential is written as 

\begin{subequations}

   \begin{align}
	V_{\textmd{mon-mon}}^{(s=0)}(r) = V_{\textmd{stat.}}^{(s=0)}(r) + \Delta V_{\textmd{mon-mon}}^{(s=0)}(r)\, ,
	\end{align}
\text{where}
	\begin{align}
	V_{\textmd{stat.}}^{(s=0)}(r) =
	-\frac{\kappa^2 m_1 m_2}{24 \pi \,r} \left( e^{-\mu_2 r} - \frac{1}{4} \, e^{-\mu_0 r} \right)
    \label{potencial_CDT_spin0_monopolo_termo_estáico}
    \end{align}
\text{and}
	\begin{align}
	\Delta V_{\textmd{mon-mon}}^{(s=0)}(r) =
	\frac{\kappa^2 m_1 m_2}{192 \pi \,r} \bigg(\frac{1}{m_1^2} + \frac{1}{m_2^2}\bigg) \left( \mu_2^2 \, e^{-\mu_2 r} + \frac{5}{4} \mu_0^2 \,e^{-\mu_0 r} \right) \, ,
    \label{potencial_CDT_spin0_monopolo_termo_extra}
    \end{align}
\label{potencial_CDT_spin0_monopolo_completo}
\end{subequations}

\noindent
in which the suppressed behaviour of $\Delta V_{\textmd{mon-mon}}^{(s=0)}(r)$ is passive of being demonstrated. 

Assuming the Newtonian potential is a suitable theory to describe interactions within the bound of the Solar System perimeter, being aware that the experimental data evidence points out that the Newtonian theory, for large distances, presents deviation from the observed data within the galaxy and extra-galaxy scales \cite{Milgrom_AJ,livro_Overduin,Famaey_LRR,Hameeda_PDU}. To define the Solar System range or area is not a trivial task, however, considering, in rough way, its radius ($r_s$) as the semi-major axis of the Pluto's orbit, one finds approximately 39 $\textrm{A.U.}$ \cite{Williams_AJ}. Rescuing the Pioneer anomaly \cite{Anderson_PRL,Turyshev_LRR} that registered from Pioneer 10 and 11 spacecrafts the presence of a small anomalous blue-shifted frequency among the distances of 20 to 70 $\textrm{A.U.}$ from the Sun and kept an unsolved question, it is adopted the distance $r_s \approx$ 10 \, $\textrm{A.U.}$, whereby the Newtonian physics is assumed valid. Since is provided that $\mu_i r_s \ll 1$, then $\mu_i \ll 10^{-25} \textrm{MeV}$, what incurs in $(\mu_i^2/m_j^2) \rightarrow{0} \,$ -- even to particle masses (order $\sim \textrm{MeV}$) -- and vanishes the spin-0 extra term in monopole potential sector, recovering the Newtonian one.

The velocity sector keeps the same $r$-dependency order ($r^{-1}$) as the monopole one. As already argued, under the static limit scenario, where $\, \vec{p}\rightarrow{\vec{0}}$ and $\, m_{i}\rightarrow{\infty}$, this sector is suppressed in both types of particles. The same happens upon the non-relativistic limit, once the characteristic factor of the sector, $\, \vec{p}^{\,\, 2}/m_{i}m_{j}$, tends to a value far smaller than the unity, what practically annuls the velocity sector influence in these gravitational potentials.

The $r$-dependency assumes different orders in the spin-orbit and spin-spin sectors. In the spin-orbit (\ref{potencial_CDT_spin1/2_spin_orbita}) there are terms in $r^{-1}$ and $r^{-2}$, noticing that the angular momentum $\vec{L}$ is proportional to $r$. In its turn, $V_{\textmd{spin-spin}}^{(s=1/2)}(r)$ contains orders from $r^{-1}$ to $r^{-3}$. In these two cases, the $r^{-1}$ terms are suppressed by the same reasons that leaded the extra monopole term $\Delta V_{\textmd{mon-mon}}^{(s=0)}(r)$ to be suppressed, namely, $(\mu_i^2/m_j^2) \rightarrow{0} \,$. 

Therefore, it is notable the predominance of $V_{\textmd{mon-mon}}^{(s)}(r) \sim r^{-1}$, restricted to the static one $V_{\textmd{stat.}}^{(s=0)}(r)$ in the spin-0 case (see eqs. (\ref{potencial_CDT_spin0_monopolo_completo})), over the others in the long-range assumption. Things are quite different when the interactions of short-range distances are analysed. Firstly, taking the spin-0 potential, one observes that

\vspace{0.5 cm}
$\bullet$ Monopole-monopole sector

\begin{subequations}
	\begin{align}
	V_{\textmd{mon-mon}}^{(s=0)}(r) \sim 
	&-\frac{\kappa^2 m_1 m_2}{32 \pi \,r} \bigg( 1 -\sum_{i=0,2} \, \sum_{j=1,2} c_{ij} \frac{\mu_i^2}{m_j^2} \bigg) \, .
	\label{potencial_CDT_spin0_monopolo_curtadistância}
	\end{align}
	
\vspace{0.5 cm}
$\bullet$ Velocity-velocity sector	
	\begin{align}
	V_{\textrm{vel-vel}}^{(s=0)}(r) \sim  -\frac{\kappa^2 m_1 m_2}{32\pi r} \, \vec{p\,\,}^2 \sum_{i,j=1,2} \frac{c_{ij}}{m_im_j} \, .
	\label{potencial_CDT_spin0_velocidade_curtadistância}
	\end{align}
\label{potencial_CDT_spin0_curtadistância}
\end{subequations}

\vspace{0.5 cm}

\noindent
and, then, the gravitational potential for spin-1/2 externals particles

\vspace{0.5 cm}
$\bullet$ Monopole-monopole sector

\begin{subequations}
	\begin{align}
	V_{\textmd{mon-mon}}^{(s=1/2)}(r) \sim  
	-\frac{\kappa^2 m_1 m_2}{32 \pi \,r} \, .
	\label{potencial_CDT_spin1/2_monopolo_curtadistância}
	\end{align}
	
\vspace{0.5 cm}
$\bullet$ Velocity-velocity sector	

	\begin{align}
	V_{\textrm{vel-vel}}^{(s=1/2)}(r) \sim  -\frac{\kappa^2 m_1 m_2}{32\pi r} \, \vec{p\,\,}^2 \sum_{i,j=1,2} \frac{c_{ij}}{m_im_j} \, .
	\label{potencial_CDT_spin1/2_velocidade_curtadistância}
	\end{align}
	
\vspace{0.5 cm}
$\bullet$ Spin-orbit sector	

	\begin{align}
	V_{\textrm{spin-orbit}}^{(s=1/2)}(r) &\sim \frac{\kappa^2 m_1 m_2}{32\pi r^3} 
	\sum_{i,j=1,2} \bigg( \frac{c_{ij}}{m_i m_j} \vec{S}_i \cdot\vec{L} \bigg) \, . 
	\label{potencial_CDT_spin1/2_spin_orbita_curtadistância}
	\end{align}
	
\vspace{0.5 cm}
$\bullet$ Spin-spin sector	

	\begin{align}
	V_{\textrm{spin-spin}}^{(s=1/2)}(r) \sim &-\frac{\kappa^2 m_1 m_2}{32\pi \,r^3} \Bigg( \frac{\vec{S}_1 \cdot \vec{S}_2}{m_1 m_2} \Bigg) \, .
	\label{potencial_CDT_spin1/2_spin_curtadistância}
	\end{align}
\label{potencial_CDT_spin1/2_curtadistância}
\end{subequations}
\vspace{0.5 cm}

\noindent
in which, $c_{ij}$ stands for particular constants to each one of the eqs. \eqref{potencial_CDT_spin0_curtadistância} and \eqref{potencial_CDT_spin1/2_curtadistância}. The potentials lose their damping effect, which tends to the unity as $r$ goes to zero. The expressions presented above (eqs. \eqref{potencial_CDT_spin0_curtadistância} and \eqref{potencial_CDT_spin1/2_curtadistância}) are the major contributions for the gravitational potentials in both external particles under short-range restrictions. In spin-0 situation, the monopole sector, eq. \eqref{potencial_CDT_spin0_monopolo_curtadistância}, still holds as the predominant term, since the velocity part, eq. \eqref{potencial_CDT_spin0_velocidade_curtadistância}, is suppressed in the non-relativistic limit. 

The behaviour of the potential in the static limit (eq. \eqref{potencial_estático_limites}) is straight forward. The condition $\mu_i r_s \ll 1$ is guaranteed because $r$ already tends to zero. Considering the static limit premise $m_j \rightarrow{\infty} \,$, one has $(\mu_i^2/m_j^2) \ll 1\,$, what rises the static potential $V_{\textrm{stat.}}^{(s=0)}(r) = -\frac{\kappa^2 m_1 m_2}{32\pi r} \,$. In contrast, taking a look at the spin-1/2 potential equations \eqref{potencial_CDT_spin1/2_curtadistância}, one notices an inversion of the relevance among the sectors when compared to long-range results manifested in eqs. \eqref{potencial_CDT_spin1/2}. The spin interaction potential, expressed in eq. \eqref{potencial_CDT_spin1/2_spin_curtadistância}, assumes the leading contribution due to its order $r^{-3}$, followed by $V_{\textrm{spin-orbit}}^{(s=1/2)}(r)$ (eq. \eqref{potencial_CDT_spin1/2_spin_orbita_curtadistância}), which carries an order $r^{-2}$. The monopole-monopole term equation \eqref{potencial_CDT_spin1/2_monopolo_curtadistância} tends naturally to the Newtonian potential, but occupying a suppressed position in the gravitational potential under analysis.
In the short-range panorama, one concludes that the leading gravitational potential terms for spin-0 and spin-1/2 external particles are absent of the form factors described in eq. \eqref{fator_de_forma_CDT}. This finding may express the infrared nature of them.     


\section{Partial Conclusion}
\label{seção_conclusão_grav}
\indent

The analysis aims a quantum linearized gravity model, sustained by effective field theory, structured with form factors associated to squared curvature terms, from which is obtained quantum corrections to non-relativistic inter-particle potentials involving scattering of scalars and spinors. The form factors are capable of revealing the relevant information attributed to the graviton propagator in a framework of metric perturbative fluctuations around a flat background. The work contemplates tree level vertices and graviton propagator, for which is obtained quantum corrections. The research is passive of further investigations including loop corrections.

The comparison of spin-0 and $-\frac{1}{2}$ potential sectors, points the presence of the universal potential contribution in both monopole-monopole interactions (eqs. \eqref{potencial_spin0_monopolo} and \eqref{potencial_spin1/2_monopolo}) and a presence of extra sub-leading correction for the scalar one. The velocity-velocity sectors \eqref{potencial_spin0_velocidade} and \eqref{potencial_spin1/2_velocidade} are identical for both particles. The spin-orbit one \eqref{potencial_spin1/2_spin-orbita} manifests contributions of both form factors $F(\Box)$ and $W(\Box)$, while the spin-spin \eqref{potencial_spin1/2_spin} only from $W(\Box)$. In spite of the particularities for each sector, it is observed that, in the static limit (eq. \eqref{potencial_estático_limites}), both gravitational potentials (for spin-0 and $-\frac{1}{2}$ particles) converge to $V_{mon-mon}^{(s=1/2)}(r)$ (eqs. \eqref{potencial_spin1/2_monopolo} and \eqref{potencial_estático}).

An investigation confronting the potentials results of the currently gravitational system with the modified effective electrodynamics \cite{Gustavo_PRD} unveils similar structures in each sector contribution of the spin-0 and -1/2 inter-particles potentials.

The results are submitted to a specific model, in order to evaluate its performance. The model is grounded in a non-perturbative method combined to Casual Dynamics Triangulation to establish a quantum gravity effective action, which subsidizes the information to determine the form factors. The non-relativistic potentials for spin-0 and -1/2 particles in long-range regime present exponential formulation $\sim e^{-\mu r}$ characterizing the damping effect and the universal term leading the contribution for both scenarios. In short-range distances the damping effect disappears, anyway the major spin-0 contribution keep coming from the universal term in monopole sector. To spin-1/2, the spin-spin sector assumes the main contribution, once it is proportional to $1/r^3$ and the monopole-monopole becomes a suppressed correction.

Besides the possibility of expanding the calculations of the quantum corrections present in this study beyond the tree-level propagator and vertices, there is also the opportunity of developing the spin-1 inter-particle potential in the non-relativistic scattering process exchanging a graviton. In ref. \cite{Ross_Arx}, the authors achieved contributions of velocity, spin and involving quadrupole nature for massive spin-1 scattered particle, what endorses the motivation to explore this theme in the future.


\chapter {Modified Maxwell-Higgs Model in an Effective Scenario \\ \it Work in Progress}
\label{cap_vortice}

\section{Introduction}
\indent

Since the phenomenological formulation of superconductivity by Landau and Ginz\-burg in 1950 \cite{Ginzburg_ZET} and its extension for Type-II superconductors by Abrikosov \cite{Abrikosov}, the investigation on topological systems presenting vorticity has been increased and reached several areas of physics, specially in condensate matter, particles physics and cosmology. In the begin of the 70's, Nielsen and Olesen \cite{Nielsen_Olesen} generalized the Landau-Ginzburg-Abrikosov description through a relativistic Abelian Maxwell-Higgs model revealing a quantized magnetic flux electrically neutral and obtained approximate solutions for the equations of motion. Three years later, Bogomol'nyi \cite{Bogomol} introduced an algebraic manipulation that, stipulating some specific values to the potential coupling constant -- the Bogomol’nyi-Prasad-Somerfield (BPS) limit \cite{Prasad_Sommerfield}--, was capable of converting the second order differential equations of the system in first ones. Along the history of physics, planar topological defects have been unfolded in a large variety of research branches. Some of these branches widely developed are the fractional statistics and anyons (see, for instance, \cite{Wilczek_PRL1,Wilczek_PRL2,Wilczek_livro,Khare_livro}). In condensed matter systems are observed giant vortices \cite{Marston_PRL,Engels_PRL,Cren_PRL}. Supersymmetric models \cite{Hindmarsh_RPP} and cosmic strings \cite{Tong_ArX} integer prominent branches as well, while topological arrangements are directly related to the early Universe period during phases transitions \cite{Vilenkin_livro}. Topological theories involving different dimensionalities with unusual kinetic terms are explored in the context of inflationary phase of the Universe \cite{Babichev_JHEP,Garriga_PLB,Rendall_CQG}, dark \cite{Armendariz-Picon_JCAP,Scherrer_PRL} and tachyonic matter \cite{Sen_JHEP}, gravitational waves \cite{Mukhanov_JCAP}, alternatives proposals for Yang-Mills theory in infrared regime \cite{Faddeev_PRL}, strongly interacting particles \cite{Skyrme_PRSA}, solitons mapped by Hopf index \cite{Nicole_JPG,Aratyn_PRL,Adam_JMP} and defects in D-dimensional systems \cite{Bazeia_PRL}.

The vortex description has multi faces. Diversified formulations structuring models to describe the phenomenology of topological defects has been developed since Nielsen and Olesen in 1973. The Maxwell-Higgs model in Abelian version was exactly solved \cite{Vega_PRD}, adopted for two votices interaction \cite{Jacobs_PR}, extended to generate the Chern-Simons term from spontaneous symmetry breaking \cite{Paul_PL}, altered to include non-conventional kinetic terms (see, for example, \cite{Bazeia_EPJ_2011,Bazeia_EPL} and references therein), added Born-Infeld non-linear Electrodynamics (see \cite{Casana_EPJ} and references therein), generalized such that presented compact vortices \cite{Bazeia_EPJ_2017} and non-minimally coupled in non-Abelian version \cite{Cugliandolo_PRD}. Models formed just by Higgs for two-component superconductors \cite{Babaev_PRL} and associated to Born-Infeld \cite{Shiraishi_IJMP} are derivations from the 70's too.

In the manufacturing process of construction of models, the Chern-Simons term stands out due to the possibility of obtaining vortex solutions in Lagrangian possessing its presence summed to the Higgs (kinetic and potential scalar terms). These vortex solutions carry magnetic flux and, contrarily to the Nielsen-Olesen solution, electric charge. In this way, theorists have been explored models in different configurations with the Chern-Simons term. One of them is the pure Chern-Simons theory \cite{Jackiw_PRL}, with multivortex solution \cite{Hong_PRL}, supporting topological and non-topological solitons solutions \cite{Jackiw_PRD}, there is also non-relativistic model \cite{Jackiw-Pi_PRL} and non-minimally coupled in Abelian \cite{Torres_PRD} and non-Abelian \cite{Antillon_PLB} theory generating non-topological magnetic flux. Other configurations investigated are the Maxwell-Chern-Simons-Higgs models \cite{Paul_PLB}, with addition of two neutral fields \cite{Lee_PLB,Bazeia_PRD} to enable the development of Bogomol'nyi prescription \cite{Bogomol}, in non-Abelian theory \cite{Kumar_PLB,Vega_PRL}, carrying non-minimal coupling (see refs. \cite{Antillon_PRD,Anacleto_PLB,Chandelier_PRD} and references therein) and structured in generalized models with unconventional kinetic term \cite{Andrade_PRD} and modified Maxwell term \cite{Ghosh_PLB,Ghosh_PRD}. There are researches segments involving supersymmetry \cite{Helayel_IJMP}, sigma model (see, for instance, refs. \cite{Mendes_MPL,Cunha_PRD,Izquierdo_JHEP}), Lorentz symmetry violation \cite{Casana_PLB} (see also references therein from paper 21 to 24), fermionic theory \cite{Fridberg_PRD,Kim_PRD} and diversified geometry \cite{Ahmed_CTP} (see references therein).

The present work aims to analyse an Abelian Maxwell-Higgs model, minimally coupled and involved by a strong electromagnetic background field. A cosmic star like a neutron star (pulsars or magnetars) could assume the background field source role, perturbing the vacuum structure, spontaneously breaking its symmetry. This perturbation is manifested in the Lagrangian through a modification in the (Minkowski) metric of the kinetic scalar term, similarly practiced in the gravitational theory in eq. \eqref{métrica gravitacional covariante}, originating a new metric which absorbs the field strength representing the Maxwell Electrodynamics of the background field. It is taken into account the scenario with the new metric constant in relation to space-time derivatives. The Bogomol'nyi \cite{Bogomol} equations are determined in the energy ground state condition and vortex solutions are implemented, what leads to quantized magnetic flux coming from the propagating electromagnetic wave. The chapter starts with the Lagrangian containing the constant metric, in a matter that firstly is developed the field equations and achieved the energy (density) of the system (Sec. \ref{Lagrangean 1-modelo_eqs_de_campo}). Then, in subsection \ref{Lagrangean 1-Energia_eq_Bogomol'nyi}, the energy equation is diagonalized, the Bogomol'nyi equations are calculated and the Bogomol'nyi limits obtained. In the sequence \ref{defeitos_topológicos}, the scalar and vector fields are achieved, where is analysed their structure in terms of quantization, anisotropy and continuity. The charge of the system is also found. One last analysis around the energy in the lower bound is executed, where, through the equations system formed by the self-dual type expressions, are formulated the gauge potential $a(r)$ and  the Higgs potential. The partial conclusions in Sec. \ref{Conclusao_3} close the chapter.   


\section{Maxwell-Higgs Model with a Strong Electromagnetic Background Field}
\label{Lagrangean 1}

\subsection{The Model and Field Equations}
\label{Lagrangean 1-modelo_eqs_de_campo}
\indent

The first model proposed is the Maxwell-Higgs Lagrangian (density), similar to the one studied, for example, in refs. \cite{Nielsen_Olesen,Vega_PRD}, containing the particularity of modified metric in relation to the kinetic scalar term, what could classify it as theory of modified kinetic term as pointed above (refs. \cite{Bazeia_EPJ_2011,Andrade_PRD} and references therein). The Maxwell term representing the background field responds for the modification in the Minkowski metric, equivalently to practice in linearized gravitation (eq. \eqref{métrica gravitacional covariante}), inciting the interpretation of the background field as perturbation of elevate intensity in the system. As intensity as capable of perturbing the propagating electromagnetic wave and the vacuum and promoting alteration in the physical properties of the propagating wave as frequency, incurring in birefringence or some anisotropy, as well as collapsing the vacuum through a spontaneous symmetry breaking, what could alter the masses of the gauge field. Such a strong field occurs at the proximity of neutron stars, which could suggest a(n) (ultra)superconducting behavior and topological defects formation.  

The conventions adopted are: $\eta_{\mu\nu} = (+,-,-)$, $x_{\mu} = (x_0, x_i) = (t, -\vec{x})$ and natural unities $c = \hbar = 1$, observing that latin index means spatial components and greek space-time ones. Along the text, $\varphi$ stands for a charged complex scalar field, $f_{\mu\nu}$ is the usual field strength for the gauge field $a_\mu = (\phi, -\vec{a}) $, $f_{\mu\nu} = \partial_{\mu} \, a_{\nu} - \partial_{\nu} \, a_{\mu}$, and the covariant derivative associated to the minimal coupling is identified by $D_{\mu}$ where $D_{\mu} \varphi =  ( \partial_\mu + ig\, a_\mu ) \varphi $ with $g$ being a constant. $G_{\mu\nu}$ is a symmetric tensor playing the role of effective modified Minkowski metric 

\begin{equation}
  G_{\mu\nu} = \eta_{\mu\nu} + \zeta F_{\mu\kappa}F^{\kappa}_{\,\,\,\nu} \,\, ,
\label{métrica_G}
\end{equation}

\noindent
where the modification is due to the presence of $\zeta F_{\mu\kappa}F^{\kappa}_{\,\,\,\nu}$, which represents the field strength for an external strong electromagnetic background field, and  $\zeta$ a constant. This term carries no mass dimension, following $\eta_{\mu\nu}$. In first moment, $G_{\mu\nu}$ is considered to be space-time constant. At last, the Higgs potential $U$ is left unspecified. The first Maxwell-Higgs model evaluated is 

\begin{eqnarray}
  \mathcal{L} = -\frac{1}{4}f_{\mu\nu}^{\,2} + G_{\mu\nu} D^{\mu} \varphi^{*} D^{\nu} \varphi - U \, .
\label{lag_GD_const} 
\end{eqnarray}

\noindent 
The Lagrangian conserves the gauge invariance in relation to the transformations 

\begin{eqnarray}
   \varphi (x) \rightarrow \varphi(x) e^{i\Lambda(x)} \, , \, 
   \varphi^* (x) \rightarrow \varphi^*(x) e^{-i\Lambda(x)} \, , \,
   a_\mu (x) \rightarrow a_\mu (x) - \frac{1}{g} \partial_\mu \Lambda(x) \, .
\label{inv_calibre} 
\end{eqnarray}

\noindent
The field equations that arise from \eqref{lag_GD_const} are

\begin{subequations}
\begin{align} 
  ig \, G_{\mu}^{\,\,\,\, \nu} \,( \varphi D^{\mu} \varphi^{*} - \varphi^{*} D^{\mu} \varphi)  + \partial_{\mu} f^{\mu\nu} = 0\, ,
\label{eq_campo_corrente} 
\end{align}
\begin{align}
  \, - G_{\mu\nu} \, D^{\mu}D^{\nu} \varphi =  \delta_{\varphi^*} U \, ,
\label{eq_campo_dmu1}
\end{align}
\begin{align}
  \, - G_{\mu\nu} \, D^{\mu}D^{\nu} \varphi^{*} = \delta_{\varphi} U  \, ,
\label{eq_campo_dmu2}
\end{align}
\label{eq_campo_lag_GD_const}
\end{subequations}

\noindent
noticing that $\delta_{\varphi^{(*)}} U $ denotes the functional variational principle applied on the Higgs potential in relation to the scalar field $\varphi^*$ in \eqref{eq_campo_dmu1} and $\varphi$ in \eqref{eq_campo_dmu2}.


\subsection{Energy and Bogomol'nyi Type Equations}
\label{Lagrangean 1-Energia_eq_Bogomol'nyi}
\indent

Moving toward the energy-momentum tensor $\Theta^{\mu}_{\,\,\,\nu}$, it presents the structure

\begin{align}
  \Theta^{\mu}_{\,\,\,\,\nu} = G^{\, \mu\kappa}( D_{\kappa} \varphi^{*} D_{\nu} \varphi + D_{\kappa} \varphi D_{\nu} \varphi^{*}) - f^{\mu\kappa} f_{\nu\kappa} - \eta^{\mu}_{\,\,\, \nu} \mathcal{L} \, .
\label{TEM_GD_const} 
\end{align}

\noindent
The component $\Theta^{0}_{\,\,\,0}$ is passive of being algebraically manipulated. The energy assumes the equation

\begin{eqnarray}
  \mathcal{E} &=& \int d^{\, 2} x ( 1 + \zeta E_{i}^{\,2} ) D_{0} \varphi^{*} D_{0} \varphi + \, [(1 - \zeta B^{\,2}) \, \delta_{i j} + \zeta E_i E_j ] \, D_{i} \varphi^{*} D_{j} \varphi  \nonumber \\
  && + \frac{1}{2} (e_{i}^{\,2} + b^{\,2}) + U \,,
\label{energia_GD_const2}
\end{eqnarray}

\noindent
which is exempted of crossed spatio-temporal terms. $E_i$ and $B$ stand for the electric and magnetic background fields, respectively, as well as $e_i$ and $b$ for the propagating wave fields. This configuration suggests a matrix diagonalization trial through rotation of the elements. The eigenvalues, which compound the diagonalized Hamiltonian (density), are $1 + \zeta E^2$ and $ 1 $, turning the equation \eqref{energia_GD_const2} in

\begin{eqnarray}
  \mathcal{E}  &=& \int d^{\, 2} x ( 1 + \zeta \vec{E}^{\,2} ) \bar{D}_{0} \varphi^{*} \bar{D}_{0} \varphi + ( 1 + \zeta \vec{E}^{\,2} ) \, \bar{D}_{1} \varphi^{*} \bar{D}_{1} \varphi + \bar{D}_{2} \varphi^{*} \bar{D}_{2} \nonumber \\
  &+& \int d^{\, 2} x \frac{1}{2} (\vec{e}^{\,\,2} + b^{\,2}) + U \, ,
\label{energia_GD_const3}
\end{eqnarray}

\noindent
observing that $\bar{D}_{\mu}$ means the covariant derivative expressed in the rotated base. Along the diagonalization calculus, an intrinsic constrain emerges: $\zeta B^2 = 0 $. Then, the equality (dropping the "bar" sign)

\begin{eqnarray}
    && - \big| \big( a D_0 \, + b D_1 \, + c D_2 \big) \, \varphi \big|^2 + 
    \big| \big( a D_0 \, + b D_1 \big) \, \varphi \big|^2 + \big| \big( a D_0 \, + c D_2 \big) \, \varphi \big|^2  \nonumber \\
    && + \big|\big( b D_1 \, + c D_2 \big) \, \varphi \big|^2 = a^2 D_{0} \, \varphi^{*} D_{0} \, \varphi + b^2 \, D_{1} \, \varphi^{*} D_{1} \, \varphi + c^2 D_{2} \, \varphi^{*} D_{2} \, \varphi 
\label{igualdade_d}
\end{eqnarray}

\noindent
is adopted in eq. \eqref{energia_GD_const3}, where $a$, $b$ and $c$ are constants. In this case, the usual relation ($ D_i \varphi D_i \varphi^* \sim | (D_1 \pm i D_2 ) \varphi |^2 $, few examples in \cite{Bazeia_EPJ_2017,Jackiw_PRD}) is rewritten in a form including the time component. Noteworthy of the left side of eq. \eqref{igualdade_d}, which is supported by triangular inequality -- establishes that: being $\vec{M}$ and $\vec{N}$ vectors,  $|\vec{M}| + |\vec{N}| \geq |\vec{M} + \vec{N} | \, $ (see, for instance, ref. \cite{Shankar_livro}) -- and it guarantees that the right side of eq. \eqref{igualdade_d} is equal or bigger than zero (even tough $D_0$ being a scalar). Managing the second integral of the right side of eq. \eqref{energia_GD_const3} to determine the Bogomol'nyi type equations and the inferior energy bound, one finds for the electromagnetic field of the free propagating wave and Higgs potential that

\begin{subequations}
\begin{eqnarray}
  \int d^{\, 2} x \,\, \frac{1}{2} (\vec{e}^{\,\,2} + b^2) + U &=& 
  \int d^{\, 2} x \,\, \frac{1}{2} \big[(|\vec{e} \,| \pm b) \pm \sqrt{2U} \big]^2 \mp \int d^{\, 2} x \,\, ( |\vec{e} \,| \pm b ) \sqrt{2U} \nonumber\\
  &\mp& \int d^{\, 2} x \,\, |\vec{e} \,| \, b \,
\label{campo_em_Higgs1}
\end{eqnarray}
\text{or}
\begin{eqnarray}
  \int d^{\, 2} x \,\, \frac{1}{2} (\vec{e}^{\,\,2} + b^2) + U &=& 
  \int d^{\, 2} x \,\ \frac{1}{2} (|\vec{e} \,| \pm \sqrt{U})^2 
  \mp \int d^{\, 2} x \,\, |\vec{e} \,| \sqrt{U} \nonumber\\
  &+& \int d^{\, 2} x \,\, \frac{1}{2} (b \pm \sqrt{U})^2 
  \mp \int d^{\, 2} x \,\, b\sqrt{U} \, .
\label{campo_em_Higgs2}
\end{eqnarray}
\label{campo_em_Higgs}
\end{subequations}

\noindent
The liberty of choosing the minus and plus signals in equations \eqref{campo_em_Higgs} guarantees to each term be bigger than zero and, consequently, the energy be positive. In order to obtain the ground energy value from equation eq. \eqref{energia_GD_const3}, taking into account eqs. \eqref{igualdade_d} and \eqref{campo_em_Higgs}, are determined the self-dual expressions   

\begin{subequations}
\begin{eqnarray}
   D_{0} \, \varphi^{*} D_{0} \, \varphi = - ( D_{1} \, \varphi^{*} D_{1} \, \varphi + \, \frac{1}{1 + \zeta \vec{E}^{\,\,2}} \, D_{2} \, \varphi^{*} D_{2} \, \varphi ) \, ,
\label{eq_Bogomol1}
\end{eqnarray}
\begin{align}
   (|\vec{e} \,| \pm b) - \sqrt{2U} = 0 \,\,\, \text{for eq. \eqref{campo_em_Higgs1}}
\label{eq_Bogomol2}
\end{align}
\text{or}
\begin{align}
   |\vec{e} \,| - \sqrt{U} = 0 \,\,\, \text{and} \,\,\, b - \sqrt{U} = 0 \,\,\, \text{for eq. \eqref{campo_em_Higgs2}}\, ,
\label{eq_Bogomol3}
\end{align}
\label{eq_Bogomol}
\end{subequations}

\noindent
and the Bogomol'nyi limit-type equation in the saturated scenario (eq. \eqref{eq_Bogomol}) is

\begin{subequations}
\begin{eqnarray}
   \mathcal{E}' \geq + \sqrt{2U} ( \Phi_{e} \pm \Phi_{b} ) \mp \int d^{\, 2} x \, |\vec{e} \,| b \,\,\,\, \text{for} \,\,\eqref{campo_em_Higgs1} \, ,
\label{eq_limite_Bogomol_1}
\end{eqnarray}
\text{or}
\begin{align}
   \mathcal{E}'' \geq  + \sqrt{U} ( \Phi_{e} + \Phi_{b} ) \,\,\,\, \text{for} \,\, \eqref{campo_em_Higgs2},
\label{eq_limite_Bogomol_2}
\end{align}
\label{eq_limite_Bogomol}
\end{subequations}

\noindent
 with $\Phi_{b}$ representing the magnetic flux $\int d^{\,2} x \,\, b $. One observes in both equations \eqref{eq_limite_Bogomol} the appearance of the electric flux $\, \int d^{\,2} x \, |\vec{e} \,| \, \equiv \, \Phi_{e} \,$, which is not commonly obtained and, depending on $a_0$, it will or not be quantized. Besides, in eq. \eqref{eq_limite_Bogomol_1}, one of the energy parcel`s is originated from the Poynting vector $\, \int d^{\, 2} x \, |\vec{e} \,| \, b \, $, where both are unexpected. No stationary condition is imposed and the gauge choice is opened, so far. The possibility of expressing in two formulations \eqref{campo_em_Higgs} the same energy term $\big( \int d^{\, 2} x \,\, \frac{1}{2} (\vec{e}^{\,\,2} + b^2) + U \big)$ opens an alternative to define the energy lower bound in function of the Poynting vector \eqref{eq_limite_Bogomol_1}. The self-dual type equations (eqs. \eqref{eq_Bogomol2} and \eqref{eq_Bogomol3}) should converge in some moment. Then, demanding $\mathcal{E}' = \mathcal{E}''$, the Poynting vector becomes $\,  \mp \int d^{\, 2} x \, |\vec{e} \,| \, b \, =  \sqrt{U} \, [( \Phi_{e} + \Phi_{b} ) - \sqrt{2} ( \Phi_{e} \pm \Phi_{b} )] \,$. So, the convergence of the energy formulations $\mathcal{E}'$ and $\mathcal{E}''$ enables to express the Poynting vector as function of the magnetic and electric fluxes.


\subsection{Topological Vortices Solutions}
\label{defeitos_topológicos}
\indent

The proposition of this section is to find vortices solutions to the Maxwell-Higgs model in scene, starting from generalized (reads initially unspecified) vector field carrying, in principle, radial and angular degrees of freedom in a polar coordinate system. The expectation is to determine algebraically $a_0$ and $\vec{a}$ from the field equations. Thus, is defined

\begin{subequations}
  \begin{eqnarray}
    a_0 (x,y) \rightarrow a_0 (r,\theta) 
  \label{coord_pol_a_0}
  \end{eqnarray}
  \begin{eqnarray}
    a(x,y) \rightarrow a (r,\theta)
  \label{coord_pol_a}
  \end{eqnarray}
\end{subequations}

\noindent
where, in principle, the gauge potentials $a_0$ and $a$ are assumed dependent on both coordinates ($r$ and $\theta$). Aiming vortex solution, is chosen for the scalar field a recurrent ansatz ({\it e.g.}, refs. \cite{Torres_PRD,Vega_PRD,Jackiw_PRL,Casana_PLB_2013}) 

\begin{eqnarray}
  \varphi (r, n, \theta) = f(r, n) e^{in\theta}
  \label{campo_escalar}
\end{eqnarray}

\noindent
with $n$ being an integer and $f(r, n)$ a function of the radius $r$ and the integer $n$. In this context, maintaining the raised constraint $\zeta B^2 = 0$ along the whole subsequent analysis, the temporal component of the field equation \eqref{eq_campo_corrente} given by the charge (density) $J^0 \equiv i g \, G_{\mu}^{\,\,\,\,0} \,( \varphi D^{\mu} \varphi^{*} - \varphi^{*} D^{\mu} \varphi )$, the so called Gauss equation, becomes 

\begin{eqnarray}
  \Big( \nabla^2 - (1 + \zeta \vec{E}^2) 2 g^2 f^2 \Big) \, a_0 = 0 \, ,
\label{eq_Gauss_estático}
\end{eqnarray}

\noindent
where is restricted to the stationary case ($\partial_0 \, a_\mu = 0$ and $\partial_0 \, \varphi = 0$ ). These constrains are adopted from this point on. The eq. \eqref{eq_Gauss_estático} is algebraically resolvable setting $ \big(\frac{1}{r} \frac{\partial^2}{\partial \theta^2} + \frac{\partial}{\partial r} \big) a_0 = 0$, resulting in an exponential form to $a_0 (r, n, \theta) = l \, e^{-\chi(r, n, \theta)} \,$ where $l$ is a constant and the function $\chi(r, n, \theta)$ allows a few different combinations of terms involving $r$ and $\theta$. Some matching varying the power degree of the radius $r$ and considering $\chi$ a pure real or pure imaginary, were tried. In general, in the limit $r \rightarrow 0$, they presented a divergent or vanishing behaviour. In a way that the chosen option is $\chi(r, n, \theta) = \chi(r, n) + in \theta$, where $\chi (r, n) $ is a general function of $r$ and $n$. This choice naturally leads to $\chi(r, n) \sim n^2 \ln{r}$, with $f$ from eq. \eqref{eq_Gauss_estático} being resolved explicit and presenting an unexpected quantization $n$, 

\begin{subequations}
  \begin{eqnarray}
    a_0 (r, n, \theta) = l \, r^{n^2} e^{- in \theta}  \, ,
   \label{pot_escalar}
   \end{eqnarray}
   \begin{eqnarray}
    f(r,n) = \sqrt{\frac{n^2(n^2 - 1)}{2g^2(1+ \zeta \vec{E}^{\,2})} } \frac{1}{r} \,\,\, ,
  \label{função_f}
  \end{eqnarray}
  \begin{eqnarray}
    && \lim_{r\rightarrow 0} a_0 = 0 \,\,\, \text{and} \,\,\, \lim_{r\rightarrow \infty} a_0 = \infty \, , \nonumber \\
    && \lim_{r\rightarrow 0} f = \infty \,\,\, \text{and} \,\,\, \lim_{r\rightarrow \infty} f = 0 \, ,
  \label{limites_a_0_f}
  \end{eqnarray}
\end{subequations}

\noindent
observing that $l$ absorbs the constant emergent from the integral calculation of $\chi(r, n)$. The scalar potential $a_0$ points out an anisotropic solution dependent on $\theta$. It also contains a topological feature manifesting quantization and a particular direct proportionality to $r$ and $n$. The limit of $a_0$ is divergent for long ranges, in other hand, the function $f(r, n)$ detains a singularity at $r \rightarrow 0$, which reflects in the scalar field $\varphi$ \eqref{campo_escalar}. In generalized models including effects of magnetic permeability \cite{Bazeia_EPJ_2017} (which is an extension of multivortex study in ref. \cite{Hong_PRL}) is obtained resembling quadratic dependence on $n$ for scalar functions like $f(r)$. Analysing the current (density) -- Ampère-Maxwell equation --, extracted from eq. \eqref{eq_campo_corrente},

\begin{eqnarray}
  \Big( \nabla^2 \delta_{ij} - \partial_i \partial_j + 2 f^2 g^2 (\eta_{ij} + \zeta E_i E_j) \Big) a_j = - 2f^2 g n (\eta_{ij} + \zeta E_i E_j) \partial_j \theta
\label{eq_Ampere-Maxwell_1}
\end{eqnarray}

\noindent
in which are explicit the individualized contributions of the vector potential components $a_1$ and $a_2$. Particularizing to the already mentioned conditions of stationary scenario, one finds that, firstly for $ i = 1$,

\begin{eqnarray}
     && \bigg( \sin{\theta} \frac{\partial}{\partial r} - \cos{\theta} \frac{1}{r} \frac{\partial}{\partial \theta} \bigg)  
         \bigg( \frac{\partial a}{\partial r} 
     + \frac{1}{r} a \bigg) \nonumber \\
     && = 2 g f^2 \Big( g a + \frac{n}{r} \Big) \Big( \sin{\theta} + \zeta E_1 ( \cos{\theta} \, E_2 - \sin{\theta} \, E_1 ) \Big) \, ,
\label{eq_Ampere-Maxwell_1_estático}
\end{eqnarray}

\noindent
and for $i = 2$

\begin{eqnarray}
     && \bigg( \cos{\theta} \frac{\partial}{\partial r} + \sin{\theta} \frac{1}{r} \frac{\partial}{\partial \theta} \bigg)  
     \bigg( \frac{\partial a}{\partial r} + \frac{1}{r} a \bigg) \nonumber \\
     && = 2 g f^2 \Big( g a + \frac{n}{r} \Big) \Big( \cos{\theta} + \zeta E_2 ( \sin{\theta} \, E_1 - \cos{\theta} \, E_2 ) \Big) \, .
\label{eq_Ampere-Maxwell_2_estático}
\end{eqnarray}

\noindent
Equations \eqref{eq_Ampere-Maxwell_1_estático} and \eqref{eq_Ampere-Maxwell_2_estático} reveal a potential source of anisotropy due to the background electric field components ${E}_1$ and ${E}_2$ presented in both equations. In Amperè-Maxwell equation, the stationary scenario transfers exclusively the source of the current to the magnetic field, which in this case is $ b = \big(  \frac{\partial}{\partial r} + \frac{1}{r} \big) a $. There are two extra approaches to solve the differential equations system formed by eqs. \eqref{eq_Ampere-Maxwell_1_estático} and \eqref{eq_Ampere-Maxwell_2_estático}, which consist in impose a magnetic field $b$ only in function of $r$ or $\theta$, that means $ \frac{\partial b}{\partial \theta} = 0 $ or $ \frac{\partial b}{\partial r} = 0 $, respectively. In both situations, $a(r, \theta)$ results in

\begin{eqnarray}
     a(r,n) = - \frac{n}{g} \frac{1}{r} \, \hat{\theta} \,\, .
\label{potencial_vetorial_1}
\end{eqnarray}

\noindent
The solution leads $b$ to zero, has a topological feature with vorticity $n$ and is dependent only on $r$. It also presents the expected angular direction of movement $\hat{\theta}$, perpendicular to the radius direction $\hat{r}$. Noticing the bound limits in \eqref{potencial_vetorial_1}, when $r\rightarrow \infty$ it has a convergent behaviour with $a = 0$, while it diverges in $r\rightarrow 0$. Situations of singularities in $r \rightarrow 0$ for $a(r) \sim \frac{1}{r} $ are commonly prevented inserting a general function at the gauge potential expression, which controls the limit bound (see some examples in ref. \cite{Chandelier_PRD}, in the context of a Maxwell-Chern-Simons model coupled to an external background charge, in ref. \cite{Andrade_PRD}, which treats a generalized Maxwell-Chern-Simons-Higgs model with modified Maxwell and kinetic terms and in ref. \cite{Jackiw_PRD}, where is demonstrated that Chern-Simons solitons support topological and non-topological solutions). At the same time, exists in literature a case of non-vanishing gauge potential at large distances \cite{Ghosh_PLB}. An observation in relation to these examples is the fact that they start from an ansatz $a(r)\sim \frac{1}{r}$, while a very similar result was obtained naturally from the calculation of the field equations in the present work. A third option is to solve, probably numerically, the expression resultant from the manipulation and summing of eqs. \eqref{eq_Ampere-Maxwell_1_estático} and \eqref{eq_Ampere-Maxwell_2_estático}, 

\begin{equation}
     \Omega \, \Bigg( \frac{\partial^2}{\partial r^2} + \frac{1}{ r}\frac{\partial}{\partial r} - \bigg(1 + \frac{1}{\Omega} \bigg) \frac{1}{r^2} \Bigg) a = \frac{n}{g} \frac{1}{r^3} \, \hat{\theta} \,\,\,
\label{eq_Ampere-Maxwell_estático_soma}
\end{equation}
with
\begin{equation}
   \Omega(\theta, n, E) = \frac{1 + \tan^2{\theta}}{2 g^2 \tau^2 (1 + \tan^2{\theta} + \zeta (\tan{\theta} \, E_1 - E_2)^2 )}  \, , \nonumber
\end{equation}

\noindent
observing that $\tau (r,n) = r f(r, n)$ \eqref{função_f}. The algebraic resolution is feasible imposing $\frac{\partial a}{\partial r} - \bigg(1 + \frac{1}{\Omega} \bigg) \frac{a}{r} = 0$, so that

\begin{equation}
     a(r,\theta, n, E) = \frac{\Omega}{ \Omega + 1} \, \frac{n}{g} \frac{1}{r} \, \hat{\theta}, \nonumber 
\end{equation}
in an explicitly way
\begin{equation}
     a(r,\theta, n, E) = \frac{1 + \zeta \vec{E}^{\,2}}{ 1 + \zeta \vec{E}^{\,2} + \big( 1 + \zeta (E_1 \sin{\theta} - E_2 \cos{\theta} )^2 \big) n^2(n^2+1)} \, \frac{n}{g} \frac{1}{r} \, \hat{\theta} \,\, .
\label{potencial_vetorial_2}
\end{equation}

\noindent
$a(r,\theta, n , E)$ conserves the main structure manifested in \eqref{potencial_vetorial_1}, however the gauge potential is also in function of $\theta$ and $E$. Beyond that, it carries vorticity more elaborated and anisotropy because of $\theta$ and $E$ components dependency.  

Turning the attention to the charge $Q$, it reveals an electrically neutral flux

\begin{equation}
    Q = \int \vec{r} \, d\vec{r} d\theta \, 2 \, g^2 (1 + \zeta \vec{E}^2 ) f^2 a_0 \, = 0 \, ,
\label{carga}
\end{equation}

\noindent
in which $a_0$ is given by \eqref{pot_escalar}. Paul and Khare \cite{Paul_PL} also found neutral vortices from a model containing Chern-Simons term generated by spontaneous symmetry breaking and a line of investigation developed by a Brazilian group works with neutral vortices as consequence of the gauge-fixing adopted (see refs. \cite{Bazeia_EPL,Andrade_PRD} just as a start point for a long researching line), noticing that the particular investigation of this group on generalized Maxwell-Higgs models  \cite{Bazeia_EPJ_2017} resulted in ones with neutral charge and, in spite of it, with existence of quantized magnetic flux, as observed in the present study. Additionally, in this thesis there is the presence of the electric flux too. 

A last analysis is an inspection on the energy functional equation \eqref{energia_GD_const3} expressed in terms of polar coordinates

\begin{eqnarray}
   \mathcal{E} &&= \int \, r \, dr d\theta \,\,\, (1 + \zeta \vec{E}^{\,\,2}) \Big(\frac{\tau}{r} \, g a_{0} \Big)^2  \nonumber\\ 
   &&+ 2 \frac{\tau^2}{r^4} + \frac{\tau^2}{r^2} \Big(-\frac{1}{r^2} + \frac{n^2}{r^2} + g^2 a^2 + \frac{2n}{r} g a \Big) \nonumber \\
   &&+ \zeta \vec{E}^{\,\,2} \bigg[ \frac{\tau^2}{r^4} + \frac{\tau^2}{r^2} \sin^2{\theta} \Big( -\frac{1}{r^2} + \frac{n^2}{r^2} + g^2 a^2 + \frac{2n}{r} g a \Big) \bigg] \nonumber\\
   &&+ \frac{1}{2} \bigg[ \bigg( \frac{1}{r}\frac{\partial a_0}{\partial \theta} \bigg)^2 + \bigg( \frac{\partial a_0}{\partial r} \bigg)^2 + \bigg( \frac{\partial a}{\partial r} + \frac{a}{r} \bigg)^2 \bigg] + U \, ,
\label{energia_GD_const_coord_polar1}
\end{eqnarray}

\noindent
what evidences the topological structure associated to $\tau$ and to the gauge potentials $a_0$ and $a$, endorsing the quantized characteristic of the energy. The dependency on $\theta$ indicates an anisotropy in the energy value. Moving towards the calculation of the Bogomol'nyi limit \eqref{eq_limite_Bogomol}, one rescues the self-dual relations \eqref{eq_Bogomol} in polar coordinates,

\begin{subequations}
\begin{eqnarray}
   &-& (1 + \zeta \vec{E}^{\,\,2}) \, a_{0}^2 = \bigg( a^2 + \frac{2n}{g} \frac{a}{r} \Big) (1 + \zeta \vec{E}^{\,\,2} \sin^2{\theta}) \nonumber \\
   &+& \frac{1}{g^2 r^2} \big( n^2 - 1 \big)(1 + \zeta \vec{E}^{\,\,2} \sin^2{\theta})+ \frac{1}{g^2 r^2} ( 2 + \zeta \vec{E}^2 ) \, ,
\label{eq_Bogomol_coor_pol1}
\end{eqnarray}
\text{where one has for eqs. \eqref{campo_em_Higgs1} and \eqref{eq_Bogomol1}}
\begin{eqnarray}
   \bigg[ \bigg( \frac{1}{r}\frac{\partial a_0}{\partial \theta} \bigg)^2 + \bigg( \frac{\partial a_0}{\partial r} \bigg)^2 \bigg]^{\frac{1}{2}}  \pm \bigg( \frac{\partial a}{\partial r} + \frac{a}{r} \bigg) - \sqrt{2U} = 0 \nonumber\\
\label{eq_Bogomol_coor_pol2}
\end{eqnarray}
\text{or for eqs. \eqref{campo_em_Higgs2} and \eqref{eq_Bogomol2}}
\begin{eqnarray}
   \bigg[ \bigg( \frac{1}{r}\frac{\partial a_0}{\partial \theta} \bigg)^2 + \bigg( \frac{\partial a_0}{\partial r} \bigg)^2 \bigg]^{\frac{1}{2}} =
   \frac{\partial a}{\partial r} + \frac{a}{r} = \sqrt{U} \, .
   \label{eq_Bogomol_coor_pol3}
\end{eqnarray}
\label{eq_Bogomol_coor_pol}
\end{subequations}

\noindent
With the equations above is possible to calculate $U$. The self-dual expression \eqref{eq_Bogomol_coor_pol1} enables to set the gauge potential $a_0$ in function of $a$ and vice versa, which in association with eq. \eqref{eq_Bogomol_coor_pol2} or \eqref{eq_Bogomol_coor_pol3} forms a equations system and provide material enough to calculate $a_0$, $a$ and $U$ explicitly in the inferior bound energy regime. Then, one taking \eqref{eq_Bogomol_coor_pol1} demonstrates that

\begin{subequations}
\begin{eqnarray}
   a (r, \theta, n) = - \frac{n}{g} \frac{1}{r} \, \hat{\theta} \pm H(r,\theta, n) \, \hat{\theta}
\label{limite_energia_a}
\end{eqnarray}
\text{where}
\begin{eqnarray}
   H(r, \theta, n) = \bigg[\frac{1}{g^2r^2} \bigg( 1 - \frac{1}{1+\zeta \vec{E}^{\,\,2} \sin^2{\theta}} \bigg) + \bigg( a_0^2 - \frac{1}{g^2r^2} \bigg)\bigg( \frac{1 + \zeta \vec{E}^{\,\,2}}{1 + \zeta \vec{E}^{\,\,2} \sin^2{\theta}} \bigg) \bigg]^{\frac{1}{2}} , \nonumber\\
\label{limite_energia_H}
\end{eqnarray}
\label{limite_energia}
\end{subequations}

\noindent
and substituting eq. \eqref{limite_energia_a} in \eqref{eq_Bogomol_coor_pol3}, one writes that

\begin{equation}
    \mp \bigg( \frac{\partial H}{\partial r} + \frac{H}{r} \bigg) = \sqrt{U} \, .
\label{potencial_Higgs}
\end{equation}

\noindent
The results presented in \eqref{limite_energia} and \eqref{potencial_Higgs} are in function of $a_0$. Remember that, working with the expression for the electric field in eq. \eqref{eq_Bogomol_coor_pol3}, $a_0$ may be determined probably numerically due to the complexity of the equations system. The eq. \eqref{limite_energia_a} preserves a similar structure to eqs. \eqref{potencial_vetorial_1} and \eqref{potencial_vetorial_2}, but has no dependency on the background electric field. It contains an anisotropy generated by $\theta$ and the singularity when $r \rightarrow 0$. All the features just described for $a (r, \theta, n)$ are propagated to the Higgs potential $U$ in eq. \eqref{potencial_Higgs}, which means that it carries anisotropy and is quantized in the minimum energy bound scenario. And finally, once obtained the potential $U$, one is in condition of calculating the energy lower bound $\mathcal{E'}$ \eqref{eq_limite_Bogomol_1} or $\mathcal{E''}$ \eqref{eq_limite_Bogomol_2}. Observing that the magnetic flux (and probably the electric too) generator of $\mathcal{E'}$ and $\mathcal{E''}$ is quantized.

This closes the present section involving the Maxwell-Higgs model with a strong electromagnetic background field with constant metric. Some comments about the results achieved along the section are exposed in the Partial Conclusions \ref{Conclusao_3}. In the next section is proposed a similar model to the first one, differing by a non-constant metric.


\section{Partial Conclusion}
\label{Conclusao_3}
\indent

The exploration through the two Maxwell-Higgs models unveiled some aspects of the topological physics around these systems. Initially, it seems worthy to briefly debate about the methodology applied to approach the cases, which has few variations in relation to the usually one adopted \cite{Bogomol}. The diagonalization of the energy density equation \eqref{energia_GD_const2} allows to write it in a simplified arrangement \eqref{energia_GD_const3} carrying the differential temporal component and naturally manifesting an anisotropy associated to the energy (see eq. \eqref{energia_GD_const3} or \eqref{energia_GD_const_coord_polar1}). It should be added that the triangular inequality \eqref{igualdade_d} permits to express the differential terms from the diagonalized eq. \eqref{energia_GD_const3} in definite positive ones, keeping the temporal component, and assures that the equation \eqref{igualdade_d} is equal or bigger than zero. This procedure generates equations \eqref{eq_Bogomol} (and \eqref{eq_Bogomol_coor_pol} in polar coordinates) alike to self-dual ones, what enables to determine the energy lower bound of the system \eqref{eq_limite_Bogomol}. Then, the methodology described may be an alternative to calculate this energy value, which has no requesting of stationary condition or imposition of the gauge-fixing to achieve the energy expression. However, so far, it is tested just in one model and certainly there is room for improvements or even discarding it. Besides, the diagonalization of the energy density is a vulnerable step depending on the complexity of the Hamiltonian and maintain differential equations of second order. The unexpected constraint on the background magnetic field $\zeta B^2 = 0$, emerged from the diagonalization of eq. \eqref{energia_GD_const2}, simplified the calculation, although imposed a severe limitation to physical systems, probably turning it in an unfeasible proposition as model for an neutron star scenario.

Some comments related to the field formulations should be done, once one of the purposes of the study is to determine algebraically the vector field from the field equations, assuming, initially , undefined $a_0$ and $a$, both containing radial and angular degrees of freedom. Starting with the function $f(r)$, representing the radial component of the field $\varphi$, it is quantized and presents a divergent behaviour when $r \rightarrow 0$. Such difficult could be circumvented defining boundary conditions for $f(r)$ as practiced in refs. \cite{Bolognesi} and \cite{Casana_PRD} (in magnetic monopoles scenario), for instance, or redefining $\varphi$. Another option, which diverges from one of the premises of the working, is to consider $a_0(r)$, eliminating the angular degree of freedom, and recalculating the Gauss equation \eqref{eq_Gauss_estático}. In this case, $f(r)$ is undefined and $a_0(r)$ is in function of $f(r)$ (a similar result is also observed for an Abelian Chern-Simons-Higgs model in ref. \cite{Bolognesi}). The obtained exponential and the $\theta$ dependency resulting for the gauge scalar potential are unusual in the literature. Moving the debate to the results to $a(r)$, the structure $\sim \frac{1}{r}$ of the solutions in eqs. \eqref{potencial_vetorial_1}, \eqref{potencial_vetorial_2} and \eqref{limite_energia_a} is generally adopted as ansatz in diversified context of the literature as magnetic monopoles \cite{Prasad_Sommerfield}, Maxwell-Chern-Simons model \cite{Ghosh_PRD}, sigma model \cite{Cunha_PRD}, Lorentz violation \cite{Casana_PLB_2013}, are just few examples. In all the scenarios described, the gauge potential $a$ has vorticity and a singularity for short distances. The particular case of $a(r, \theta, n, E)$ \eqref{potencial_vetorial_2} (and also $f(r)$ \eqref{função_f}) presents a richer vorticity structure manifested in generalized Maxwell-Higgs models \cite{Bazeia_EPJ_2017}. The former gauge potential carries an anisotropy originated by the angular dependency and by the background electric field components $E_1$ and $E_2$. The background field anisotropy is not manifested in $a(r, \theta, n)$ \eqref{limite_energia_a}, what incurs in an energy lower bound expression \eqref{limite_energia} absent of them. Nevertheless, one consideration must be done: both $a(r, \theta, n)$ and the energy $\mathcal{E}$ have an anisotropy source from the presence of the $\sin{\theta}$ in their equations.

The equations \eqref{campo_em_Higgs} revealed that, besides the contribution of a magnetic flux, the one from an electric flux and a possibility of writing the lower energy bound in terms of the Poynting vector \eqref{eq_limite_Bogomol_1}. In other hand, the system in \eqref{carga} pointed out a neutral charge. The potential $U$ is algebraically calculated from the self-dual type equations \eqref{eq_Bogomol_coor_pol} and from the results for $a_0$ (not calculated here) and $a$ \eqref{limite_energia_a} in the context of the inferior energy bound. The result is unstable when $r \rightarrow 0 $, what is frequently controlled by imposition of boundary conditions and betaking auxiliary functions \cite{Bazeia_EPJ_2011,Bazeia_EPL,Bazeia_PRD_2012}. It inherit the anisotropy and the quantized characteristic from $a(r, \theta, n)$.

The motivation is to let the mathematics points whether the $\theta$ is feasible or not. In general line, to work with generalized $a_0$ and $a$ tends to a complicated equation system, what imposes a hard task to find the solutions to the equations (which frequently are determined only numerically). At the same time, work with this assumption of generalized gauge potentials, whenever possible, allows the system speak "by itself" and to manifest its inner formation.   


\chapter{Concluding Considerations}
\label{cap_conclusao}
\indent

Throughout the PhD program, an extremely deep theoretical foundation was laid by the researchers, postdocs and guest lectures in our scientific coordination, where a great variety of courses that approached the forefront themes in physics have been delivered. During this period, I collaborated in a workgroup to produce chapters \ref{cap_grav_1} and \ref{cap_grav_2}, and paper \cite{Brito_PRD}. The research for chapter \ref{cap_FW} and the publication of \cite{Ospedal_MPA} was conducted with another researcher; whilst chapter \ref{cap_vortice} was individually produced. This work took form in a complete experience of multifaceted production scenario, collaboration, and working environment. The group of researchers working at the Great "Republic of Diracstan" of CBPF also promoted an inclusive and participative workspace, conducting internal debates, collaborative studies, and experiences, as well as knowledge sharing, which were continuously supported and encouraged by the staff members. The opportunities and frequently encouraged chances of presentation of ongoing or concluded works, equally enriched the supervised researcher’s development.

Observing the theoretical contentment of the present construction, the low dimensional effective theories with non-minimal coupling explored in the chapter \ref{cap_FW} are a wide territory. In the context of the particular study approached in this thesis, there are room for investigation of terms in higher orders of $\mathcal{O}(1/m)$, for diversification of the electrodynamics which describes the external field and for testing the impact of different dimensional reduction methodologies. The calculation of higher order terms in fermionic case, apart from revealing new relativistic interactions, is a consistent checking for the validity of the dimensional reduction result obtained in eq. \eqref{Hamiltoniana_reduzida} and for verification of the matter relevance ($\vec{J} \neq \vec{0}$) in orders beyond $\mathcal{O}(1/m^2)$ in non-minimal relativistic interactions in eq. \eqref{Dirac_hamiltoniana_diagonalizada} when the external electromagnetic field is governed by Maxwell-Chern-Simons electrodynamics (see eqs. \eqref{MCS_equações_de_movimento}). Still in the chapter \ref{cap_vortice}, now turning to the scalar field configuration, considering the surging of a Chern-Simons term when a non-minimal Abelian Higgs model is submitted to a spontaneous symmetry breaking \cite{Paul_PL}, maybe to find out the Hamiltonian density of this resultant Lagrangian and FW transform it, comparing the results with the one achieved in eq. \eqref{Hamiltoniana_diagonalizada_spin0}, or even better for the situation containing terms in order superior to $\mathcal{O}(1/m^3)$, will permit to establish some notices. This analysis is extended to an inspection which quests if the critical Hamiltonian $H_c$ in eq. \eqref{Hamiltoniana_crítica} is maintained in the breaking symmetry scenario mentioned.

In chapters \ref{cap_grav_1} and \ref{cap_grav_2} around effective gravitational quantum theory structured with higher derivatives, naturally comes the perspective of advancing the results considering vertices and/or graviton propagator beyond the tree level. The work developed in the present thesis was a first step in the inter-particle potentials investigation among bosonic and fermionic particles mediated by graviton, in a sense that the calculations involving one-loop quantum corrections are the next step. Other perspective is to develop similar inter-particles potential study involving massive spin-1 particles under gravitational scattering, partially worked by Holstein and Ross \cite{Ross_Arx}. There is one more investigation, the inter-particle potential involving the scattering of two anti-fermions or one fermion and one anti-fermion, which could reveal, theoretically, whether the gravitational potentials for fermions are the same that to anti-fermions (considering that their rest energy (mass) are numerically equal).

An invitation to address a reflection about the interpretation of gravity is proposed. Does the gravity emanate from matter (as a charge) or it is there yet, independently from the matter? Maybe, this is the bifurcation where one way (the emanation concept) leads to interpret gravity as a discrete point (virtual) particle and analyzes it as the mediator of inter-particle potentials between massive (or energy holder) particles. The other way leads to the geometrical consideration (General Relativity), a continuum interpretation of the field, which is deformed in presence of energy. This impasse among discrete and continuum, commented in the presentation section of the thesis, maybe summarize the difficult in conciliate the General Relativity and the Standard Model. It seems a materialism tendency to work with point particles. The Green functions, fundamentally settle in the delta`s Dirac, perform this connection/switch from discrete to continuum description of the Universe in a great accuracy. Even though, the struggle to conciliate unitarity and renormalizability is patent. Then, one could inquiry whether the electromagnetism field is a geometric continuum occupying the Universe, which deforms in a presence of an electric charge, reading the gauge field as an affine connection. 

The chapter \ref{cap_vortice} is fully opened in relation to the directions definitions. A second model with the same Lagrangian expression but carrying a non-constant metric ($\partial_\kappa \bar{G}_{\mu\nu} \neq 0 $) is proposed. The algebraic development of the energy-momentum tensor resulted in a non-conservative system. The condition to attend the Noether's theorem is demanding that $- (D_{\kappa} \, \bar{G}_{\, \mu\nu}) D^{\mu} \varphi^{*} D^{\nu} \varphi = 0$, what violates the premise of the model (non-constant metric). There are other two local gauge invariant models to be explored, in which the modified (non-)constant metric acts on the kinetic and Maxwell terms $\,( \mathcal{L} \sim G_{\mu\nu} \big( -\frac{1}{4}f^{\mu\kappa}f_{\kappa}^{\,\,\, \nu} + D^{\mu} \varphi^{*} D^{\nu} \varphi \big) - U)$. In these cases, it is evaluated the effect of the strong electromagnetic background field on the propagating wave as well. The analysis of the Lagrangian with constant metric is already on the road with the energy density equation calculated. In this scenario, the constraint $\zeta B^2 = 0 $ is present, once the index structure of $G_{\mu\nu} D^{\mu} \varphi^{*} D^{\nu} \varphi$ is the same of the models investigated in this work. A possible alternative to detour it, is a modification in the index arrangement $\, \mathcal{L} \sim G_{\mu\kappa} \big( -\frac{1}{4}f^{\, \kappa\xi}f_{\xi}^{\,\,\, \mu} + D^{\kappa} \varphi^{*} D^{\mu} \varphi \big) - U$. The exploration of these models will probably rise insights about the structure of the gauge scalar and vector potentials, pointing their vorticity and topological behaviour. In these cases, the diagonalization methodology applied in the thesis (eq. \eqref{energia_GD_const3}) keeps validated, as well as the triangular inequality (eq. \eqref{igualdade_d}), once the kinematic scalar sector is unmodified. Then, the anisotropy manifested (see, for instance, eq. \eqref{energia_GD_const_coord_polar1}) will be present in these two models. The modifications will appear in the Bogomol'nyi type equations (see eqs. \eqref{campo_em_Higgs} and \eqref{eq_Bogomol}) and the field equations due to the metric tensor associated to the Maxwell term. 

One of the most relevant lessons and thought-provoking experiences during the PhD program was the publication of the paper \cite{Ospedal_MPA}, which was co-produced with Professor Leonardo Ospedal. We selected its publication to be in a free-of-charge journal, so the Brazilian people would not be encumbered with fees. However, upon the publication of our paper, we confirmed that we were not granted access to the journal and in order for us to be able to access our own work we would need to pay for it. 

In our group of researchers, we follow the practices of publishing every single work, from the preliminary versions up to the final manuscript, in the arXiv repositor and, at the same time, is awaited to submit the results to the appreciation of the community. This procedure grants open access for the community to the whole research developed by the group. Such attitude is widely adopted by researchers, transforming arXiv in a place where is possible to find the forefront and breakthrough works, like the one uploaded by the Russian mathematician Grigori Perelman \cite{Perelman_arx} which awarded him the Fields Medal in 2006, but he never published it in any magazine or journal or even accepted the award, or like the complete Field Theory book \textit{Fields} by Warren Siegel \cite{Siegel_livro}, which material he just published in arXiv. Therefore, one can see arXiv as a start point of the emancipation process from the private publication companies. 

Maybe we should raise the question: are private scientific magazines and journals really indispensable for the promotion of science? It seems important to promote this debate in the scientific community involving researchers and public scientific institutions (and, immediately thereafter, the State and government) with the aim to structure a centralized social system of paper review, analysis and publication that are independent from these private journals and magazines, which, in a general way, set a monopolistic and domineering stage in the world of science.

Researchers must have awareness of this situation in order not to become alienated. Those who are basically, and sometimes almost exclusively, focused on the betterment of their own curriculum vitae (which clearly is a tool of classification, distinction, and social status), may not realize how harmful these publication and impact factor processes are on scientific community, which in turn, subjects itself and abides before the magazines and journals system (that is inherently bind to the productivity system). As a result, providing to the oppressive hands the fruit of our hard labour and the societal investment, granting them the control of the global scientific work.

One of the roles of the researcher is to question and argue if the environment in which they and their labor activities enter are ethic. We should not unquestionably insert and immerse ourselves in a pre-built structure, which is presented and sold as the best and/or only one existent. We need to be aware of the social context of our country and understand that we (as researchers, professors, middle and high economic classes) live a privileged condition, in which the vast majority of our people will not have access to the scientific community due to the sequestration of social inclusion, sequestration of opportunity, sequestration of racial equality, sequestration of gender equality, sequestration of education, sequestration of basic constitutionals rights. A solid foundation of inequality supports the world of researchers and assures our position. We have to step out of this self-centered sense of accomplishment where we believe our achievements are exclusively a result of our own merits and realize that the positions we occupy as researchers only exist because they have been supported by the foundational work of the less fortunate and working class, which find themselves very distant from any possibility and chance of frequenting and accessing "our" positions, academic spaces and of spaces in science in general, those who, due to their level of disadvantage, frequently cannot merely comprehend the specificity of the craft of a researcher. We must unite ourselves, work towards modifying this injustice and we cannot consent and connive with this enormous loot of our nation and to our work as nation. Science, Nature, politics and social principles work and walk together.



\appendix

\chapter{Foldy-Wouthuysen Transformation Review}
\label{apêndice_FWT}
\indent

It expends seventy years Leslie Foldy and Siegfried Wouthuysen contemplated the literature with the transformation methodology that takes their names \cite{FW_PR}. In this paper, which was curiously followed by a José Lopes Leite paper \cite{JJLeite_PR} in the sequence of the magazine, the authors presented an approach to diagonalize Hamiltonians in approximate and, in particulars cases, exact way. The FW Transformation is described in several Chemistry and Physic text books, \textit{e.g.} \cite{Itzykson_livro,Schwabl_livro,Reiher_livro}, thus it is possible to find a wide number of discussions and developed applications. With the spirit of facilitate and make more comfortable the reading of the Thesis, the FW transformation procedure is detailed below, in a way to possibility an eventual consultation by the reader.

The central idea of the FW transformation is to interpret it as an unitary transformation on a Hamiltonian $H$ which permits to decouple equations in a linear system, in order to avoid mixed terms and, in parallel, has the potential to reveal new terms and to establish corrections of higher order. Considering the Wigner principle that demands the eigenfunctions be (anti)unitarity transformed in a way to assure the conservation of the probability amplitudes of the eigenstates, one establishes a canonical unitary transformation on an eigenfunction $\psi$

\begin{equation}
    \psi' = e^{-iS} \psi \, ,
\label{transformada_psi}
\end{equation}

\noindent
where $S$ is a Hermitian operator. Then, substituting the expression \eqref{transformada_psi} in a Schröndiger Equation

\begin{eqnarray}
     i \partial_{t} (e^{-iS} \psi) = H (e^{-iS} \psi) \nonumber \\
     i e^{-iS} \partial_{t}\psi + (i \partial_{t} \, e^{-iS}) \psi = H (e^{-iS} \psi) \nonumber \\
     (e^{iS} \, \times) \,\,\,\,\,\,\,\, i \partial_{t}\psi  = - e^{iS} ( i \partial_{t} \, e^{-iS} ) \psi + e^{iS} H ( e^{-iS} \psi) \nonumber \\
     i \partial_{t}\psi  = e^{iS}(H - i \partial_{t}) \, e^{-iS}  \psi \, ,
\label{Hamiltoniana_transformação_eq_Schröndiger}
\end{eqnarray}

\noindent
one obtains the Hamiltonian transformed $H'$

\begin{equation}
    H' = e^{iS} ( H - i\partial_{t} ) e^{-iS} \,.
\label{Hamiltoniana_transformada}
\end{equation}

\noindent
The apparent exponential in eq. \eqref{Hamiltoniana_transformada} is passive of being expanded through the Baker–Hausdorff identity 

\begin{equation}
    e^{A} \, B \, e^{-A} = B + [A,B] + \frac{1}{2} [A,[A,B]] + \frac{1}{6} [A, [A, [A,B]]] + ... + \frac{1}{n!} [A, ...[A, [A,B]]...] \, ,
\label{identidade_Baker–Hausdorff}
\end{equation}

\noindent
with $[A,B]$ meaning the commutator of the operators $A$ and $B$, resulting in

\begin{equation}
    H'= H + i[S,H] + \frac{i^2}{2} [S, [S,H]] + \frac{i^3}{6} [S, [S, [S,H]]] + ... -( \Dot{S} + \frac{i}{2} [S, \Dot{S}] + \frac{i^2}{6} [S, [S, \Dot{S}]] \, ... \, ) ,
\label{expansão_Hamiltoniana}
\end{equation}

\noindent
noticing that the dotting above the operators, like $\Dot{S}$, stands for the time derivative of the operator $S$. The substitution of the Hamiltonian $H$ in the eq. \eqref{expansão_Hamiltoniana} is preceded by one step. The Hamiltonian must be organized in a particular form

\begin{equation}
    H = \beta m + \mathcal{E} + \mathcal{O} \, ,
\label{Hamiltoniana_operadores_par_ímpar}
\end{equation}

\noindent
where $\beta$, $\mathcal{E}$ and $\mathcal{O}$ are operators which attend to the (anti)commutation relations $\beta\mathcal{E}=\mathcal{E}\beta$ and $\beta\mathcal{O}= - \mathcal{O}\beta$, with $\mathcal{E}$ meaning "even" operators and $\mathcal{O}$ "odd" ones. This arrangement (in eq. \eqref{Hamiltoniana_operadores_par_ímpar}) is fundamental to achieve the diagonilized Hamiltonian, once the products $\mathcal{E}^2$ and $\mathcal{O}^2$ result in even operators, while $\mathcal{E} \mathcal{O}$ and $\mathcal{O} \mathcal{E}$ in odd ones. $\beta$ is usually represented as a diagonal matrix and $m$ stands for the mass. The generator $S$ is defined as

\begin{equation}
    S = -i \, \frac{\beta \mathcal{O}}{2m}  \, ,
\label{operador_S}
\end{equation}

\noindent
in which the presence of $\mathcal{O}$ performs the conversion of odd operators in even ones, diagonalizing the Hamiltonian order by order, where the orders are in terms of the mass inverse. Substituting eqs. \eqref{Hamiltoniana_operadores_par_ímpar} and \eqref{operador_S} in eq. \eqref{expansão_Hamiltoniana}, one obtains the Hamiltonian

\begin{equation}
    H'= \beta m +\mathcal{E} + \beta \frac{\mathcal{O}^2}{2m}  -\frac{1}{8m^2}[\mathcal{O},[\mathcal{O},\mathcal{E}] + i\mathcal{\Dot{O}}] - \beta \frac{\mathcal{O}^4}{8m^3} +\frac{\beta}{2m}[\mathcal{O},\mathcal{E}] -\frac{\mathcal{O}^3}{3m^2} +\frac{i\beta \mathcal{\Dot{O}}}{2m} ...\, ,
\label{Hamiltoniana_diagonalizada}
\end{equation}

\noindent
in which the highest order of odd terms is $m^{-1}$. Therefore, $H$ is diagonalized until order $m^0$ and written as $H'= \beta m +\mathcal{E}' + \mathcal{O}'$, in a way that the terms present in eq. \eqref{Hamiltoniana_diagonalizada} are passive of being regrouped in one even operator $\mathcal{E}'$, holding the odd powers of $\mathcal{O}$, and one odd $\mathcal{O}'$

\begin{subequations}
  \begin{align}
  \mathcal{E}'= \mathcal{E} + \beta \frac{\mathcal{O}^2}{2m}  -\frac{1}{8m^2}[\mathcal{O},[\mathcal{O},\mathcal{E}] + i\mathcal{\Dot{O}}] - \beta \frac{\mathcal{O}^4}{8m^3} ... \, , 
  \label{operador_par_iterado}
  \end{align}
  
  \begin{align}
  \mathcal{O}'= \frac{\beta}{2m}[\mathcal{O},\mathcal{E}] -\frac{\mathcal{O}^3}{3m^2} +\frac{i\beta \mathcal{\Dot{O}}}{2m} ... \, ,
  \label{operador_ímpar_iterado}
  \end{align}
  
\noindent
and, consequently, a new $S$ operator is determined

  \begin{align}
  S' = -i \, \frac{\beta \mathcal{O}'}{2m} \, .
  \label{operador_S_iterado}
  \end{align}
\label{operadores_iterados}
\end{subequations}

\noindent
Once defined the new operators $\mathcal{E}'$, $\mathcal{O}'$ and $\mathcal{S}'$ - eqs. \eqref{operadores_iterados} -, the Hamilotnian $H'$ is submitted to the unitary transformation in eq. \eqref{Hamiltoniana_transformada}, what results in a Hamiltonian $H''$ diagonalized until order $m^{-1}$

\begin{equation}
    H''= \beta m +\mathcal{E'} + \beta \frac{\mathcal{O'}^2}{2m}  -\frac{1}{8m^2}[\mathcal{O'},[\mathcal{O'},\mathcal{E'}] + i\mathcal{\Dot{O'}}] - \beta \frac{\mathcal{O'}^4}{8m^3} +\frac{\beta}{2m}[\mathcal{O'},\mathcal{E'}] -\frac{\mathcal{O'}^3}{3m^2} +\frac{i\beta \mathcal{\Dot{O'}}}{2m} ...\, ,
\label{Hamiltoniana_diagonalizada_2}
\end{equation}

\noindent
indeed, the odd term with the lowest order in $H''$, $\,  \frac{\beta}{2m}[\mathcal{O'},\mathcal{E'}] \,$, is a term of order $O(1/m^{2})$. Then, from $H''$, new even, odd and $\mathcal{S}$ terms are defined, namely $\mathcal{E}'''$, $\mathcal{O}'''$ and $\mathcal{S}'''$, and the process may be repeated $n+1$ times till the odd components achieve the desired $(n+1)^{th}$ order to $m^{-1}$, resulting in a Hamiltonian diagonalized up to order $O(1/m^{n})$.

In the present work is considered the Hamiltonian diagonalized until the order $O(1/m^{3})$

\begin{equation}
  H_{FW} \approx \beta m +\mathcal{E} + \beta \frac{\mathcal{O}^2}{2m}  -\frac{1}{8m^2}[\mathcal{O},[\mathcal{O},\mathcal{E}] + i\mathcal{\Dot{O}}] - \beta \frac{\mathcal{O}^4}{8m^3} \, .
\label{Hamiltoniana_diagonalizada_final}
\end{equation}

\noindent
This formulation for the diagonalized Hamiltonian is the one applied to all cases investigated in the present Thesis. Since the scalar to fermionic systems or from three to four spacetime dimensions, the general structure of the $H_{FW}$ is that expressed in \eqref{Hamiltoniana_diagonalizada_final}. What varies is the definition of the operators and the dimensional adjustments, basically manifested through matrices dimensionality and their algebra.


\chapter{Integrals}
\label{apêndice_int}
\indent

The integrals indicated along the chapters \ref{cap_grav_1} and \ref{cap_grav_2} are 

\begin{equation}
 I^{(a)}_n (r) = \int \frac{d^3\vec{q}}{(2 \pi)^3} \, \frac{ e^{i \vec{q} \cdot \vec{r}}   }{  (\vec{q}^{\,2})^n \, Q_a} \, , 
\label{apêndice_I_a_n} 
\end{equation}
 
\begin{equation}
  I^{(a)}_{ij} (r) = \int \frac{d^3\vec{q}}{(2 \pi)^3} \, \frac{ e^{i \vec{q} \cdot \vec{r}}  }{  \vec{q}^{\,2} \, Q_a} \, \vec{q}_i \vec{q}_j \,,
\label{apêndice_I_ij} 
\end{equation}

\noindent
where $n \in \mathbb{N}$ and $ a=0,2$. For Fourier transform of the type eq. \eqref{apêndice_I_a_n} is possible to solve its angular part passing the function $Q_a(\vec{q}^2)$ to the spherical coordinate system, leaving only radial dependence (see, for instance, ref. \cite{Accioly_PRD_2016}). This artifice enables one to recast some integrals, for example:

\begin{equation}
  \int \frac{d^3\vec{q}}{(2 \pi)^3}  \, \frac{ e^{i \vec{q} \cdot \vec{r}}  }{ \vec{q}^{\,2} \, Q_a} \,  i \vec{A} \cdot \vec{q} =
  \vec{A} \cdot \vec{\nabla} \left[ I^{(a)}_1 (r) \right] = \vec{A} \cdot \frac{\vec{r}}{r} \, \frac{d}{dr} \, I^{(a)}_1 (r) \, ,
\label{apêndice_int_A} 
\end{equation}
being $\vec{A}$ a vector independent of $\vec{q}$.


\end{document}